\documentclass[11pt]{article}
\usepackage{amsmath,amssymb,color,epsfig,cite}
\usepackage{mathrsfs}

\usepackage{mathtools}
\usepackage{datetime}

\newdateformat{monthyeardate}{%
  \monthname[\THEMONTH], \THEYEAR}

\usepackage{amsfonts}
\newcommand{\hoch}[1]{$\, ^{#1}$}

\makeatletter
\@addtoreset{equation}{section}
\makeatother

\newcommand{\be}{\begin{equation}}
\newcommand{\ee}{\end{equation}}
\newcommand{\bea} {\begin{eqnarray}}
\newcommand{\eea}{\end{eqnarray}}
\newcommand{\nn}{\nonumber}

\def\ft#1#2{{\textstyle{\frac{\scriptstyle #1}{\scriptstyle #2} } }}
\def\fft#1#2{{\frac{#1}{#2}}}

\def\0{{\sst{(0)}}}
\def\1{{\sst{(1)}}}
\def\2{{\sst{(2)}}}
\def\3{{\sst{(3)}}}
\def\4{{\sst{(4)}}}
\def\5{{\sst{(5)}}}
\def\6{{\sst{(6)}}}
\def\7{{\sst{(7)}}}
\def\8{{\sst{(8)}}}
\def\sst#1{{\scriptscriptstyle #1}}
\def\oneone{\rlap 1\mkern4mu{\rm l}}

\def\del{{\partial}}

\def\cA{{{\cal A}}}
\def\cF{{{\cal F}}}
\def\cG{{{\cal G}}}

\def\wtd{\widetilde}

\def\bx {{\bf x}}
\def \M {M_{\rm irr}}
\def \C{ \gamma_{\rm extreme}}

\def\arctanh{{\rm \,arctanh\,}}
\def\ie{{ i.e.~}}

\def\scri{{\mathscr{I}}}

\begin{document}

\begin{flushright}
\hfill { UPR-2091-T\ \ \ MI-TH-1886
}\\
\end{flushright}

\begin{center}
{\large {\bf Killing Horizons: Negative Temperatures and
  Entropy Super-Additivity}}

\vspace{15pt}
{\large M. Cveti\v c$^{1,2}$, G.W. Gibbons$^3$, H. L\"u$^4$ and
              C.N. Pope$^{5,3}$ }

\vspace{15pt}

{\hoch{1}\it Department of Physics and Astronomy,\\
University of Pennsylvania, Philadelphia, PA 19104, USA}

\vspace{10pt}

{\hoch{2}\it Center for Applied Mathematics and Theoretical Physics,\\
University of Maribor, SI2000 Maribor, Slovenia}

\vspace{10pt}

\hoch{3}{\it DAMTP, Centre for Mathematical Sciences,
 Cambridge University,\\  Wilberforce Road, Cambridge CB3 OWA, UK}

\vspace{10pt}

\hoch{4}{\it Department of Physics,\\
Tianjin University, Tianjin 300350, China}

\vspace{10pt}

\hoch{5}{\it George P. \& Cynthia Woods Mitchell  Institute
for Fundamental Physics and Astronomy,\\
Texas A\&M University, College Station, TX 77843, USA}

%\vspace{30pt}

%\underline{ABSTRACT}

\end{center}

%\today

%\vfill {\footnotesize Emails: }

%\thispagestyle{empty}

\begin{abstract}
Many discussions in the literature
of spacetimes with more than one Killing horizon note
that some horizons have positive and some have negative surface gravities,
but  assign to all a positive
temperature. However, the first law of thermodynamics then takes a
non-standard form. We show that if one regards
the Christodoulou and Ruffini formula for the total energy or enthalpy
as
defining the Gibbs surface, then the rules of Gibbsian thermodynamics imply
that negative temperatures arise inevitably on inner horizons, 
as does the conventional
form of the first law.  We provide many new examples of this phenomenon, 
including
black holes in STU supergravity.  We also give a discussion of left and right
temperatures and entropies, and show that both the left and right 
temperatures are non-negative.  The left-hand sector contributes
exactly half the total energy of the system, and the right-hand sector
contributes the other half.  Both the sectors satisfy conventional
first laws and Smarr formulae.  For spacetimes
with a positive cosmological constant, the cosmological horizon is
naturally assigned a negative Gibbsian temperature.  We also explore 
entropy-product formulae and a novel entropy-inversion formula,
and we use them to test whether the entropy is a super-additive
function of the extensive variables.  We find that super-additivity
is typically satisfied, but we find a counterexample for dyonic
Kaluza-Klein black holes.

\end{abstract}

\pagebreak

\tableofcontents
\addtocontents{toc}{\protect\setcounter{tocdepth}{2}}

%%%%%%%%%%%%%%%%%%%%%%%%%%%%%%%%%%%%%%%%

%\newpage
%%%%%%%%%%%%%%%%%%%%%%%%%%%%%%%%%%%%%%%%
%\documentclass[preprint,showpacs,preprintnumbers,
%  amsmath,amssymb,nofootinbib]{revtex4}

\def\half{\frac{1}{2}}
\def\ben{\begin{equation}}
\def\bea{\begin{eqnarray}}
\def\een{\end{equation}}
\def\eea{\end{eqnarray}}
\def \bp {{\bf p}}
\def \bv {{\bf v}}
\def \bs {{\bf s}}
\def\bt{{\bf t}}

\def \p {\partial}
\def \cL {{\cal  L}}
\def \cG {{\cal  G}}
\def \cLEG {{\cal LEG}}

\def\ft#1#2{{\textstyle{\frac{\scriptstyle #1}{\scriptstyle #2} } }}
\def\fft#1#2{{\frac{#1}{#2}}}

%%%%%%%%%%%%%%%%%%%%%%%%%%%%%%%%%%%%%%%%%%%%%%%%%%%%%%%%%%%%%%%%%%%%%%%%%%%%%%
\section{Introduction}

Since the early days of black hole thermodynamics
there have been suggestions that the thermodynamic of
the inner, Cauchy,  horizons of charged and or  rotating black holes
should be taken more seriously than it has been
\cite{Curir,CFrancaviglia1,CFrancavilia2,CalvaniFrancaviglia,CFrancaviglia3,Curir:1985wn,Curir:1986irp,CurirC,Curir:2011zza,OK}.
With the development of String Theory approaches these suggestions
have become more insistent
\cite{CYI,CYII,CveticTseytlin,Larsen,CLII,CLIII,Castro:2010fd}.
This interest increased considerably with the observation
that the product of the areas and hence entropies of the
inner and outer horizon takes in many examples a universal
form which should be quantised at the quantum level
\cite{Cvetic:2010mn,Visser:2012wu,Cvetic:2013eda,Page:2015gia,Castro:2012av}.
Some of these papers, and others, for example 
 \cite{sqwuneg,huan,mipark1,mipark2,mipark3,Castro:2012av}, encountered
the same feature first noticed in \cite{Curir}:
the  fact that with a conventional first law of thermodynamics the
temperature of the inner horizon would be negative.  The authors of
\cite{Castro:2012av}  chose to
resolve this issue by defining the temperature of the inner horizon
to be the absolute value of the ``thermodynamic'' temperature, and
proposing an appropriately-modified first law on the inner horizon to
compensate for this.    In this paper we shall explore the consequences
of adhering to the standard first law of thermodynamics for inner
horizons, with the inevitable result that the temperature will be
negative there.

   In the derivation of the
first law of black hole dynamics one finds, integrating in the
region between the inner and outer horizons, that
%%%%%
\be
0 = \fft{\kappa_+}{8\pi}\, dA_+ -\fft{\kappa_-}{8\pi}\, dA_-
+\cdots\,,\label{fl}
\ee
%%%%%
where $\kappa_\pm$ are the surface gravities and $A_\pm$ the areas of
the outer and inner horizons respectively.
(The contributions from the angular momentum and charge(s) are represented
by the ellipses in this equation.)  If, as turns out to be the case in the
examples we consider, the signs of $dA_+$ and $dA_-$ are opposite for a
given change in the black-hole parameters, then the signs of the surface
gravities at $\kappa_+$ and $\kappa_-$ must be opposite too.
The surface gravity is defined by
evaluating
%%%%%
\be
\ell^\mu\,\nabla_\mu\, \ell^\nu = \kappa\, \ell^\nu\label{kappadef}
\ee
%%%%%
on the horizon,
where $\ell^\mu$ is the future-directed null generator of the horizon,
which coincides with a Killing vector $K^\mu$
on the horizon.
One then finds that whilst $\kappa$ is positive on the outer horizon,
it is negative on the inner horizon.\footnote{For example, in a static
metric $ds^2=-h(r)\, dt^2 + dr^2/h(r) + r^2\, (d\theta^2 + \sin^2\theta\,
d\phi^2)$ one finds (after changing to a coordinate system that
covers the horizon region) from (\ref{kappadef}) that if $K=\del/\del t$ then
$\kappa=\ft12 dh/dr$, which is of the form of the negative of the gradient of
the gravitational potential, evaluated on the horizon.  If
$h=(r-r_+)(r-r_-)/r^2$, as in the
Reissner-Nordstr\"om metric, then $\kappa_+=(r_+ -r_-)/(2 r_+^2)>0$,
while $\kappa_-= -(r_+-r_-)/(2 r_-^2) <0$. In general, of course, the slope of
$h(r)$ must always have opposite signs at two adjacent zeros,
and thus the surface gravities must have opposite signs.} Hawking showed that
for an isolated event horizon in an asymptotically flat spacetime (for
which in fact $\kappa$ is positive), the temperature is
$\kappa/(2\pi)$.  We shall discuss the extension of Hawking's calculation
to the case of inner horizons in the concluding section of this paper.
In what follows, however, we shall frequently make reference to the formula
%%%%%
\be
T= \fft{\kappa}{2\pi}\,,\label{Tfromkappa}
\ee
%%%%%
with the understanding that $T$ may not be a temperature measured by a
physical thermometer, but rather, as we shall explain shortly, a
``Gibbsian'' temperature.

%Now entropy, being $\log\Omega$ (where $\Omega$ is the number of microstates
%consistent with a given macrostate), is, like the horizon area $A$,
%necessarily non-negative.

   The occurrence of
a negative $\kappa$ on an inner horizon is
somewhat obscured in many discussions in the literature by the fact that the
surface gravity is commonly calculated by evaluating
%%%%%
\be
\kappa^2 =-\fft{g^{\mu\nu}\, (\del_\mu K^2)(\del_\nu K^2)}{
  4 K^2}\label{kappa2}
\ee
%%%%%%
in the limit on the horizon.
This formula is derivable from (\ref{kappadef}), but the
information about the sign of $\kappa$ is lost, and commonly the positive
root is assumed when calculating $\kappa$ from (\ref{kappa2}). A guaranteed
correct procedure is to use the formula (\ref{kappadef}), working in a
coordinate system that covers the horizon region.

Another situation where one  encounters two horizons
is when a positive cosmological constant $\Lambda$ is involved
and one has both a black hole event horizon and a  cosmological event horizon
bounding a static or stationary region \cite{Gibbons:1977mu}.
A number of recent studies have pointed out that the surface gravities
of the black hole horizon $\kappa_B$
and the cosmological event horizon
$\kappa_C$ again have opposite signs
\cite{Frolov:2002va,Dolan:2013ft,Gregory:2017sor}.
Most have followed the procedure adopted  in \cite{Gibbons:1977mu} and taken
the physical temperature
to be $\frac{|\kappa|}{2 \pi}$ (for example, see \cite{Bousso:1996au}).
A similar situation arises in the case of the C-metrics, which contain
both a black-hole horizon and an acceleration horizon.  Their surface
gravities are of opposite signs.

%While this seems the physically correct
%it nevertheless leads at the formal level to some awkward looking
%formulae.

   In order to assess the status of these suggestions, in this paper we
shall re-examine the foundations of classical black hole thermodynamics from
the viewpoint of the approach to classical thermodynamics advocated by
Gibbs \cite{Gibbs}.  The central idea of this approach
is that the physical properties of a substance are encoded into the shape
of its Gibbs surface, i.e. the surface given by regarding the height of
the surface as given by the total energy, regarded as a function of the
remaining extensive variables. From this point of view, the temperature is
given by the slope of the curve of energy versus entropy.  To this end, we
shall need explicit Christodoulou-Ruffini formulae, and a major goal of
this paper is to obtain these for a variety of black hole solutions.  As we
shall see, it is a common feature of these examples that the
``Gibbsian temperature'' thus defined, while positive for black hole event
horizons, is negative for inner horizons (i.e.~Cauchy horizons) and for
cosmological horizons.  We shall return to a discussion of the physical
consequences for spacetimes with two horizons in the conclusions.

  Letus recall that the formalism  of  thermodynamics, applied to 
classical black holes,
began with two independent discoveries:

\begin{itemize}

\item Christodoulou's concept of reversible and irreversible
transformations such that the energy $E$ of a rotating black hole
of angular momentum ${\bf J}$ and momentum ${\bf P}$   may be expressed as
\ben
M^2 = M_{\rm irr}^2 + {\bf P}^2  + \frac{{\bf J} ^2} {M^2_{\rm irr}}\,,
  \label{Christ}
\een
where the irreducible mass $M_{\rm irr}$ is non-decreasing
\cite{Christodoulou}
\item
Hawking's Theorem \cite{Hawking:1971tu,Hawking:1971vc} that the area $A$ of
the event horizon is non-decreasing.
\end{itemize}

In fact
%%%%%
\ben
A= 16 \pi M_{\rm irr}^2\,,
\een
%%%%%
and for charged rotating Kerr-Newman black holes and dropping
the momentum contribution and setting
$J=|{\bf J}|$,  one has \cite{Christodoulou:1972kt}
the Christodoulou-Ruffini formula:
%%%%%
\ben
M^2 = \Bigl( M_{\rm irr} + \frac{Q^2}{4 M_{\rm irr}}
\Bigr )^2 +  \frac{ J ^2} {M^2_{\rm irr}} \,. \label{CRFormula}
\een
%%%%%
The obvious analogy of some multiple of the area of the horizon
with entropy became even more striking with
the discovery by Smarr \cite{Smarr:1972kt} of an analogue of the Gibbs-Duhem
relation for homogeneous substances. For Kerr-Newman black holes, this
reads
%%%%%
\ben
M= \frac{1}{4 \pi} \kappa A + 2\Omega J + \Phi Q \label{Smarr} \,,
\een
%%%%%
where $\kappa$ is the surface gravity,
$\Omega$ the angular velocity and $\Phi$ the electrostatic
potential of the event horizon.  The analogy
became almost complete
with the the formulation of three laws of black hole mechanics,
including the first law
%%%%%
\ben
dM=  \frac{1}{8 \pi} \kappa d A + \Omega d J + \Phi dQ  \label{firstlaw} \,,
\een
%%%%%
by Bardeen, Carter and Hawking \cite{Bardeen:1973gs}.
Note that the Smarr relation (\ref{Smarr})
follows from the first law (\ref{firstlaw})
by differentiating the weighted homogeneity relation
%%%%%
\ben
M(\lambda^2 A,\lambda ^2 J, \lambda Q)= \lambda M \label{scale}
\een
%%%%%
with respect to $\lambda$ and then setting $\lambda=1$.

    The existence of a ``first law'' is not in itself surprising,
nor does it, in itself, imply any thermodynamic consequences.  Whenever
one has a problem involving varying a function subject to some constraints,
and considering the value of the function at critical points, one has
a formula analogous to (\ref{firstlaw}).  In the case of black hole
solutions of the Einstein equations, they are known to satisfy a variational
principle in which the mass is extremised keeping the horizon area,
angular momenta
and charges fixed (see, for example, \cite{hawkingvar,dain,ancvpa}).
Similar formulae arise in the theory of rotating stars (see,
for example, \cite{Bardeen:1970vja}). The study of these variations is
sometimes referred to as comparative statics.

For homogeneous substances with pressure $P$, volume $V$ and
internal energy $U$, it is well known that
the Gibbs-Duhem relation is equivalent to the statement
that the Gibbs free energy, or thermodynamic potential,
%%%%%
\ben
G=U-TS+PV\,,
\een
%%%%%
vanishes identically.
For black holes the Smarr relation (\ref{Smarr}) implies that
%%%%%
\ben
G=TS+\Omega J \,.
\een
%%%%%

\emph{Classically}, a number of
arguments led to the conclusion that the laws of black hole mechanics
were just analogous to the laws of thermodynamics.
One argument was that as perfect absorbers, classical black holes should
have vanishing temperature
and hence the entropy should be infinite (cf. \cite{Geroch,Bekenstein:1972tm}).
Another was based on dimensional reasoning.  In units where Boltzmann's
constant is taken to be unity, entropy is dimensionless,
but in classical general relativity it is not obvious how to achieve this
without introducing a unit of length. The obvious guess
for entropy would be some multiple of the area $A$, but why not
some monotonically increasing function of the area?  Despite these
doubts it was conjectured by Bekenstein \cite{Bekenstein:1972tm}
that when quantum mechanics is taken into account
some multiple of $\frac{A}{l_p^2}$ should
correspond to the physical entropy of a black hole.
This conjecture was subsequently confirmed at the semi-classical level by
by Hawking \cite{Hawking:1974rv,Hawking:1974sw},
using quantum field theory in a curved
background. Given this, one recognises the Christodoulou-Ruffini formula
(\ref{CRFormula})
in the form
%%%%%
\ben
M=M(S,J,Q)
\een
%%%%%
as an explicit expression for the analogue of the
Gibbs surface $U=U(S,V)$ for a homogeneous substance.

  To summarise, the purpose of the present paper is to re-examine these issues
systematically, based on Gibbs's geometric viewpoint
of the mathematical formalism  of thermodynamics \cite{Gibbs}.
This starts with
a choice of pairs of extensive and intensive variables
and an expression  for some sort of  ``energy,''  which is regarded
as a function of the extensive variables. For the black holes in
asymptotically
flat spacetimes that we shall consider, the energy is taken to be
the ADM mass $M$,
and the extensive variables $S^\mu $ are usually taken to be
$S^\mu=(S,J,Q_i, P^i) =(S,\bs)$, where
$S = \frac{1}{4} A$ and
$A$ is  the area of the event horizon, $J$ is the total angular momentum
and $Q_i$ and $P^i$  are  $2N$ electric and
magnetic charges.\footnote{We shall not consider scalar charges and moduli
\cite{Gibbons:1996af} in this paper.}
Thus we have
%%%%%
\ben
M=M(S,J, Q_i, P^i) = M(S^\mu)  \,.  \label{Gibbs}
\een
%%%%%
The intensive variables are taken to be
$T_\mu =\frac{\p M}{\p S^\mu}=(T,\Omega, \Phi^i, \Psi_i) = (T,\bt)$
where  $T$ is the temperature,
$\Omega$ is the angular velocity of the horizon,
and $\Phi^i$  and $\Psi_i$ are the electrostatic and magnetostatic
potentials.

   The organisation of this paper is as follows.  In section 2, we review
the theory of Gibbs surfaces, and the various thermodynamic metrics with
which they may be equipped.  Section 3 forms the core of the paper.  In it, we
give many new results for the thermodynamics of a wide range of
asymptotically-flat black holes.
We begin in subsections 3.1, 3.2 and 3.3 by reviewing how the well-known
Reissner-Nordstr\"om, Kerr and Kerr-Newman black holes fit into the Gibbsian
framework.  Subsection 3.4 then provide a extensive discussion of the
thermodynamics of families of black holes in four-dimensional STU
supergravity. In particular, we give a systematic discussion of the 
notion of the decomposition of the system into left-handed and right-handed 
sectors, and their associated thermodynamics. 
Subsection 3.5 has analogous results for five-dimensional
STU supergravity black holes.  Subsections 3.6 and 3.7 give similar
results for the general family of four-dimensional
Einstein-Maxwell-Dilaton (EMD)
black holes, and a two-field generalisation.  Included in the discussion of
these two-field EMD black holes, we exhibit a new area-product formula.

   A rather general feature of many
asymptoticaly flat black holes with two horizons is
that the product of the areas of the two horizons is independent of the
mass, and given in terms of conserved charges and angular momenta, which
may plausibly be quantised at the quantum level.  In section 4, we use
this area-product formula to exhibit an intiguing $S\rightarrow 1/S$
inversion symmetry of
the Christodoulou-Ruffini formulae for such black holes.  This symmetry
of the Gibbs surface interchanges the positive and negative temperature
branches.

   In section 5 we extend our discussion to black holes that are asymptotically
AdS, or black holes with positive cosmological constant.  In the AdS case the
situation for inner and outer horizons is broadly similar to that for the
asymptotically flat case.  For positive cosmological constant, the
black hole event horizon continues to have positive Gibbsian temperature, but
that of the cosmological horizon is negative.

   In section 6, we revisit an old observation, that the entropy of the
Kerr-Newman solution is a super-additive function of the extensive variables,
and we its relation to Hawking's area theorem
for black-hole mergers.
We find that super-additivity  
holds also for a wide variety of the asymptotically-flat
examples that we considered in section 3.  However, we find that
Kaluza-Klein dyonic black holes provide a counterexample, and we speculate on
the reason for this.

  The paper ends with conclusions and future prospects in section 7.

\section{The Gibbs Surface and Thermodynamic Geometry}

\subsection{The Gibbs surface}

In this section we shall briefly summarise
those aspects of the Gibbs surface which are relevant for the
latter part of the paper.
If we think of $(S^\mu, M)$ as coordinates   in $\mathbb{R}^{3+2N}$
then (\ref{Gibbs}) defines a hypersurface ${\cal G} \subset \mathbb{R}^{3+2N}$
whose co-normal
is $  (T_\mu , -1)$. Since in our case  $M$ is  a unique function
of the extensive variables,  the intensive variables are
unique functions of the extensive variables: $T_\mu=T_\mu(S^\nu)$.
The converse need not be true. If the function $M(S^\mu)$
were  convex, then for fixed  co-normal $(T_\mu , -1)$  the plane
%%%%%
\ben
T_\mu  S^\mu   -M =0
\een
%%%%%
would touch the surface at a unique point $(S^\mu,M)$ .
For a smooth Gibbs surface ${\cal G}$,  convexity requires that
the Hessian
\ben
 g^W_{\mu \nu}= \frac{\p^2 M}{\p S^\mu \p S^\nu} \label{Weinhold}
\een
be positive definite and one may then define a positive definite metric
\ben
ds^2 =  g^W_{\mu \nu} dS^\mu dS^\nu \,,
\een
called the Weinhold metric \cite{Weinhold}. Because one of the components
of the Weinhold metric (\ref{Weinhold})  is related to  the
heat capacity\footnote{We use the term ``heat capacity'' rather than
``specific heat''
 because  the latter is  defined to be per unit mass.}
at constant $J$ and $Q^i$ and $P^i$, namely
\ben
g^W_{SS} = T C^{-1}_\bs =   \frac{\p T}{\p S} \Bigr |_ \bs\,,
\een
and  neutral black holes or black holes with
small charges or angular momentum  have negative heat capacities,
the Gibbs surface ${\cal G}$ is typically not convex and the Weinhold
metric for black holes is typically Lorentzian \cite{Pagepd}.

If one defines a  totally symmetric co-covariant tensor of rank three
by
\ben
C_{\lambda \mu \nu }= \frac{\p^3 M}{\p S^\lambda \p S^\mu \p S^\nu}\,,
\een
the Riemann and Ricci tensors  and the Ricci scalar of the Weinhold metric
are given by
\bea
{R^\alpha \,}_{\beta \mu \nu} &=& -\frac{1}{4} \Bigl
[{C^\alpha\,}_{\mu \lambda} {C ^\lambda\,}_{\nu \beta} -
  { C^\alpha\,}_{\nu \lambda} {C ^\lambda\,}_{\mu \beta} \Bigr ]\,,
\nonumber \\
R_{\beta \nu}  &=&   -\frac{1}{4} \Bigl [ {C^\alpha \,}_{\alpha \lambda }
  {C^\lambda} _{\beta \nu} -{C^{\alpha \lambda}\,}_\nu  C_{\lambda \alpha \beta}
\Bigr ]\,, \nonumber
\\
R &= & - \frac{1}{4} \Bigl [ {C^\alpha \,}_{\alpha \lambda } {C^{\lambda \nu}\,}_\nu
    - C^ {\alpha \nu \lambda }  C_ {\alpha \nu \lambda }  \Bigr ] \,,
\label{Hessian}
\eea
all indices being raised with  $g_W ^{\mu \nu}$,
the inverse of $g^W_{\mu \nu}$.  Divergences in $R$ are sometimes held to
be a diagnostic for phase transitions.

  The geometry of the Gibbs surface is essentially the geometry behind the
first law of thermodynamics.  As we remarked previously, this fits into a
pattern that is more general than just the theory of black holes, and arises
whenever one is considering a variational problem with constraints.  Since
this is not as widely known as it deserves to be, we shall pause to describe
the general situation, and then we shall restrict attention to its
application to black hole theory.   Consider a real-valued function $f(x)$
on some space $X$ with coordinates $x$, subject to the $n$ constraints that
certain functions $C^a(x)=c^a$, $1\le a\le n$, where the $c^a$ are constants.
Adopting the method of Lagrange multipliers, we require that
%%%%%
\be
 \delta f - \lambda_a \, \delta C^a =0\,,\label{fvar}
\ee
%%%%%
for all variations in $X$.  Suppose the solutions of these equations lie
in an $n$-dimensional sub-manifold $S$ of $X$, parameterised by the values of
the constraints, $c^a$.  One may restrict the variations in (\ref{fvar})
to directions within the solution space $S$, in which case we obtain the
formula
%%%%%
\be
\delta f(c) = \lambda_a\,\delta c^a\,.\label{fvarc}
\ee
%%%%%
Geometrically, we can think of this situation as follows.  We construct a
$(2n+1)$-dimensional space with coordinates $(f,\lambda_a, c^a)$.  Since,
locally at least, $f$ and $\lambda_a$ may be thought of as functions of the
$c^a$, we obtain an $n$-dimensional surface in this space.
  From (\ref{fvarc}), it follows that the Lagrange multipliers $\lambda_a$
are determined by the tangent planes to the surface.  From now on, we
shall restrict attention to the thermodynamic case, with $f$ being the
total energy, or mass, $M$, and the $c^a$ being $(S,J,Q_i,P^i)$.

   The Gibbs surface ${\cal G}$ can be lifted
to give a $(3+2N)$-dimensional submanifold
${\cal LA} : T_\mu = T_\mu (S^\nu)$  of
the thermodynamic phase space , i.e. ${\cal LA} \subset \mathbb{R}^{6+2N} $
with
coordinates $(T_\mu, S^\nu)$,  equipped
with the symplectic form
\ben
\omega= dT_\mu \wedge d S^\mu \,.
\een
Since, when pulled back to ${\cal LA } $ we have $T_\mu dS^\mu =d M(S^\mu)$,
the pull-back of $\omega $ to ${\cal LA}$ vanishes,
\ben
\omega \Bigl | _{{\cal LA}} =0 \,.
\een
In other words,
${\cal LA} $ is a Lagrangian submanifold of $\mathbb{R}^{6+3N}$.

One may go a step further and lift ${\cal LA}$ to $\mathbb{R}^{7+2N}$ with
coordinates $(P_\mu,S^\nu, M)$, equipped
with the contact form
\ben
\eta = T_\mu dS^\mu-dM\,,
\een
as a Legendre submanifold ${\cal LE}$,  i.e. one for which
\ben
\eta \Bigl | _ {{\cal LE}} =0 \,.
\een

In most of the cases we shall be considering, for dimensional reasons
$M(S,J,Q_i,P^i)$ satisfies
the weighted homogeneity relation
\ben
M(\lambda^2 A,\lambda ^2 J, \lambda Q^i, \lambda P^i)= \lambda M \,.
\label{scale2}
\een
Differentiating with respect to $\lambda$ and then setting $\lambda=1$ yields
the Smarr relation \cite{Smarr:1972kt}
\ben
M= 2 TS + 2 \Omega J + \Phi^iQ_i  + \Psi _i P^i \,.\label{smarrrel}
\een
%%%%%
The Gibbs function, or thermodynamic potential, $G$, is
the total Legendre transform of the mass with respect to
the extensive variables.  It
satisfies
%%%%%
\ben
dG = - S^\mu d T_\mu\,,
\een
%%%%%
where
%%%%%
\ben
G(T_\mu)  = M-T_\mu S^\mu =
M-TS-  \Omega J- \Phi^i Q_i  -  \Psi _i P^i = TS + \Omega J \,.\label{gibbsbh}
%%%%%
\een
Note that $G$ is not necessarily a single-valued function of the
intensive variables $T_\mu$, unless the Gibbs surface ${\cal G}$ is convex.
The Hessian of the Gibbs function with respect to the intensive
variables is related to the Weinhold metric by the
easily-verified formula
%%%%%
\be
\fft{\del^2 G}{\del T_\mu\del T_\nu} \,
\fft{\del^2 M}{\del S^\nu\del S^\lambda} = -\delta_\lambda^\mu\,.
\ee
%%%%%
It provides a metric on the space of intensive variables.

It is important to realise that from the point of view
of the symplectic and contact structures described above,
the coordinates $(S^\mu, T_\mu, M)$ have a privileged status
and it makes little physical sense to consider arbitrary
coordinate transformations even if they preserve the
symplectic or contact structures. Only a limited number
of Legendre transformations are of physical relevance.
It is physically meaningful to consider positive linear combinations
of the vectors $S^\mu$, i.e. sending $S^\mu\rightarrow D^\mu{}_\nu\,
 S^\nu$, where $D^\mu{}_\nu$ is a constant diagonal matrix,
and also to reverse the sign
of any but the first component (i.e. the first diagonal component of
$D^\mu{}_\nu$, associated with scaling the entropy itself,
should be positive). In other words, physical states
are future directed with respect to the first component.

   In the literature on thermodynamic metrics, much discussion has focused
on whether or not the Ricci scalar is a good indicator of phase
transitions.  Because, as explained above, general coordinate
transformations do not have physical significance, it is not obvious
that one should be concerned with a scalar such as the Ricci scalar.  In fact,
what is more relevant is the behaviour of the Hessian, i.e. the
thermodynamic metric.  If this is not invertible then a divergence of the
Ricci scalar will occur, but the value of the Ricci scalar itself does
not appear in general to have any physical significance.

\subsection{Thermodynamic metrics}\label{hongthermosec}

   It has been traditional in the  literature to focus on the Ruppeiner
and Weinhold metrics, and this is especially convenient if one has available
an explicit Christrodoulou-Ruffini formula. However, as in standard
text books on thermodynamics,
it is frequently convenient to introduce a variety of other thermodynamic
potentials
related by Legendre transformations, depending upon what quantities are
being held fixed.
In the context of black hole thermodynamics this corresponds to what
boundary conditions
are being considered. The consequent uniqueness or ``No Hair'' properties
will depend in general on precisely what is to be held fixed. This lack of
uniqueness
is what is often referred to as a ``phase transition,'' but as in
standard thermodynamics
it is important to specify the physical conditions under which the phase
transition takes place.

From the point of view of the Gibbs surface ${\cal G}$, geometrically this
should really be thought of
as an $n$-dimensional Legendrian sub-manifold of the $(2n+1)$-dimensional
Legendre manifold
whose coordinates consist of the the total energy and the the $n$ pairs of
intensive and extensive variables . Given a choice of $n$ coordinates chosen
from these $2n$ variables, one may
\emph{locally} describe the surface in terms of the associated
thermodynamic potential, and from that compute
the associated Hessian metric. But \emph{globally}, it is not in general
true that the Gibbs surface
equipped with the choice of Hessian is a single-valued non-singular graph
over the $n$-plane spanned by the chosen set of
$n$ coordinates. It should also be remembered that although the Hessian
metrics may be thought of as the
pull-back to $\cal G$  of a flat metric on the $2n$-dimensional flat
hyperplane spanned by the choice $n$ pairs of intensive and extensive
variables, the signature of that flat metric depends upon that choice.

  Here we review some key results on the general classes of
thermodynamic metrics
that were presented in \cite{hongthermo}.  Consider first
the energy $M=M(S^\mu)$, which obeys the first law
%%%%%
\be
dM= T_\mu\, dS^\mu = T dS + \Omega dJ + \Phi^i dQ_i +\cdots\,.\label{Mfirstlaw}
\ee
%%%%%
One can define from this the metric
%%%%%
\be
ds^2(M) = dT_\mu \otimes_s dS^\mu\,,\label{Mmetric0}
\ee
%%%%%
where $T_\mu$ are viewed as functions of the $S^\mu$ variables, with
%%%%%
\be
T_\mu=\fft{\del M}{\del S^\mu}\,,\label{TfromS}
\ee
%%%%%
and $\otimes_s$ denotes the symmetrised
tensor product.  In the usual parlance of general relativity we may simply
write (\ref{Mmetric0}) as
%%%%%
\be
ds^2(M) = dT_\mu \, dS^\mu\,.\label{Mmetric1}
\ee
%%%%%
In view of (\ref{TfromS}) we have
%%%%%
\be
ds^2(M)= \fft{\del^2 M}{\del S^\mu\, \del S^\nu}\, dS^\mu\, dS^\nu\,,
\label{Weinmet}
\ee
%%%%%
which is nothing but the Weinhold metric.

   One can obtain a set of
conformally-related metrics by dividing (\ref{Mfirstlaw}) by any one of
the intensive variables $T_\mu$ for $\mu=\bar\mu$ where $\bar\mu$ denotes
the associated specific index value of the chosen intensive variable, and
then constructing the thermodynamic metric $ds^2(S^{\bar \mu})$ for the
conjugate extensive variable by using the same procedure as before
\cite{hongthermo}.  Thus,
for example, if we choose $\bar\mu=0$, so that $T$ is the chosen intensive
variable and $S$ its conjugate, then we rewrite (\ref{Mfirstlaw}) as
%%%%%
\be
dS= \fft{dM}{T} - \fft{1}{T}\, T_a\, dS^a\,,
\ee
%%%%%
where we have split the $\mu$ index as $\mu=(0,a)$, and then write the
associated thermodynamic metric
%%%%%
\bea
ds^2(S) &=& -\fft1{T^2}\, dT dM + \fft1{T^2}\, dT\, dS^a -
     \fft1{T}\, dT_a\, dS^a\nn\\
&=& -\fft1{T}\, (dT dS + dT_a\, dS^a)\nn\\
&=& -\fft1{T}\, ds^2(M)\,.\label{RupWein}
\eea
%%%%%
The second line was obtained by using (\ref{Mfirstlaw}), and the third
line follows from (\ref{Weinmet}).  Thus $ds^2(S)$, which is the
Ruppeiner metric, is conformally related by the factor $-1/T$ to the
Weinhold metric.  Weinhold and Ruppeiner metrics were introduced
into black hole physics in \cite{Pagepd,Ferrara:1997tw}.  The literature is
by now quite extensive. For a recent review see \cite{Aman:2015wsa}.
Other conformally-related metrics can be defined by
dividing (\ref{Mfirstlaw}) by any other of the intensive variables and
the repeating the analogous calculations.  For example, if there is a
single charge $Q$ and potential $\Phi$, then dividing the first law
$dM= T dS + \Phi dQ + \cdots$ by $\Phi$ and
calculating the metric $ds^2(Q)$, one obtains
%%%%%
\be
ds^2(Q) = -\fft1{\Phi}\, ds^2(M)\,.\label{Qmetric}
\ee
%%%%%

   Further thermodynamic metrics that are not merely conformally related
to the Weinhold metric can be obtained by making Legendre transformations
to different energy functions before implementing the above procedure
\cite{hongthermo}.  For
example, if one make the Legendre transform to the free energy $F=M-TS$,
for which one has the first law
%%%%%
\be
dF= -S\, dT + T_a\, dS^a\,,\label{Ffirstlaw}
\ee
%%%%%
then the associated thermodynamic metric will be
%%%%%
\be
ds^2(F)= - dT dS + dT_a\, dS^a\,,\label{Fmetric}
\ee
%%%%%
where $S$ and $T_a$, which are now the intensive variables, are viewed
as functions of $T$ and $S^a$.  The metric components in $ds^2(F)$ are
therefore given by the Hessian of $F$.  As observed in \cite{hongthermo},
the metric $ds^2(F)$ has the property that, unlike the Weinhold or
Ruppeiner metrics, its curvature is singular on the so-called Davies
curve where the heat capacity diverges.

   Clearly, by making different Legendre transformations, one can
construct many different thermodynamic metrics, which take the form
%%%%%
\be
ds^2 = \sum_{\mu\ge 0} \eta_\mu\, dT_\mu\, dS^\mu\,,
\ee
%%%%%
where each $\eta_\mu$ can independently be either $+1$ or $-1$.  The overall
sign is of no particular importance, and so metrics related by making a
complete Legendre transformation of all the intensive/extensive pairs in
a given energy definition really yields an equivalent metric.  For
example, the Gibbs energy $G=M- T_\mu S^\mu$ gives the metric
%%%%%
\be
ds^2(G) = -dT_\mu\, dS^\mu\,,
\ee
%%%%%
which is just the negative of the Weinhold metric $ds^2(M)$ in (\ref{Weinmet}).

  One further observation that was emphasised in \cite{hongthermo} is that
one is
not, of course, obliged when writing a thermodynamic metric to use the
associated extensive variables as the coordinates.  It is sometimes the
case, as we shall see in later examples, that although one can calculate
the thermodynamic variables in terms of the metric parameters,
one cannot explicitly invert these relations.  In such cases, one can always
choose to use the metric parameters as the coordinates when writing
the thermodynamic metrics.  Geometric invariants such as the Ricci
scalar of the thermodynamic metric will be the same whether written using the
thermodynamic variables or the metric parameters, since one is just making
a general coordinate transformation.  Thus even in cases where the
relations between the thermodynamic variables and metric parameters are too
complicated to allow one to find an explicit Christodoulou-Ruffini formula
to define the Gibbs surface, one can still study the geometrical properties
of the various thermodynamic metrics.

\section{Asymptotically Flat Black Holes}

  In this section, we shall illustrate the issues raised in the previous
section by listing the cases of asymptotically-flat
black holes for which we have explicit formulae.  Whilst the formulae for
the Kerr-Newman family of black holes are well known, we first review these
in some detail in preparation for our discussion of much less well known
black holes, such as those that occur in supergravity or Kaluza-Klein
theories.

\subsection{The Gibbs surface for Reissner-Nordstr\"om}

The Gibbs surface ${\cal G}$ for the
Reissner-Nordstr\"om solution is given by the
Christodoulou-Ruffini formula
\be
M=  \sqrt \frac{S}{4 \pi}  + \frac{Q^2}{4}  \sqrt{\frac{4 \pi}{S}} =
\M + \frac{Q^2}{4\M}  \,,
\ee
%%%%%
where $\M = \sqrt{\frac{S}{4 \pi}}$.
It is convenient to
envisage $(M,Q, S)$ as  a right-handed  Cartesian
coordinate system with $M>0$ taken vertically and $-\infty <Q < \infty$
and $S>0$
spanning  a  horizontal half-plane.
In $(M,Q,S)$ coordinates the surface is part of the quadratic cone
%%%%
\be
M^2 =  \Big(\fft{S}{2\pi} - M\Big)^2  + Q^2
\,.\label{MS1}
\ee
%%%%%
We have
%%%%%
\be
M \ge |Q| \,,
\ee
%%%%%
with $M>|Q|$ being sub-extremal black holes.
Rewriting (\ref{MS1}) as
%%%%%
\be
S^2 -2\pi (2M^2-Q^2)\, S + \pi^2\, Q^4=0\,,
\ee
%%%%%
the two solutions for $S$ at fixed $M$ and $Q$ are given by
%%%%%
\be
\fft{S_\pm}{\pi} = 2M^2 -Q^2 \pm 2M \sqrt{M^2-Q^2}\,,
\ee
%%%%%
with these corresponding to the entropies (\ie one quarter the area) of
the outer ($S_+$) and inner ($S_-$) horizons respectively. It is
straightforward to see that the temperature $T=\del M/\del S$ is
positive on the outer horizon and negative on the inner horizon.

    Equality, $M=|Q|$, corresponds to extreme black holes.
They lie  on the space  curve $\C$
given by the the intersection of the two  surfaces
\be
M=|Q|\,, \qquad M= \sqrt{\frac{S}{\pi}} \,.
\ee
The first is a plane orthogonal to the $Q$ plane, and the second
a parabolic cylinder with generators parallel to the $Q$ axis.
The projection of $\C$ onto the $Q-S$ plane  is given by the parabola
\be
S=  \pi Q^2 \,.
\ee
Roughly speaking, the Gibbs surface ${\cal G}$
is folded over the space curve $\C$.
Now the Weinhold metric, or equivalently the Hessian of $M(S,Q)$, is
given by
%%%%%
\be
ds_W ^2 = \sqrt{\frac{4 \pi}{S}}
\Bigl  \{
\fft12  dQ^2  - \frac{Q}{2S}  dQ dS
+ \frac{1}{16S^2} \bigl (3 Q^2 - \frac{S}{ \pi} \bigr ) dS^2
\Bigr \} \,.
\ee
%%%%%
Note that $\frac {\p^2 M}{\p S^2}$ changes sign, passing through zero,
along the space curve $\gamma_{Davies}$, given by
%%%%%
\be
S = 3 \pi Q^2 =  \frac{9}{ 4} M ^2   \,.   \label{SQphase}
\ee
%%%%%
Since the  heat capacity at constant charge, $C_Q$,
is given by
\be
C_Q =  T\,  \Big(\frac{\p ^2 M }{\p S^2}\Big)^{-1}_Q\,,
\ee
it also changes sign across the curve $\gamma_{Davies}$, on which it diverges \cite{Davies:1978mf}.
This is often taken as a sign of a phase transition.  In support of this
interpretation, it has been shown \cite{Monteiro:2008wr} that
the single negative
mode of the Lichnerowicz operator passes through zero and becomes positive as
$Q$ is increased across $\gamma_{Davies}$.

  The curve $\gamma_{Davies}$ is an example of what, in the literature
on phase transitions, is often referred to as a spinodal curve, and
is usually defined in terms of the vanishing of a
diagonal element of the Hessian of the Gibbs function.  In the present case,
the Gibbs function is
%%%%%
\be
G=M-TS -\Phi Q = \fft{(1-\Phi^2)^2}{16\pi T}\,,
\ee
%%%%%
and the Hessian is given by
%%%%%
\be
\begin{pmatrix} \fft{\del^2 G}{\del T^2} & \fft{\del^2 G}{\del T \del \Phi}\\
\fft{\del^2 G}{\del \Phi \del T} & \fft{\del^2 G}{\del\Phi^2}
\end{pmatrix} =
\begin{pmatrix} \fft{(1-\Phi^2)^2}{8\pi T^3} &
                \fft{(1-\Phi^2)\, \Phi}{4\pi T^2}\\
  \fft{(1-\Phi^2)\, \Phi}{4\pi T^2} & -\fft{(1-3\Phi^2)}{4\pi T}
\end{pmatrix}\,.
\ee
%%%%%
The spinodal curve is thus given by $\Phi^2=\pm\ft1{3}$, which, in terms
of $S$ and $Q$, coincides with (\ref{SQphase}).

   The Weinhold metric may written as
\be
ds_W ^2 = \sqrt{\frac{4 \pi}{S} }
\Bigl  \{
\fft12 ( dQ  - \frac{Q}{2S}  dS )^2
- \frac{1}{16S^2} \bigl ( \frac{S}{ \pi} - Q^2  \bigr ) dS^2
\Bigr \} \,,
\ee
and hence the Gibbs surface for sub-extremal black holes has a Hessian,
or equivalently  a Weinhold metric, that is non-singular
but Lorentzian.  Moreover the Gibbs surface for non-extreme black holes
is non-convex.  Expressed in terms of $S$ and the electrostatic potential
\be
\Phi = \Phi(S,Q) =  Q\, \sqrt{\frac{ \pi}{S}}\,,\label{PhiSQ}
\ee
%%%%%
the Weinhold metric becomes
%%%%%
\be
ds_W^2 = \fft{1}{8 \sqrt{\pi}\, S^{3/2}}\, \Big[
  - (1-\Phi^2)\, dS^2 + 8 S^2\, d\Phi^2\Big]\,.
\ee
%%%%%
Note that the metric is non-singular when either $\Phi^2<1$, corresponding to
the outer horizon, or $\Phi^2>1$, corresponding to the inner horizon.
It changes signature from $(-+)$ to $(++)$ as $\Phi$ goes from $\Phi^2<1$
to $\Phi^2>1$.  The heat capacity passes through infinity at $\Phi^2=\ft13$.

Expressed in terms of $\Phi$ and $S$, the temperature is
given by $T= (1-\Phi^2)/(4\sqrt{\pi S})$, and so the
Ruppeiner metric is given by
%%%%%
\bea
ds_R^2 = -\fft1{T}\, ds_W^2 &=& -\fft{dS^2}{2S} +
  4 S\, \fft{d\Phi^2}{1-\Phi^2}\nn\\
&=& -d\tau^2 + \tau^2\, d\sigma_+^2\,,\label{milne}
\eea
%%%%
where we have defined, for the outer horizon,
%%%%%
\be
S=\ft12 \tau^2\,,\qquad \Phi= \sin\fft{\sigma_+}{\sqrt2}\,.
\ee
%%%%%
The metric in the second line of (\ref{milne}) is the Milne metric on a wedge
of Minkowski spacetime inside the light cone.  This is made apparent by
introducing new coordinates according to
%%%%%
\be
  t= \tau\, \cosh \sigma_+\,,\qquad x= \tau\, \sinh\sigma_+\,,
\ee
%%%%%
in terms of which the Ruppeiner metric becomes
%%%%%
\be
ds_R^2 = -dt^2 + dx^2\,, \qquad S=S_+ = \ft12 (t^2-x^2)\,.
\ee
%%%%%
Since the range of $\sigma_+$ is $-\ft{\pi}{\sqrt2} \le \sigma_+
  \le \ft{\pi}{\sqrt2}$, the extremal solutions lie on the timelike
geodesics $t= \pm \arctanh \ft{\pi}{\sqrt2}$.  The heat capacity changes
sign at $\Phi^2=\ft13$.

  If $\Phi^2>1$, corresponding to the inner horizon, then, if $Q>0$,
substituting
%%%%%
\be
\Phi= \cosh\fft{\sigma_-}{\sqrt2}
\ee
%%%%%
(or $\Phi=-\cosh(\sigma_-/\sqrt2)$ if $Q<0$)
into (\ref{milne}) gives
%%%%%
\be
ds^2 = -(d\tau^2 + \tau^2\, d\sigma_-^2)\,.
\ee
%%%%%
The metric in brackets is the flat metric on Euclidean space in polar
coordinates, except that the range of the coordinate $\sigma_-$ is
$0\le\sigma_-\le\infty$, so we have an infinitely branched covering of the
Euclidean plane, with the branch point at the origin.  The Weinhold metric
is itself positive definite.  Thus the Gibbs surface is convex and the
entropy surface is concave for the inner horizon.

 The flatness of the Ruppeiner metric for Reissner-Nordstr\"om has
 given rise
to much comment, because singularities of the Ruppeiner metric are
expected to reveal the occurrence of phase transitions.  However,
the geometrical significance of the change in sign of the
heat capacity is that for fixed charge $Q$, there is a maximum temperature.
In fact
\be
T=T(S,Q)=  \frac{1}{2S} \sqrt{\frac{S}{4 \pi}}   - \frac{ Q^2}{8 S}
\sqrt{\frac{4 \pi }{S}}     \,,
\ee
so for given $|Q|$ and  positive $T$ less than  $\frac{\sqrt3}{8 \pi |Q|}$,
there are two positive values of $S$ and hence two non-extreme
black holes. By contrast, since the electrostatic potential $\Phi$
satisfies (\ref{PhiSQ}),
there is a unique positive value of $S$ and hence a
unique black hole for given $Q$ and $\Phi^2 <1$.

  Every two-dimensional metric is conformally flat.  Therefore it is
not surprising that both the Weinhold and Ruppeiner metrics for
Reissner-Nordstr\"om are conformally flat.  It is, however, nontrivial that
the Ruppeiner metric is flat.  It has recently been pointed out
\cite{1411.2582} that
one can also consider the Hessian of the charge $Q$, considered as
a function of the mass and entropy, as a metric $ds_Q^2$. In fact
$ds_Q^2 = -\Phi^{-1}\, ds_W^2$, as in (\ref{Qmetric}).
Geometrically, there is no reason to give a
preference to
any of the metrics $ds_W^2$, $ds_R^2$ or $ds_Q^2$.
Since $T$ and $\Phi$ are both non-singular on the
curve along which the heat capacity diverges, none of the three metrics
is capable of detecting the associated ``phase transition.''

   As was shown in \cite{hongthermo}, and we reviewed in section
\ref{hongthermosec},
the thermodynamic metric (\ref{Fmetric})
constructed from the free energy $F=M-TS$ does
exhibit a singularity on the Davies curve where the heat capacity diverges.
For the Reissner-Nordstr\"om metric (\ref{Fmetric}) is the restriction
of $ds^2(F)= -dT dS +d\Phi dQ$ to the Gibbs surface, and hence we
find
%%%%%
\be
ds^2(F) = \sqrt{\fft{\pi}{S}}\, dQ^2 + \fft{1}{8\sqrt{\pi}\, S^{5/2}}\,
  (S-3\pi Q^2)\, dS^2\,.
\ee
%%%%%
A straightforward calculation shows that its Ricci scalar is given by
%%%%%
\be
R_F = \fft{4\sqrt{\pi}\, S^{3/2}}{(S-3\pi Q^2)^2}\,,
\ee
%%%%%
which does indeed diverge on the Davies curve $S=3\pi Q^2$.

\subsection{The Gibbs surface for Kerr}

This is qualitatively very similar to the Reissner-Nordstr\"om case.
To begin with, we shall summarise, in our notation, some results first
presented by Curir \cite{Curir}.
One has
\ben
M^2 = \frac{S}{4\pi } + \frac{\pi J^2}{S}\,,\label{KerrCR}
\een
%%%%%
and
$M(S,J) $ at fixed $J$ has a minimum value when
\ben
S=2\pi |J| \,, \qquad  M= \sqrt{|J|} \,.
\een
%%%%%
This is the extreme case and, as before,
the inner horizon has a negative temperature, a point made first by
Curir \cite{Curir}. Explicitly one has
\ben
T= \frac1{8\pi M}\, \Big(1 - \frac{4\pi^2 J^2}{S^2}\Big)\,.  \label{KerrT}
\een
%%%%%
 For any given values of $J$ and of $M>0$, there are two positive solutions,
$S_+$ and $S_-$, of (\ref{KerrCR}), where $S_+\ge 2\pi |J|$ corresponds to
one quarter of the area of the outer horizon of a sub-extremal black hole and
$S_-\le 2 \pi |J|$ corresponds to one quarter of the area of the inner
horizon of
a sub-extremal black hole.  From (\ref{KerrCR}), they obey the
entropy product formula
%%%%%
\be
  S_-\, S_+ = 4 \pi^2\, J^2\,.\label{SprodKerr}
\ee
%%%%%
By (\ref{KerrT}), the outer horizon has
a positive temperature, which we label $T_+$, and the inner horizon has
a negative temperature, which we label $T_-$.
One has \cite{Curir}
%%%%
\ben
T_\pm= \frac{S_\pm - S_\mp}{8 \pi M S_\pm}  \,,\qquad
\Omega_\pm = \frac{\pi J}{MS_\pm} \,,
\label{spin}
\een
%%%%%
where $\Omega_\pm =(\del M/\del J)_{S_\pm}$. Note that it follows from the
first equation in (\ref{spin}) that
%%%%%
\be
T_+\, S_+ + T_-\, S_- =0\,.\label{TSsum}
\ee
%%%%%
Note also that $M$ and $J$, which are conserved quantities defined in terms of
integrals at infinity, are universal and do not carry $\pm$ labels.

  In terms of $S_+$ and $S_-$, one has, from (\ref{KerrCR}),
%%%%%
\be
M^2 = \fft{S_+}{4\pi} + \fft{S_-}{4\pi}\,.
\ee
%%%%%
Therefore
%%%%%
\be
M = \fft{\sqrt{S_+ + S_-}}{\sqrt{4\pi}} \le
   \sqrt{\fft{S_+}{4\pi}} + \sqrt{\fft{S_-}{4\pi}}\,.
\ee
%%%%%
If one varies $M$, one has
%%%%%
\be
dM= T_\pm\, dS_\pm + \Omega_\pm\, dJ\,.
\ee
%%%%%
There is also a modified Smarr formula
%%%%%
\be
M = T_+\, S_+ + T_-\, S_- + \Omega_+\, J + \Omega_-\, J =
 (\Omega_+ + \Omega_-)\, J\,,\label{modSmarrKerr}
\ee
%%%%%
where the second equality follows from (\ref{TSsum}). This way of writing the first law of thermodynamics was employed in \cite{Huang:2016fks} for deriving a simple formula for holographic complexity. These results were interpreted in \cite{Curir} as indicating that the
total energy of a
rotating black hole may be regarded as receiving contributions from
two thermodynamic systems; one associated with the outer horizon and
the other with the inner horizon.  The negative temperature was interpreted
in terms of Ramsey's
account of the thermodynamics of isolated spin systems
\cite{Ramsey}.

  Okamoto and Kaburaki \cite{OK} introduced the
dimensionless parameter $h=\frac{a}{M+\sqrt{M^2-a^2}}$
in their discussion of the energetics of Kerr black holes
and noticed that it
satisfies the quadratic equation
%%%%%
\ben
h^2 - \frac{2h M^2}{|J|} + 1 =0 \,. \label{quadratic}
\een
%%%%%
It was initially assumed that only the solution of (\ref{quadratic})
satisfying $0\le h\le 1$ has physical significance.
However Abramowicz \cite{Abram}
drew their attention to \cite{Curir,CFrancaviglia1},
and they realised that the other root of  (\ref{quadratic}),
which satisfies  $1\le h \le \infty$ and is given by
$h=\frac{a}{M-\sqrt{M^2-a^2}}$,  is associated with the inner horizon
\cite{OK}. Expressing the thermodynamic variables in terms of $h$ they
established
(\ref{TSsum}) if $T_-$ is taken to be negative, and they also obtained
the formula
%%%%%
\ben
\frac{\Omega_+}{T_+} + \frac{\Omega _-}{T_-} =0\,.
\een
%%%%%

\subsection{Kerr-Newman black holes}

Kerr-Newman black holes may have both electric and magnetic charges.
By electric-magnetic
duality invariance one may set the magnetic charge $P$ to zero.
To restore electric-magnetic duality invariance
it suffices to replace $Q^2$ by   $Q^2 + P^2$ in all formulae
thus producing a manifestly $O(2)$ invariant Gibbs surface.

The mass of the Kerr-Newman black hole is given by
%%%%
\ben
M=  \Bigl [\fft{\pi}{4S}\,
  (\frac{S}{\pi}+Q^2 \Bigr)^2  +\frac{ \pi J^2} {S}   \Bigr]^\half \,,
\label{massKN}
\een
%%%%%
and therefore it satisfies
%%%%%
\ben
M \ge \sqrt{ \sqrt{J^2 + \frac{Q^4}{4} } +  \frac{Q^2}{2}  } \,,\label{bound}
\een
%%%%%
acquiring its least value on the surface $\gamma_{extreme}$
in the three dimensional
space of extensive variables given by
\ben
S= \pi \sqrt{ 4J^2 + Q^4 } \,,
\een
on  which the temperature
\ben
T= \Big(\frac{\p M}{\p S}\Big)_{J,Q} =  \frac{1}{8 \pi M}
\Big[1-\frac{\pi^2}{S^2} (4J^2+ Q^4) \Big]     \label{temp}
\een
vanishes. If $J=0$, then (\ref{bound}) is the usual Bogomolnyi bound
\cite{Gibbons:1982fy}.
One also has
\ben
\Omega = \frac{\pi J}{MS} \,, \qquad \Phi= \frac{\pi Q}{2MS}
\, \Big( Q^2 + \frac{S}{\pi} \Big) \,. \label{pots}
\een
The explicit formulae
(\ref{massKN}), (\ref{temp}) and (\ref{pots}) allow a lift of the Gibbs surface
$\cal G$ to a Lagrangian submanifold  $\cal L$ in $\mathbb{R}^6
$ and a Legendrian submanifold in $\mathbb{R} ^7$.
The entropy product law becomes
\ben
S_+\, S_- = \pi^2 \, (4 J^2 + Q^4)\,,\label{entopyprodKN}
\een
where the $-$ refers to the inner and $+$ to outer horizon.
%The  (\ref{bound} follows easily from (\ref{entropyprodKN}).

   The temperatures and angular velocities of the two horizons are given by
\ben
T_\pm= \frac{S_\pm -S_\mp}{8 \pi M S_\pm} \,, \qquad
\Omega_\pm = \frac{\pi J}{M S_\pm}\,,
\een
%%%%%
and one has
%%%%%
\ben
S_+T_+ + S_-T_-=0\,. \label{STsumKN}
\een
%%%%%
There is a conventional first law for both horizons:
\ben
dM=T_\pm dS_\pm + \Omega_\pm dJ + \Phi_\pm dQ \,,
\een
%%%%
and a modified Smarr formula
%%%%%
\be
M = T_+\, S_+ + T_-\, S_- + \Omega_+\, J + \Omega_-\, J +
   \ft12 \Phi_+\, Q + \ft12 \Phi_-\, Q =
 (\Omega_+ + \Omega_-)\, J + \ft12 (\Phi_+ + \Phi_-)\, Q\,.
\label{modSmarrKN}
\ee
%%%%%

\subsection{STU black holes}

   Four-dimensional black holes in string theory or M-theory can be described
as solutions of ${\cal N}=8$ supergravity.  The most general black holes
are supported by just four of the 28 gauge fields, in the Cartan subalgebra
of $SO(8)$. The black holes can therefore be described just within
the ${\cal N}=2$ STU supergravity theory, which is a consistent truncation of
the ${\cal N}=8$ theory whose bosonic sector comprises the metric, the
four gauge fields, and six scalar fields.
Black holes of the STU model are parameterised by mass $M$, angular
momentum $J$ and  four electric $Q_i$ ($i=1,2,3,4$) and four magnetic
charges $P^i$ ($i=1,2,3,4$).
The most general black hole solution was obtained by Chow and Comp{\` e}re
\cite{Chow:2013tia} by solution generating techniques.

  We shall follow the usual conventions for STU supergravity, in which the
normalisation of the gauge fields $F^{(i)}$ is such that if the scalar
fields are turned off, the Lagrangian will take the form
${\cal L}= \sqrt{-g}\, [R -\ft14 \sum_i (F^{(i)})^2+\cdots]$ (see appendix
\ref{STUlagsec} for a presentation of the bosonic sector of the 
STU supergravity Lagrangian). This contrasts
with the conventional normalisation ${\cal L}=\sqrt{-g}\, (R-F^2)$, in
Gaussian units, which we
use when describing the pure Einstein-Maxwell theory.  Since this means that
the charge normalisation conventions will be different in the two
cases, we shall briefly summarise our definitions here.  If we consider
the Lagrangian
%%%%%
\be
{\cal L}= \sqrt{-g}\, (R - \gamma F^2)\,,
\ee
%%%%%
one can derive by considering variations of the associated Hamiltonian that
black holes will obey the first law
%%%%%
\be
dM = \fft{\kappa}{8\pi} \, dA + \Phi\, dQ+ \Omega\, dJ\,,
\ee
%%%%%
where $\kappa$ is the surface gravity, $\Phi$ is the potential
difference between the horizon and infinity (with the potential being
equal to $\xi^\mu A_\mu$, where $\xi^\mu$ is the future-directed
Killing vector
that is null on the horizon and is normalised such that
$\xi^\mu\,\xi_\mu\rightarrow -1$ at infinity).
The electric charge $Q$ is given by
%%%%%
\be
Q= \fft{\gamma}{4\pi}\, \int {*F}\,.
\ee
%%%%%
Thus in Einstein-Maxwell theory, with ${\cal L}= \sqrt{-g}(R-F^2)$,
we shall have
%%%%%
\be
Q= \fft{1}{4\pi}\, \int {*F}\,,\label{emcharge}
\ee
%%%%%
while in STU supergravity we shall have (neglecting the scalar fields for
simplicity\footnote{In general, including the scalar fields, and writing
the Lagrangian as a 4-form, we shall have ${\cal L}=R\, {*\oneone} -
 \ft12  M_{ij}(\Phi)\, {*F^{(i)}}\wedge F^{(j)} -\ft12 N_{ij}(\Phi)\,
F^{(i)}\wedge F^{(j)} +\cdots$, where $F^{(i)}=d A^{(i)}$.  The electric
charges can be written as
%%%%%
\bea
Q_i= -\fft1{16\pi}\, \int  \fft{\delta{\cal L}}{\delta F^{(i)}}\,.\nn
\eea
%%%%%
(Here the variational derivative is defined by $\delta X =(\delta X/\delta F)
\wedge \delta F$.  For example if $X=u\, {*F}\wedge F+ v\, F\wedge F$ then
$\delta X/\delta F= 2 u\,{*F} + 2v\, F$.)  The magnetic charges are given
by $P^i=\ft1{16\pi}\, \int F^{(i)}$.})
%%%%%
\be
Q_i = \fft1{16\pi}\, \int {*F^{(i)}}\,.\label{stucharge}
\ee
%%%%%

 The black hole solutions have two horizons, with the
the product of the horizon entropies quantised:
%%%%%
\ben
S_+S_-=4 \pi ^2 \Big|J^2 + \Delta \Big| \,, \label{PCC}
\een
%%%%%
where $\Delta$ is the Cayley hyperdeterminant  $\Delta( Q_i ,P^i)$:
%%%%%%
\ben
\Delta = 16\,\Bigl[ 4 \Bigl( Q_1Q_2Q_3Q_4 + P^1P^2P^3P^4)
  + 2 \sum _{i<j} Q_iQ_j P^i P^j - \sum_i (Q_i)^2 (P^i)^2 \Bigr ]\,.
\label{Deldef}
\een
%%%%%%
Note that eqn (\ref{PCC}) has previously appeared in the literature
without the absolute value symbol (for example, in \cite{Chow:2013tia}).
We have written (\ref{PCC}) with an absolute value sign 
since $\Delta$, and hence $\Delta + J^2$, can be negative; for example 
for a static Kaluza-Klein dyonic black hole.  (In \cite{Chow:2013tia}
it was proposed that $S_-$ is negative when $\Delta+J^2<0$, but this 
would contradict the fact that, for example, the area of
the inner horizon of the static Kaluza-Klein
dyonic black hole is positive.)     

   It should be noted 
that if $J$ vanishes and $\Delta=0$, then $S_-$ will vanish also.  In this 
case there is no non-singular inner horizon.

   The entropy formulae (\ref{PCC}) can be cast in the form
%%%%%
\ben
S_+ =S_L +  S_R\,,\qquad S_- = |S_L - S_R|\,,  \label{SLSRCC}
\een
%%%%%
with
%%%%%
\ben
S_L= 2 \pi  \sqrt{F+ \Delta}\,, \quad \quad S_R= 2\pi \sqrt{F-J^2}
  \,, \label{SCC}
\een
%%%%%
where $F$  is another complicated expression that is a function of
$M$, $Q_i$ and $P^i$ only \cite{Chow:2013tia}.  Note that
it follows from (\ref{SLSRCC}) that $S_+\ge S_-$. Unlike \cite{Chow:2013tia},
we have put an absolute value sign around $(S_L-S_R)$ in the expression
for $S_-$, since, for the reasons discussed above, there can be circumstances
where $S_L<S_R$, but $S_-$ should be non-negative. Note that $F+\Delta$ is
always non-negative, and $F-J^2$ is non-negative provided that
the black hole is not over-rotating \cite{Chow:2013tia}.  The quantities
$S_L$ and $S_R$ are both non-negative.
In the extremal limit $F-J^2=0$, one gets the
extremal value for the entropy
$S_+=S_- = 2\pi \sqrt{|\Delta|}$.  This was seen for the
BPS solutions ($F=0$ and $J^2=0$) in \cite{CveticTseytlin}.

   Note from (\ref{SCC}) that while the right-moving entropy 
$S_R$ is a function of all the extensive 
variables $(M,Q_i,P_i,J)$, the left-moving entropy $S_L$ is a function of
$(M,Q_i,P^i)$ but not $J$ \cite{Chow:2013tia}.  This was noted 
previously in the special case of the four-charge black holes 
characterised by $(M,Q_i,J)$ in \cite{CYI,horlowmal}.
   The expressions (\ref{SCC}) may in principle be inverted to give
{\it two} different Christodoulou-Ruffini formulae:
%%%%%
\be
M=M(S_L,Q_i,P^i)\,,\qquad \hbox{and}\qquad M=M(S_R,Q_i,P^i,J)\,.\label{LRCR}
\ee
%%%%%

The structure (\ref{SCC}) ensures that the two entropies $S_+$ and $S_-$ 
are solutions of
the quadratic equation
%%%%%
\ben
S^2-S\,\Sigma +
 4\pi^2 \big|J^2 +\Delta \big|=0\,, \label{SquadCC}
\een
%%%%%
where $\Sigma= S_L+S_R +|S_L-S_R|$, and
we employed (\ref{SLSRCC}), (\ref{SCC}) and (\ref{PCC}). Note that
$\Sigma=2S_L$ if $S_L>S_R$, which corresponds to $J^2+\Delta>0$, whilst
$\Sigma=2 S_R$ if $S_L<S_R$, corresponding to $J^2+\Delta<0$.
 From (\ref{SquadCC}) we can deduce
%%%%%
\ben
\frac{\partial M}{\partial S}\frac{\partial\Sigma}{\partial M}
\Big|_{(Q_i,P^i,J)}
=\Bigl[1-\frac{4\pi^2 \big|J^2 +\Delta \big|}{S^2}\Bigr]=
\fft1{S}\, \Big[S - \fft{S_+\, S_-}{S}\Big]\,.
\label{SLDCC}
\een
%%%%%%
Since $S_+\ge S_-$, the final expression in (\ref{SLDCC}) is non-negative
for  $S=S_+$, and
non-positive for  $S=S_-$.  Since 
$\frac{\partial M}{\partial S}\Big|_{(Q_i,P^i,J)}=T$, and 
since $\frac{\partial\Sigma}{\partial M}
\Big|_{(Q_i,P^i,J)}$ is independent of whether one takes $S=S_+$ or
$S=S_-$, it then follows that 
\ben
S_+T_+ + S_- T_- = 0\,.  \label{STsum}
\een
%%%%%
In particular, this implies that $T_+$ and $T_-$ must have opposite signs.

   As well as considering the left-moving and right-moving entropies
$S_L$ and $S_R$, one can also introduce left-moving and right-moving
temperatures $T_L$ and $T_R$, defined by \cite{CLII}
%%%%%
\be
\fft{1}{T_L} = \fft1{T_+} + \fft1{T_-}\,,\qquad 
   \fft{1}{T_R} = \fft1{T_+} - \fft1{T_-}\,.\label{TLTRdef}
\ee
%%%%%
These definitions are motivated by the fact that when one calculates
scattering amplitudes for test fields propagating in the black-hole
background, one finds that they factorise into the product of
thermal Boltzmann factors for the temperatures $T_L$ and $T_R$ respectively
\cite{CLII}.  Using (\ref{STsum}), together with the expressiona for
$S_+$ and $S_-$ in terms of $S_L$ and $S_R$ in (\ref{SLSRCC}), it follows 
from (\ref{TLTRdef}) that
%%%%%%
\bea
S_L\ge S_R: && \qquad \qquad \fft {S_L}{T_L} = \fft{S_R}{T_R}\,,\nn\\
S_L\le S_R: && \qquad \qquad \fft {S_R}{T_L} = \fft{S_L}{T_R}\,,
\label{SLSRTLTR}
\eea
%%%%%
for the two cases that we described previously.  From its definition,
$T_R$ is obviously non-negative since $T_+\ge0$  and $T_-\le0$.  
It is then evident from (\ref{SLSRTLTR}) that $T_L$ is non-negative also,
since we already know that $S_L$ and $S_R$ are non-negative.

We can also derive, from 
%%%%%
\be
\Omega = \fft{\del M}{\del J}\big |_{(Q_i,P^i,S)} =
  \fft{\del M}{\del S} \fft{\del S}{\del J} \big |_{(Q_i,P^i,S)}\,,
\ee
%%%%%
and using either (\ref{SquadCC}) or else simply writing $S_+$ and $S_-$
in terms of $S_L$ and $|J^2+\Delta|$ by using (\ref{PCC}),  that
in the two cases $S_L\ge S_R$ and $S_L\le S_R$ we have
%%%%%
\bea
S_L\ge S_R:&& \qquad \Omega_+ S_+ =\Omega_- S_-\,, \qquad 
\frac{\Omega_+}{T_+} =- \frac{\Omega_-}{T_-}\,,\nn\\
S_L\le S_R:&& \qquad \Omega_+ S_+ = -\Omega_- S_-\,,\qquad
\frac{\Omega_+}{T_+} = \frac{\Omega_-}{T_-} \,.
\label{OSsum}
\eea
%%%%%
Note that when $S_L<S_R$, i.e.~when $J^2+\Delta<0$, the angular 
velocities of the inner and outer horizons are opposite.  Note also that
the two cases in (\ref{OSsum}) can be expressed in the single universal
formula
%%%%
\be
(S_L+S_R)\, \Omega_+  = (S_L-S_R)\, \Omega_-\,.\label{OSsum2}
\ee
%%%%%

\subsubsection{Thermodynamics of the left-moving and right-moving sectors}

   The introduction of the left and right temperatures
and entropies suggested the possibility of viewing the black hole as 
being composed of excitations in left-moving and right-moving sectors in
a string or D-brane description, associated with degrees of 
freedom of a weakly coupled two-dimensional conformal quantum field theory.  
The total entropy $S_+$ of the
outer horizon is viewed as the sum of the entropies $S_L$ and $S_R$ of
the left-moving and right-moving sectors.  
It is then natural to expect that there
should exist thermodynamic descriptions for these sectors, with first laws  of
the form\footnote{The analysis the thermodynamics of asymptotically-flat 
black holes in terms of left-moving and right-moving degrees of freedom 
was first addressed in \cite{CLII} for general STU black holes in 
five dimensions, and  briefly in \cite{CLIII} for four charge STU black holes.}
%%%%%
\bea
dE_L &=& T_L\, dS_L + \Omega_L\, dJ + \Phi^i_L\, dQ_i +
   \Psi_{L, i}\, dP^i \,,\nn\\
dE_R &=& T_R\, dS_R + \Omega_R\, dJ + \Phi^i_R\, dQ_i +
   \Psi_{R, i}\, dP^i \,.\label{LRfirst}
\eea
%%%%%
For now, we shall focus for simplicity on the regime where $S_L\ge S_R$, 
i.e.~$(J^2+\Delta)\ge 0$.  

 Let us first consider processes where $dJ=0$ and $dQ_i=dP^i=0$.  From the
definitions of $T_L$, $T_R$, $S_L$ and $S_R$ given in (\ref{SLSRCC}) and
(\ref{TLTRdef}), it straightforward to see from the first laws 
%%%%%
\be
dM =T_\pm\, dS_\pm + \Phi^i_\pm\, dQ_i + \Psi_{\pm, i}\, dP^i +
    \Omega_\pm\, dJ\label{pmfirst}
\ee
%%%%%
on the outer and inner horizons that we must have
%%%%%
\be E_L=E_R = \fft12 M\,.\label{ELER}
\ee
%%%%% 
In other words, the left-moving and right-moving sectors contribute equally
to the mass of the black hole.  (This was observed in the case of
Kerr-Newman black holes in \cite{sqwuneg,huan}.)  
Dividing the first laws (\ref{pmfirst}) 
by $T_\pm$ respectively and then taking the plus and minus combinations,
one finds that these match with (\ref{LRfirst}) provided that we define
the left-moving and right-moving quantities as 
%%%%%
\bea
\Phi^i_L\!\!\!&=&\!\!\! 
  T_L\Big(\fft{\Phi^i_+}{2T_+} + \fft{\Phi^i_-}{2T_-}\Big)\,,\
\Psi_{L\,, i}= T_L\Big(\fft{\Psi_{+,i}}{2T_+} + \fft{\Psi_{-, i}}{2T_-}\Big)
  \,,\
\Omega_L= T_L\Big(\fft{\Omega_+}{2 T_+} + \fft{\Omega_-}{2 T_-}\Big)\,,\nn\\
\Phi^i_R\!\!\!&=& \!\!\!
T_R\Big(\fft{\Phi^i_+}{2T_+} - \fft{\Phi^i_-}{2T_-}\Big)\,,\
\Psi_{R\,, i}= T_R\Big(\fft{\Psi_{+,i}}{2T_+} - \fft{\Psi_{-, i}}{2T_-}\Big)
  \,,\
\Omega_R= T_R\Big(\fft{\Omega_+}{2 T_+} - \fft{\Omega_-}{2 T_-}\Big)
\,,\label{LRdefs}
\eea
%%%%%
and so we have the first laws
%%%%%
\bea
\fft12 dM &=&T_L\, dS_L + \Omega_L\, dJ + \Phi^i_L\, dQ_i +
   \Psi_{L, i}\, dP^i \,,\nn\\
\fft12 dM &=& T_R\, dS_R + \Omega_R\, dJ + \Phi^i_R\, dQ_i +
   \Psi_{R, i}\, dP^i \label{LRfirstlaws}
\eea
%%%%%
for the left-moving and right-moving sectors.  

   In a similar fashion, we can then see that the Smarr relations
%%%%%
\be
M= 2 T_\pm\, S_\pm + 2 \Omega_\pm\, J + \Phi_\pm^i\, Q_i +
  \Psi_{\pm,i}\, P^i
\ee
%%%%%
on the outer and inner horizons imply the Smarr relations
%%%%%
\bea
\fft12 M &=& 2 T_L\, S_L + 2 \Omega_L\, J + \Phi_L^i\, Q_i +
   \Psi_{L,i}\, P^i\,,\nn\\
\fft12 M &=& 2 T_R\, S_R + 2 \Omega_R\, J + \Phi_R^i\, Q_i +
   \Psi_{R,i}\, P^i\label{LRsmarr}
\eea
%%%%%
for the left-moving and right-moving sectors.

  It should be noted that, from (\ref{OSsum}) and (\ref{LRdefs}), 
the left-moving angular velocity is in fact zero:
%%%%%
\be
\Omega_L=0\,,\qquad \Omega_R = \fft{T_R}{T_+} \, \Omega_+
\,.
\ee
%%%%%

  If we now turn to the regime where $S_L < S_R$, we find that the roles
of $S_L$ and $S_R$ are exchanged in both the first laws and the Smarr 
relations for the left-moving and right-moving sectors, so that we have
%%%%%
\bea
S_L< S_R:\qquad \fft12 dM &=&T_L\, dS_R + \Omega_L\, dJ + \Phi^i_L\, dQ_i +
   \Psi_{L, i}\, dP^i \,,\nn\\
\fft12 dM &=& T_R\, dS_L + \Omega_R\, dJ + \Phi^i_R\, dQ_i +
   \Psi_{R, i}\, dP^i \label{LRfirstlaws2}\,,\\
\fft12 M &=& 2 T_L\, S_R + 2 \Omega_L\, J + \Phi_L^i\, Q_i +
   \Psi_{L,i}\, P^i\,,\nn\\
\fft12 M &=& 2 T_R\, S_L + 2 \Omega_R\, J + \Phi_R^i\, Q_i +
   \Psi_{R,i}\,. P^i\label{LRsmarr2}
\eea
%%%%%
Furthermore, it follows from (\ref{OSsum}) and (\ref{LRdefs}) that
it is now $\Omega_R$, rather than $\Omega_L$, that vanishes.  
One possible way to make the formulae more uniform for the $S_L<S_R$
regime would be exchange the L and R labels in the definitions of 
all the intensive thermodynamic variables, $T,\, \Phi^i,\, \Psi_i,\, \Omega$,  
when $S_L<S_R$.  This would have the merit that, with the relabelling,
the left-moving angular velocity would vanish in all cases, while
still retaining the property that $S_L$ is independent of $J$ in all cases.
The left-moving and right-moving first laws and Smarr relations would then
take the same forms as in (\ref{LRfirstlaws}) and (\ref{LRsmarr}) for 
both $S_L\ge S_R$ and $S_L<S_R$, in terms of the relabelled variables.

\subsubsection{Four-charge STU black holes}

  The prospects for obtaining an explicit Christodoulou-Ruffini
formulae for the general 8-charge black hole solutions are not good. 
The main problem is the $F$-invariant that
appears in the expressions for $S_L$ and $S_R$ in eqn (\ref{SCC}),
whose evaluation in terms of physical charges and mass appears to be quite
intractable \cite{Sarosi:2015nja}. 
In order to obtain more explicit, concrete expressions, we shall now  
focus on the specialisation to black-hole solutions carrying just 
four electric charges, which were found in \cite{CYI}.

   These black holes are parameterised in terms of the non-extremality
parameter $m\ge 0$ (Kerr mass parameter), the ``bare'' angular momentum $a$
(Kerr rotation parameter)
and four boost parameters
$\delta_i\ge 0$ ($i=1,2,3,4$) \cite{CYI} (see also \cite{Chong} for compact
expressions for the metric and the other fields).  
In terms of these, the physical
mass, charges and angular momentum are given by
%%%%%
\bea
M &=& \frac{m}{4} \sum_{i} \cosh 2 \delta_i\,, \nn\\
Q_i&=&\frac{1}{4}\, m \sinh 2 \delta_i\,,\nn \\
J&=& m a(\Pi_c-\Pi_s)\,.
\eea
%%%%%
The black hole entropies, associated with the inner and the outer horizon,
are given by \cite{CYI,CLIII}:
%%%%%
\bea
S_{\pm}\equiv \frac{A_\pm}{4}& =& 2 \pi m \bigr [m (\Pi_c+\Pi_s) \pm (\Pi_c-\Pi_s) \sqrt{m^2-a^2}{\bigl ]}\\
  &= &2\pi {\bigr [}m^2 (\Pi_c+\Pi_s) \pm \sqrt{m^4(\Pi_c-\Pi_s)^2 -J^2}{
\bigl ]}\,.
  \eea
%%%%%
The  temperatures $T_{\pm}$, related to surface gravities $\kappa_\pm$ by
$T_{\pm}=\frac{\kappa_{\pm}}{2\pi}$,  and angular velocities $\Omega_\pm$,
which are associated with the inner and out horizon respectively, are
given by \cite{CLIII}:
%%%%%
\bea
\frac{1}{T_{\pm}}=\frac{2\pi}{\kappa_\pm} &=&\frac{4\pi m}{\sqrt{m^2-a^2} }\bigr [\pm m (\Pi_c+\Pi_s) + (\Pi_c-\Pi_s) \sqrt{m^2-a^2}
  \bigr ] ,\\
\Omega_{\pm}&=&\pm \frac{2\pi a\, T_{\pm}}{\sqrt{m^2-a^2} } \,, \label{explTO}
\eea
%%%%%
where
\ben
\Pi_c = \prod_{i} \cosh \delta_i \,,\qquad \Pi_s = \prod_{i}\sinh \delta_i\,.
\een
Note that $T_{-}$ is negative.\footnote{Note that in \cite{CLIII} the value
of $T_-$ was taken to be positive, and equal to the absolute value of
the $T_-$ given in (\ref{explTO}).}
From the above expressions one also finds
%%%%%
\be
S_{\pm}=\pm\frac{\sqrt{m^2-a^2}}{2T_{\pm}}\, .\label{TSprod}
%\Omega_{\pm} &=& \frac{\kappa_\pm}{\sqrt{m^2-a^2 }} \,.
\ee
%%%%%
It can easily be verified that the entropies $S_\pm$, temperatures $T_\pm$ and
angular veocities $\Omega_\pm$ satisfy equation (\ref{STsum}) and
the $S_L\ge S_R$ equations in (\ref{OSsum}).  

The entropies and  the inverses of the surface gravities,
associated with the outer and inner horizons, have a suggestive form in
terms of the left-moving and right-moving entropy and inverse temperature
excitations of a weakly coupled 2-dimensional conformal field theory
(2D CFT), given in \cite{CLIII}:
%%%%%
\bea
S_L&=& \frac{1}{2}\bigl ( S_++S_-\bigr )=2 \pi m^2 (\Pi_c+\Pi_s)\,, \nn \\
S_R&=& \frac{1}{2}\bigl ( S_+-S_-\bigr )=2\pi m \sqrt{m^2-a^2} (\Pi_c-\Pi_s)\, ,  \label{SLR}
\eea
\bea
\frac{1}{T_L} &=& \frac{1}{T_+}+\frac{1}{T_-} = 8 \pi m \bigl (\Pi_c - \Pi _s   \bigr )\, , \\
\frac{1}{T_R}&=&  \frac{1}{T_+}-\frac{1}{T_-}= \frac{8 \pi m^2}{\sqrt{m^2-a^2 }}
\bigl(\Pi_c + \Pi _s  \bigr )\, .
\eea
Note that these solutions with four electric charges have $\Delta\ge0$,
as can be seen from (\ref{Deldef}), and so they have $S_L\ge S_R$, as is
evident from (\ref{SLR}).
In this suggestive form the central charges $C_{L,R}$ of the left-moving and right-moving sector of the the 2D CFT, related to $S_{L,R}$ and $T_{L,R}$ via Cardy relation $S_{L}=\frac{\pi^2}{3}C_{L}T_{L}$ and  $S_{R}=\frac{\pi^2}{3}C_{R}T_{R}$, respectively,  turn out to be the same and equal to:
\be
C_{L}=\frac{3\, S_{L}}{\pi^2\, T_{L}}=48m^3(\Pi_c^2-\Pi_s^2)=\frac{3\, S_{R}}{\pi^2\, T_{R}}=C_{R}\, .
\ee
%As a consequence of these relations, there are number interesting relations:

Again the product of outer and inner horizon entropies is quantized  in
terms of $J$ and $Q_{i}$ ($i=1,2,3,4$) only \cite{Cvetic:2010mn}:
\ben
S_+S_- = S_L^2-S_R^2=4 \pi ^2 \Bigl(J^2 + 64 \prod_{i}
Q_i \Bigr )  \,, \label{SpSmSTU}
\een
%%%%%
which agrees with the result for Kerr-Newman black hole %(\ref{productRN})
after equating $Q_1=Q_2=Q_3=Q_4=\ft14 Q$:
\be
S_+S_- = 4 \pi ^2 \Bigl(J^2 +  \frac{1}{4}Q^4 \Bigr )\,.
\ee

The main challenge here is to obtain the formulae $M=M(S,J,Q_i)$ and
  $S=S(M,J,Q_i)$.
%%
%One deduces that
%\ben
%8 \pi m T_\pm( \Pi_c -\Pi_s )  = \frac{S_+ -S_- }{S_\pm}
%\een
%{\bf which differs from  the first equation of (\ref{spin}) even if}
%$\delta_i =0 $.
As an initial step, we observe the solutions for $S_{\pm}$, due to
relation (\ref{SLR}),  satisfy a quadratic equation:
\be
S^2-2S\,S_L+ 4\pi ^2 \Bigl(J^2 +64 \prod_{i=1}^{4}  Q_i \Bigr )=0\,,
\label{Squad}
\ee
where $S_L$, defined in (\ref{SLR}), depends on $M$ and $Q_i$ ($i=1,2,3,4$)
only. %One also has \cite {CLIII}
Furthermore as $S_L\ge S_R$,  $S_+\ge S_-\ge 0$, where the extremal value $S_+=S_-$ is achieved for $S_R=0$.
The extremal the case either corresponds to the BPS solution $\delta_i\to\infty$, $m\sim a\to 0$ and $Q_i=\frac{m}{2} \exp (2
\delta_i)$ - finite, or to the extremal rotating solution with $m=a$.

Eq. (\ref{Squad})  (which is a special case of (\ref{SquadCC}) implies again  that $T_+$ and $T_-$ have opposite signature. By having an explicit expression for $S_L$ we can actually obtain an explicit expression for the temperatures. Namely, we can express  $S_L$ in terms of $m$ and $Q_i$, by employing:
\be
4m^2\bigl(\Pi_c\pm\Pi_s\bigr)=\Bigl(\prod_{i=1}^{4}
\sqrt{ \sqrt{m^2 + 16 Q_i^2} +m  } \pm  \prod_{i=1}^{4}
\sqrt{ \sqrt{m^2 + 16 Q_i^2} -m }\Bigr)  \,, \label{sumanddiff}
\ee
and
\ben
M= \frac{1}{4} \sum_{i=1}^{4} \sqrt{m^2 + 16 Q_i^2 }\, . \label{mass}
\een
From (\ref{Squad}) we obtain:
\be
\frac{\partial S_L}{\partial S}=\frac{1}{2}\Bigl[1-\frac{ 4\pi ^2 \bigl(J^2
 + 64 \prod_{i=1}^{4}  Q_i \bigr )}{S^2}\Bigr]\, ,
\ee
Furthermore, employing (\ref{sumanddiff}) and (\ref{mass}) we obtain:
\be
\frac{\partial S_L}{\partial S}\big |_{Q_i}=\frac{\partial S_L}{\partial m}\,\frac{\partial m}{\partial M}\,\frac{\partial M}{\partial S}=4\pi m(\Pi_c-\Pi_s)\frac{\partial M}{\partial S}\, ,
\ee
which leads to the explicit expression for the temperature:
\be
T= \frac{\partial M}{\partial S}=\frac{1}{8\pi m\bigl(\Pi_c-\Pi_s\bigr)}
\Bigl[1-\frac{ 4\pi ^2 \bigl(J^2 + 64\prod_{i=1}^{4}
         Q_i \bigr )}{S^2}\Bigr]\,,
\ee
and angular velocity:
\be
\Omega= \frac{\partial M}{\partial J}=\frac{1}{m\bigl(\Pi_c-\Pi_s\bigr)}
\frac{\pi J}{S}=\frac{a\pi}{S}\,.
\ee
These expressions are in agreement (\ref{STsum}) and (\ref{OSsum}), and  explicitly determine  $T_+>0$, $T_-<0$ and $\Omega_{\pm}$, in agreement with direct calculations at the horizons  (\ref{explTO}).

The technical difficulty in obtaining an explicit Christodoulou-Ruffini mass expression is due to the fact that an explicit expression for $S_L$ in terms of $M$ and $Q_i$ is  cumbersome, in general. However, we succeeded in the following special cases.

\subsubsection{Pairwise-equal charges}

The four-charge black-hole 
solutions simplify considerably in the special case of 
pair-wise equal charges (see, for example, \cite{Chong}) 
$Q_1=Q_3$ and $Q_2=Q_4$ where  (\ref{Squad})  can  be solved
explicitly for M:
%%%%%
\ben
M^2 = \frac{\pi}{4S}\Bigl[\Bigl(\frac{S}{ \pi} + 16 Q_1^2  \Bigr )
\Bigl(\frac{S}{ \pi} + 16 { Q_2^2 } \Bigr) +4 J^2\Bigr]\, .
%
%\frac{1}{4}\bigl(\sqrt{\frac{S_{\pm}}{ \pi} } + \sqrt{\frac{\pi}{S_{\pm}}}{ q_1^2 }  \bigr ) \bigl(\sqrt{\frac{S_{\pm}}{ \pi} } + \sqrt{\frac{\pi}{S_{\pm}}}{ q_2^2 }  \bigr )    +\frac{  \pi J^2}{S_\pm} \, .
%
\label{CRq1q2}
\een
%%%%%
 Furthermore (\ref{CRq1q2}) and (\ref{SpSmSTU})  implies:
%%%%%
\be
M^2=\frac{S_+}{4\pi}+\frac{S_-}{4\pi}+ 4 Q_1^2 + 4 Q_2^2 \,.
\ee
%%%%%
For $Q_2=0$ the result reduces to the example of rotating dilatonic black hole
with the dilaton coupling $a=1$.\footnote{Note, however, that when the
black hole is rotating, an axion in the STU supergravity is also turned
on when $Q_1$ and/or $Q_2$ is non-zero (except in the case $Q_1=Q_2$).}
The result reduces to the Kerr-Newman (or Reissner-Nordstr\"om) black
hole expression when $Q_1=Q_2= \ft14 Q$.

It becomes straightforward that the differentiation of  (\ref{CRq1q2})
with respect to $S_{\pm}$ (with  $J$ and $Q_{1,2}$ fixed), produces the expected expressions for $T_{\pm}$, including the sign.

\subsubsection{Three equal non-zero charges}

  It turns out that for the example of three equal non-zero charges,
i.e. $Q_1=Q_2=Q_3=q$ and $Q_4=0$, which  corresponds to the rotating
dilatonic black hole with the dilaton coupling $a=\fft1{\sqrt{3}}$, one can
again obtain an explicit expression for the  the Christodoulou-Ruffini mass:
%%%%%
\ben
M^2 =\frac{\Bigl[16 q^2+\sqrt{64 q^4 +\Bigl(\ft{S_\pm}{\pi}+\ft{4\pi}{S_\pm}
J^2\Bigr )^2}\, \Bigr]^2}{32{q^2}+4\sqrt{ 64 q^4 +\Bigl(\ft{S_\pm}{\pi}
+\ft{4\pi}{S_\pm}J^2\Bigr )^2}\, }\, . \label{CRq3}
\een
%%%%%
(As in the pairwise-equal charge case above, here too an axion is also
turned on if the black hole is rotating.)

\subsubsection{One non-zero charge}

We also note in the case of only one non-zero charge
(say, $Q_1=q=\ft14 m\sinh2\delta$), which corresponds to the
rotating dilatonic black hole with the dilaton coupling $a=\sqrt{3}$,
the Christodoulou-Ruffini mass  can be expressed in the following form:
%%%%%
\be
M^2=\frac{S_L}{8\pi}\bigl(3\cosh\delta+\frac{1}{\cosh\delta}+y\bigr)\, ,
\ee
%%%%%
where $y=\frac{32\pi}{S_{L}} q^2$,
$S_L=\frac{1}{2}\bigl(S_{\pm}+\frac{4\pi^2 J^2}{S_{\pm}}\bigr)$, and
$\cosh\delta$ is a solution of the cubic equation
$\cosh^3\delta-\cosh\delta-y=0$:
%%%%%
\be
\cosh\delta=A^{\frac{1}{3}}+\frac{1}{3A^{\frac{1}{3}}}\, , \quad \quad A=\frac{y}{2}+\sqrt{\frac{y^2}{4}-\frac{1}{27}}\, .
\ee
%%%%%

\subsubsection{Dyonic Kaluza-Klein black hole}\label{kkdyonsec}

  In all the explicit STU supergravity black holes we have discussed so far,
each of the four field strengths carries a charge of a single complexion
(which could be pure electric or pure magnetic).  The most general
possibility is where each field strength carries independent electric
and magnetic charges, as described in the general 8-charge case that was
constructed by Chow and Comp\`ere.  Although explicit, these general
solutions are rather unwieldy.  Here, we discuss a much simpler case,
which is still rather non-trivial, and that goes beyond what we have explicitly
presented so far.  We consider the case where just one of the four field
strengths is non-vanishing, but it carries independent electric and
magnetic charges.  For simplicity we shall restrict attention to the
case of static black holes.  The Lagrangian (in the normalisation we
are using for the STU supergravities) is given by\footnote{This
Lagrangian can also be obtained by means of a circle reduction of
five-dimensional pure Einstein gravity.  For this reason, the black hole
solutions are sometimes referred to as Kaluza-Klein dyons.}
%%%%%
\be
{\cal L}_4 = \sqrt{-g}\, \Big[R -\ft12 (\del\phi)^2 - \ft14 e^{-\sqrt 3\phi}\,
   F^2\Big]\,,\label{KKemd}
\ee
%%%%%
and a convenient way \cite{Lu:2013ura} to present the static dyonic 
black hole solutions is
%%%%%
\bea
ds_4^2 &=& -(H_1 H_2)^{-\ft12}\, f\, dt^2 +
  (H_1 H_2)^{\ft12}\, \big(f^{-1}\, dr^2 +
                 r^2 (d\theta^2 + \sin^2\theta d\varphi^2)\big)\,,\nn\\
\phi&=& \fft{\sqrt3}{2}\, \log \fft{H_2}{H_1}\,,\qquad
    f= 1 - \fft{2\mu}{r}\,,\nn\\
A &=& \sqrt2\, \Big[\fft{(1-\beta_1\, f)}{\sqrt{\beta_1\, \gamma_2}\, H_1}\,
     dt + \fft{2\mu\,\sqrt{\beta_2 \, \gamma_1}}{\gamma_2}\, 
\cos\theta\, d\varphi
    \Big]\,,\nn\\
H_1 &=& \gamma_1^{-1}\, (1-2\beta_1\, f + \beta_1\,\beta_2\, f^2)\,,\qquad
H_2 = \gamma_2^{-1}\, (1-2\beta_2\, f + \beta_1\,\beta_2\, f^2)\,,\nn\\
\gamma_1&=& 1-2\beta_1 +\beta_1\, \beta_2\,,\qquad
\gamma_2= 1-2\beta_2 +\beta_1\, \beta_2\,,
\eea
%%%%%
where $m$, $\beta_1$ and $\beta_2$ are constants that parameterise the
physical mass $M$, electric charge $Q$ and magnetic charge $P$, with
%%%%%
\bea
 M&=& \fft{(1-\beta_1)(1-\beta_2)(1-\beta_1\, \beta_2)\, \mu}{
           \gamma_1\, \gamma_2}\,,\nn\\
Q&=& \fft{\sqrt{\beta_1\, \gamma_2}\, \mu}{\sqrt2 \gamma_1}\,,\qquad
P= \fft{\sqrt{\beta_2\, \gamma_1}\, \mu}{\sqrt2 \gamma_2}\,.\label{MQPdyon}
\eea
%%%%%
A necessary condition for regularity of the black hole is $0\le \beta_i\le 1$.
The entropy of the outer horizon, located at $r=2\mu$, is given by 
%%%%%
\be
S_+= \fft{4\pi \mu^2}{\sqrt{\gamma_1\, \gamma_2}}\,,\label{Sdyon}
\ee
%%%%%
whilst the entropy of the inner horizon, located at $r=0$, is given by
%%%%%
\be
S_- = \fft{4\pi \beta_1\beta_2 \mu^2}{\sqrt{\gamma_1\, \gamma_2}}\,.
\label{Smdyon}
\ee
%%%%%
The product of the entropies on the outer and inner horizons is given by
%%%%%
\be
S_+\, S_- =  64 \pi^2\, P^2\, Q^2\,.
\ee
%%%%%
Note that $S_-$ vanishes if $Q$ or $P$ vanishes.  Note also that 
the dyonic black hole is an example where the invariant $\Delta$, defined
in (\ref{Deldef}), is negative.  Of course since the solutions we
are considering here are static, $(J^2+\Delta)$ is negative too, and so
we are in the regime where $S_L< S_R$ for these black holes, and in fact we
have
%%%%%
\be
S_L= \fft{2\pi\mu^2\, (1-\beta_1\beta_2)}{\sqrt{\gamma_1\gamma_2}}\,,
\qquad 
S_R= \fft{2\pi\mu^2\, (1+\beta_1\beta_2)}{\sqrt{\gamma_1\gamma_2}}\,.
\ee
%%%%% 
 One can straightforwardly calculate the temperatures on the
oouter and inner horizons, finding as usual that the temperature $T_+$ is 
positive and $T_-$ is negative.  The left-moving and right-moving 
temperatures, defined by (\ref{TLTRdef}), then turn out to be
%%%%%
\be
T_L= \fft{\sqrt{\gamma_1\gamma_2}}{8\pi\mu\, (1-\beta_1\beta_2)}\,,\qquad
T_R=\fft{\sqrt{\gamma_1\gamma_2}}{8\pi\mu\, (1+\beta_1\beta_2)}\,.
\ee
%%%%%
These are both non-negative.

   A special case is when the black hole is
extremal, which is achieved in this parameterisation by taking a limit
in which $m$ goes to zero and the $\beta_i$ go to 1.  The result is that
in the extremal case
%%%%%
\be
M_{\rm ext} =  \Big(Q^{\ft23} +P^{\ft23}\Big)^{\ft32}\,,\qquad
S_{\rm ext} = 8 \pi Q P\,.
\ee
%%%%%

   By a straightforward, although somewhat intricate, procedure, one can
eliminate the metric parameters $m$, $\beta_1$ and $\beta_2$ from the
four equations (\ref{MQPdyon}) and (\ref{Sdyon}) that define the
physical mass, charges and entropy, thereby arriving at a
Christodoulou-Ruffini type formula relating these quantities.  If we
first define
%%%%%
\be
   \wtd S= \fft{S}{\pi}\,,
\ee
%%%%%
we find that they and $M$ obey the relation $W(\wtd S,M,Q,P)=0$
where
%%%%%
\bea
&&W(\wtd S, M, Q, P) = \nn\\
&& 4096 M^8+
  \frac{16 M^6 \left(P^2+Q^2\right) \left(P^2 Q^2-8 P Q \wtd S+4 \wtd S^2\right)
    \left(P^2 Q^2+8 P Q \wtd S + 4 \wtd S^2\right)}{P^2 Q^2 \wtd S^2}+\nn\\
&&\frac{M^4}{16 P^2 Q^2 \wtd S^4}\, \Big(P^8 Q^8-48
    P^8 Q^4 \wtd S^2-400 P^6 Q^6 \wtd S^2+1152 P^6 Q^2 \wtd S^4-
    48 P^4 Q^8 \wtd S^2\nn\\
&&-2208 P^4 Q^4
    \wtd S^4 -  768 P^4 \wtd S^6+1152 P^2 Q^6 \wtd S^4-6400 P^2 Q^2 \wtd S^6-
             768 Q^4 \wtd S^6+256
    \wtd S^8\Big)    -\nn\\
&&\frac{M^2(P^2+Q^2)}{64 P^2 Q^2 \wtd S^4}\Big(5 P^8
    Q^8-12 P^8 Q^4 \wtd S^2+ 40 P^6 Q^6 \wtd S^2+160 P^6 Q^2 \wtd S^4
        -12 P^4 Q^8 \wtd S^2\nn\\
&& -352 P^4 Q^4
    \wtd S^4-192 P^4 \wtd S^6+160 P^2 Q^6 \wtd S^4+640 P^2 Q^2 \wtd S^6-
   192 Q^4 \wtd S^6+1280
    \wtd S^8\Big)  -   \nn\\
&&  \frac{\left(P^4+4 \wtd S^2\right)^2 \left(Q^4+4
    \wtd S^2\right)^2 (P^2 Q^2-4 \wtd S^4)^2}{4096 P^2 Q^2 \wtd S^6}
\,.\label{dyonCR}
\eea
%%%%%
This defines a multinomial of 12th order in $\wtd S$,
and $W$ is
invariant under the inversion transformation
$\wtd S\rightarrow Q^2 P^2/(4\wtd S)$.
Note that because $M$ is invariant under the inversion, the coefficients
of each separate power of $M$ in (\ref{dyonCR}) are invariant under
the inversion.

\subsection{Five-dimensional STU supergravity}

Here, we consider black hole solutions in five-dimensional STU
supergravity.  General solutions with mass $M$,
two angular momenta $J_\phi$ and $J_\psi$, and three charges $Q_i$ were
constructed in \cite{CYII} by employing solution generating techniques.
We use principally the conventions of \cite{CLII}, except that
we shall use the labels $\uparrow$ and $\downarrow$ to denote the
sum and difference combinations of the angular momenta and angular velocities 
associated with the $\phi$ and $\psi$ azimuthal coordinates, reserving
$L$ and $R$ to denote the combinations of inner and outer horizon 
quantities, analogous to the definitions used previously for the
four-dimensional STU black holes.   The physical
mass, charges and angular momenta are given by \cite{CLII}
%%%%%
\bea
M &=& {m} \sum_{i=1}^{i=3} \cosh 2 \delta_i\,, \qquad
Q_i=\, m \sinh 2 \delta_i\,,\nn \\
J_{\downarrow}&=& m (l_1 - l_2)(\Pi_c + \Pi_s) \,,\qquad
J_{\uparrow}= m (l_1 + l_2)(\Pi_c - \Pi_s) \,,
\eea
%%%%%
where  $\Pi_c= \prod_{i=1}^{i=3}\cosh\delta_i$, 
$\Pi_s= \prod_{i=1}^{i=3}\sinh\delta_i$, and  
$J_{\downarrow}=\frac{1}{2}(J_\phi- J_\psi)$, 
$J_{\uparrow}=\frac{1}{2}(J_\phi+ J_\psi)$.  
Here the five-dimensional Newton constant is taken to be
$G_5=\frac{\pi}{4}$.  We shall, without loss of generality,
take the rotation parameters $l_1$ and $l_2$ and the charge boost parameters
$\delta_i$ to be non-negative in what follows.

   These black holes have many analogous properties to those of the
four-dimensional STU black holes, except, of course, that they can carry
only electric charges but not magnetic.
  In particular, they have two horizons, with the inner and outer horizon
entropies expressed as \cite{CLII}:
%%%%%
\ben
S_+= S_L + S_R\,,\qquad S_- = S_L-S_R\,,\label{SpSm5}
\een
%%%%%
where
%%%%%
\bea
S_L& =& 2\pi \sqrt{2m^3(\Pi_c+\Pi_s)^2 -J_\downarrow^2}\,, \\
S_R&=& 2\pi \sqrt{2m^3(\Pi_c-\Pi_s)^2 -J_\uparrow^2}\,.
\eea
%%%%
The product of the inner and outer horizon entropies is again
quantised as:
\ben
S_+S_-= 4 \pi ^2 \Bigl(J_\phi\, J_\psi + \prod_{i=1}^{i=3}Q_i \Bigr )
  =  4 \pi^2\Bigl(J_\uparrow^2 - J_\downarrow^2 +
  \prod_{i=1}^{i=3}Q_i \Bigr )\,.\label{entprod5}
\een
Note that as in the case of the four-dimensional STU black holes, here
it would in general be necessary to use an absolute value in the 
expression for $S_-$ in
(\ref{SpSm5}), and on the right-hand side of (\ref{entprod5}), 
since $S_-$ must be non-negative while $S_L$ and
$S_R$, which are both non-negative, could obey either $S_L>S_R$ or $S_L<S_R$
depending on the relative values of the charge and angular momentum 
parameters.  However, our non-negativity assumptions stated above for the
charge and rotation parameters imply that in fact $S_L\ge S_R$ in this
case, and so we can omit the absolute value in the expression for $S_-$, as
we have done in (\ref{SpSm5}), and in (\ref{entprod5}).
 
From the above expressions it follows that $S$ (either $S_+$ or $S_-$)
again obeys a quadratic equation,
%%%%%
\be
S^2 -2S\, S_L +
  4 \pi ^2 \Bigl(J_\uparrow^2 - J_\downarrow^2 
+ \prod_{i=1}^{i=3}Q_i \Bigr )= 0\,.
\ee
%%%%%
Furthermore one can analogously derive the general result that
$T_+$ and $T_-$ have opposite signs, with:
%%%%%
\ben
S_+ T_+ + S_- T_-  =0\,,
\een
%%%%%
and similarly
%%%%%
%%%%%
\ben
\frac{\Omega^{\uparrow}_+}{T_+}+ \frac{\Omega_-^{\uparrow}}{T_-}=0\,,\qquad
\frac{\Omega^{\downarrow}_+}{T_+} - \frac{\Omega_-^{\downarrow}}{T_-}=0\,,
\label{OmT5}
\een
%%%%%
where $\Omega_\pm^\uparrow=\ft12(\Omega_\pm^\phi +\Omega_\pm^\psi)$ and
$\Omega_\pm^\downarrow=\ft12(\Omega_\pm^\phi -\Omega_\pm^\psi)$.
(The relative signs between the terms in these two equations in (\ref{OmT5}) 
are the opposite
of those given in \cite{CLII}, because in that paper $\kappa_-$ was taken to
be positive.)

   The black holes obey the usual first laws on the outer and inner horizons:
%%%%%
\be
dM =T_\pm\, dS_\pm + \Omega_\pm^\uparrow dJ_\uparrow +
  \Omega_\pm^\downarrow dJ_\downarrow + \Phi_\pm^i dQ_i\,.
\ee
%%%%%
As in the four-dimensional case, the calculation of scattering amplitudes
in the black-hole background shows that they factorise into left and right
sectors with Boltzman factors corresponding to temperatures $T_L$ and
$T_R$ given by (\ref{TLTRdef}) \cite{CLII}.  Together with the normalisation of
$S_L$ and $S_R$, such that $S_+=S_L+S_R$ in accordance with the
interpretation of the entropy as the sum of left-moving and right-moving
contributions, one can then establish by rewriting the first
laws $dM=T_\pm \, dS_\pm+\cdots$ in terms of left and righ-moving quantities
that $\ft12 dM = T_L\, dS_L+\cdots$ and $\ft12 dM = T_R\, dS_R+\cdots$, and
so each
of the sectors contributes one half the total mass of the black hole. 
Matching the first laws for arbitrary variations of the parameters then
allows one to read off the appropriate definitions of the left-moving
and right-moving angular momenta and electric potentials.  Thus one finds
the first laws
%%%%%
\bea
\fft12 dM &=& T_L\, dS_L + \Omega_L^\uparrow\, dJ_\uparrow + 
   \Omega_L^\downarrow\, dJ_\downarrow + \Phi^i_L\, dQ_i\,,\nn\\
\fft12 dM &=& T_R\, dS_R + \Omega_R^\uparrow\, dJ_\uparrow +
   \Omega_R^\downarrow\, dJ_\downarrow + \Phi^i_R\, dQ_i\,,
\eea
%%%%%
where
%%%%%
\bea
\Phi^i_L \!\!\!&=& \!\!\! 
 T_L\Big(\fft{\Phi^i_+}{2T_+} + \fft{\Phi^i_-}{2T_-}\Big)\,,\
\Omega_L^\uparrow= T_L\Big(\fft{\Omega^\uparrow_+}{2 T_+} + 
    \fft{\Omega^\uparrow_-}{2 T_-}\Big)\,,\
\Omega_L^\downarrow= T_L\Big(\fft{\Omega^\downarrow_+}{2 T_+} 
   + \fft{\Omega^\downarrow_-}{2 T_-}\Big)\,, \nn\\
\Phi^i_R\!\!\! &=& \!\!\! T_R
  \Big(\fft{\Phi^i_+}{2T_+} - \fft{\Phi^i_-}{2T_-}\Big)\,,\
\Omega_R^\uparrow= T_R\Big(\fft{\Omega^\uparrow_+}{2 T_+} - 
    \fft{\Omega^\uparrow_-}{2 T_-}\Big)\,,\
\Omega_R^\downarrow= T_R\Big(\fft{\Omega^\downarrow_+}{2 T_+} - 
    \fft{\Omega^\downarrow_-}{2 T_-}\Big)\,.
\eea
%%%%%
In view of the relations (\ref{OmT5}), one finds
%%%%%
\be
\Omega^\uparrow_L=0\,,\qquad 
\Omega^\downarrow_L = \fft{T_L}{T_+}\, \Omega^\downarrow_+\,;\qquad\qquad
\Omega^\uparrow_R=\fft{T_R}{T_+}\, \Omega^\uparrow_+\,,\qquad 
\Omega^\downarrow_R=0\,.
\ee
%%%%%
Thus we see that the angular momentum $J_\uparrow$ and the associated
angular velocity $\Omega^\uparrow$ enters only in the right-moving
first law and in $S_R$, while the angular momentum $J_\downarrow$
and associated angular velocity $\Omega^\downarrow$ enters only in
the left-moving first law and in $S_L$.  Note that as in four dimensions, 
$T_L$ and $T_R$ are both non-negative.

  The Smarr formulae for
the left-moving and right-moving sectors agree with the ones derived
in \cite{CLII}:
%%%%%
\be
\fft 12 M = \fft32 T_L\, S_L + \fft32 \Omega_L^\downarrow\, J_\downarrow
+ \Phi^i_L\, Q\,,\qquad
\fft 12 M = \fft32 T_R\, S_R + \fft32 \Omega_R^\uparrow\, J_\uparrow
+ \Phi^i_R\, Q\,.
\ee
%%%%%

 The expression for the Christodoulou-Ruffini formula in terms
solely of the conserved charges, angular momenta, mass and entropy are too
cumbersome to present explicitly. Even in the case of three equal charges,
the mass is determined by a cubic equation.  

\subsection{Einstein-Maxwell-Dilaton black holes}\label{EMDsec}

There exists a more general class of black holes in the theory of
Einstein-Maxwell gravity with an additional dilatonic
scalar field, which is coupled to
the Maxwell field with a dimensionless coupling constant
$a$, with
the Lagrangian
%%%%%
\be
{\cal L}= \sqrt{-g}\, \Big(R - 2(\del\phi)^2 - e^{-2a\phi}\,
       F^2\Big)\,,
\ee
%%%%%
The electrically-charged black-hole solution can be written 
as \cite{gibbemd,Gibbons:1987ps,gahost}
%%%%%
\bea
ds^2&=&- \Big(1-\fft{r_+}{r}\Big) \Big(1-\fft{r_-}{r}\Big)^b\, dt^2 +
  \Big(1-\fft{r_+}{r}\Big)^{-1}\, \Big(1-\fft{r_-}{r}\Big)^{-b}\, dr^2\nn\\
&& +\, r^2 \Big(1-\fft{r_-}{r}\Big)^{1-b}\, d\Omega^2\,,\nn\\
e^{2a\phi} &=& \Big( 1- \fft{r_-}{r}\Big)^{1-b}\,,\qquad 
A= \fft{Q}{r}\, dt\,,
\eea
%%%%%
where
%%%%%
\be
b = \fft{1-a^2}{1+a^2}\,.
\ee
%%%%%
The relevant thermodynamic quantities for these 
black holes in this theory are given by
%%%%%
\bea
S &=&  \pi r_+^2 \left (1- {\fft{ r_-}{r_+}} \right )^{1-b}\,,
  \qquad T ={ \fft1{4 \pi r_+ }} \left ( 1- {\fft{r_-}{r_+}} \right ) ^{b}\,,
\nn\\
Q &=& \sqrt { \fft{r_+ r_-}{1 + a ^2} }\,,\qquad M = \ft12 ( r_+ + b\, r_-)\,,  
\qquad\Phi =
{ \fft1{ \sqrt { 1 + a^2 } }}     \sqrt { \fft{r_-}{r_+}}\,,\label{emdthermo}
\eea
%%%%%
where $r_+$ is the radius of the outer horizon, and $r_-$ is 
a singular surface unless $a=0$.  Since by assumption $r_+\ge r_-$, it follows
that
%%%%%
\ben
M > \frac{|Q|}{\sqrt{1+a^2}}\,.
\een
%%%%%
This is consistent with the BPS bound derived in \cite{gikalototr} using ``fake 
supersymmetry.''

The Smarr relations continue to hold and the Gibbs free energy
is again given by
\ben
G= TS= \ft14 (r_+ - r_-).
\een
The coordinates $\{r_+, r_- \}$ are now
related to the coordinates $\{ T, \Phi \}$
by
\ben r_+= { \fft1{ 4 \pi T}} \Big( 1- (1+a^2) \Phi ^2\Big)^b
\een
and
\ben
r_-= { \fft{(1+a^2) \Phi ^2 }{ 4 \pi T}} \Big( 1 - ( 1+ a^2 )
\Phi ^2 \Big)^b\,.
\een
%%%%%
Thus the Gibbs energy as a function of $\{ T, \Phi\}$ is given by
%%%%%
\ben
G= {\fft1{16 \pi T} } \Big( 1 - ( 1+ a^2) \Phi ^2 \Big)^{1+b}\,.
\een
 %%%%%

  As discussed in section \ref{hongthermosec}, the Ricci scalar of the 
Helmholtz free energy metric $ds^2(F)= -dS \,dT+ d\Phi \,dQ$ 
will be singular on the Davies curve where the heat capacity at 
constant charge changes sign.  It is easiest to use $r_+$ and $r_-$ as
the coordinate variables in this calculation, which gives
%%%%%
\be
R= \fft{4(1+a^2)^2\, r_+}{[(1+a^2) r_+ -(3-a^2) r_-]^2}\,.
\ee
%%%%%
Thus the Davies curve is given by
%%%%%
\be
{\fft{r_-}{r_+}}  = {\fft{1+a^2}{3-a^2}}\,,\label{daviesemd}
\ee
%%%%%
which implies
%%%%%
\be
\fft{Q^2}{M^2}= \fft{3-a^2}{(2-a^2)^2}\,.
\ee
%%%%%
Since we must have $r_- < r_+$, a solution for (\ref{daviesemd}) exists 
only for $a^2<1$.  
The spinodal curve thus projects down to the parabola in the
$S-Q$ plane given by
%%%%%
\be
S = \big( 3-a ^2\big)^{ \fft{1-a^2}{1+a^2} }
 ~~ 2 ^ {\fft{ 2 a^2 }{ 1+a ^2 } } ~~
\big( 1-a ^2\big)^{ \fft{ 2a^2}{ 1+a ^2}} ~~ \pi Q^2\,.
\ee
%%%%%

   From (\ref{emdthermo}), one can in general solve for $r_+$ and $r_-$ in
terms of $M$ and $Q$, obtaining
%%%%%
\be
r_+= M+ \sqrt{M^2-(1-a^2) Q^2}\,,\qquad
r_-= \fft1{b}\, \Big(M-\sqrt{M^2-(1-a^2) Q^2}\Big)\,,
\ee
%%%%% 
and hence express $S$ in terms of $M$ and $Q$ \cite{ped}:
%%%%%
\ben
\frac{S}{\pi} =
\Bigl (M + \sqrt{M^2 - (1-a^2) Q^2} \Bigr )^2 \Bigl(1 -
\frac{ (1+a^2) Q^2 }{\bigl(M + \sqrt{M^2 - (1-a^2) Q^2}\bigr )^2  }
\Bigr ) ^{\frac{2a^2}{1+a^2}} \,.\label{mess}
\een
%%%%%
If $a^2>0$ the entropy vanishes at extremality, namely $r_+=r_-$ and hence
$|Q|=\sqrt{1+a^2}\, M$.  Then $r=r_+=r_-$ is a point-like singularity
and there is no inner horizon. One can also, in general, express the
entropy in terms of $r_+$ and $Q$, using
%%%%%
\ben
\Bigl( \frac{S}{\pi r_+^2} \Bigr)^{\frac{1+a^2}{2a^2}} = 
  1- \frac{(1+ a^2)  Q^2}{r_+^2} \,.
\label{mess2}
\een
%%%%%%

Particular  cases include the following, which also arise as
special cases of STU Black holes:

\begin{itemize}
\item $a=0$ is the Reissner-Nordstr\"om case.

\item $a^2  =\frac{1}{3}$ is a  reduction
 of Einstein-Maxwell in 5 dimensions.

\item $a^2=1$ is the so-called string case. We have
\ben
\frac{S}{\pi}=4M^2 -2Q^2 \,,\qquad M = \half \sqrt{\frac{S}{\pi} + 2Q^2} \,.
 \een
%%%%%
The spinodal curve coincides with the $Q$-axis and the Gibbs surface is
{\sl nowhere} convex.
It is a hyperbolic paraboloid for which the
Ruppeiner metric is flat \cite{ped}.  
The temperature is given by
\ben
T= \frac{1}{4\pi \sqrt{\frac{S}{\pi}+2Q^2} }= \fft1{8\pi M}\,,
\een
and is always positive.  It  goes to a non-vanishing value
at extremality.
The heat capacity at constant charge is given by
\ben
C_Q= - \frac{1}{ 8\pi^2( \frac{S}{\pi} + 2 Q^2) ^{\frac{3}{2} }} =
  -\fft1{64\pi^2 M^3}
\een
and is always negative, and is also non-vanishing at extremality
\cite{Gibbons:1987ps}.

\item $a^2=3$ is the Kaluza-Klein black hole.
\end{itemize}

\subsection{Two-field dilatonic black holes}

   Here we review a class of theories \cite{Lu:2013eoa} which are similar to the
Einstein-Maxwell-Dilaton (EMD) theory of the previous subsection, but with
two field strengths rather than just one. The Lagrangian,
in an arbitrary dimension $D$, is given by
%%%%%
\be
{\cal L}_D = \sqrt{-g}\, \Big( R -\ft12(\del\phi)^2 - \ft14 e^{a_1\, \phi}\,
F_1^2 -\ft14 e^{a_2\, \phi}\, F_2^2\Big)\,.\label{2fieldlag}
\ee
%%%%%
The advantage of considering this extension of EMD theory is that by
choosing the coupling constants $a_1$ and $a_2$ appropriately, we can
find general classes of static black hole solutions with two
horizons, and one can study the thermodynamic properties at both the
outer and inner horizon.

  If we turn on both the gauge fields $A_i$ independently, the theory for
general $(a_1,a_2)$ does not admit explicit black hole solutions.
We shall determine the condition on $(a_1,a_2)$ so that the system will
give such explicit solutions.  It is advantageous for later purpose that
we reparameterize these dilaton coupling constants as
%%%%%
\begin{equation}
a_1^2=\ft{4}{N_1} - \ft{2 (D-3)}{D-2}\,,\qquad
a_2^2 = \ft{4}{N_2}- \ft{2(D-3)}{D-2}\,.\label{Nidef}
\end{equation}
%%%%%
(Note that $N_1$ and $N_2$ are not necessarily integers.)  For the
$a_i$ to be real, we must have
%%%%%
\begin{equation}
0<N_i \le \ft{2(D-2)}{D-3}\,.
\end{equation}
%%%%%
(If both $N_i$ are outside the range, the Lagrangian could still be made
real by sending $\phi\rightarrow {\rm i}\phi$, corresponding to having
a ghost-like dilaton.  We shall not consider this possibility here.)

   Here we shall consider the case where $a_1$ and $a_2$ obey the constraint
%%%%%
\be
a_1 a_2 = -\fft{2(D-3)}{D-2}\,,\label{a1a2id}
\ee
%%%%%
which implies the identities
%%%%%
\begin{equation}
N_1 a_1 + N_2 a_2 =0\,,\qquad N_1 + N_2 = \fft{2(D-2)}{D-3}\,.\label{Niident}
\end{equation}
%%%%%
It follows from the second identity in (\ref{Niident}) that both $N_i$ can
take integer values only in four and five dimensions, with $N_1 + N_2=3$
and 4 respectively. The solutions with positive integers for $N_i$ are
known black holes in relevant supergravities.

  With $a_1$ and $a_2$ obeying (\ref{a1a2id}), one can find
black hole solutions, given by \cite{Lu:2013eoa}
%%%%%
\begin{eqnarray}
ds^2 &=& -(H_1^{N_1} H_2^{N_2})^{-\fft{(D-3)}{D-2}} f dt^2 +
(H_1^{N_1} H_2^{N_2})^{\fft{1}{D-2}} (f^{-1} dr^2 + r^2 d\Omega_{D-2}^2)\,\nn\\
A_1&=& \fft{\sqrt{N_1}\,c_1}{s_1}\, H_1^{-1} dt\,,\qquad
A_2 = \fft{\sqrt{N_2}\,c_2}{s_2}\, H_2^{-1} dt\,,\nn\\
\phi &=& \ft12 N_1 a_1 \log H_1 + \ft12 N_2 a_2 \log  H_2\,,\qquad
f=1 - \fft{\mu}{r^{D-3}}\,,\nn\\
H_1&=&1 + \fft{\mu s_1^2}{r^{D-3}}\,,\qquad
H_2 = 1 + \fft{\mu s_2^2}{r^{D-3}}\,,\label{solution1}
\end{eqnarray}
%%%%%
where we are using the standard notation where $s_i=\sinh\delta_i$ and
$c_i=\cosh\delta_i$.
The mass and charges are given by
%%%%%
\begin{eqnarray}
M&=&\fft{(D-2)\mu\,\omega_{\sst{D-2}}}{16\pi}\Big(1 +
\fft{D-3}{D-2}\left(N_1\, s_1^2 + N_2\, s_2^2\right)\Big)\,,\nn\\
Q_i &=&\fft{(D-3)\mu\,\omega_{\sst{D-2}}}{16\pi} \sqrt{N_i} \,c_i s_i\,,
\label{masscharges}
\end{eqnarray}
%%%%%
where $\omega_{\sst{D-2}}$ is the volume of the unit $(D-2)$-sphere.
The outer horizon is located at $r_0=\mu^{1/(D-3)}$, and the entropy is
given by
%%%%%
\begin{equation}
S=S_+\equiv\ft14 \omega_{\sst{D-2}}\, \mu^{\fft{D-2}{D-3}}\,
c_1^{N_1} c_2^{N_2} \,.
\end{equation}
%%%%%
The inner horizon is located at $r=0$, and we have
%%%%%
\be
S_-\equiv \ft14  \omega_{\sst{D-2}}\, \mu^{\fft{D-2}{D-3}}
s_1^{N_1} s_2^{N_2} \,.
\ee
%%%%%

   Multiplying the two entropies gives the product formula
%%%%%
\be
S_+ S_- = S_{\rm ext}^2\,,
\ee
%%%%%
where
%%%%%
\be
S_{\rm ext} = 4^{\fft{D-1}{D-3}}\, \Big(\fft{\pi}{D-3}\Big)^{\fft{D-2}{D-3}}
\, \omega_{D-2}^{-\fft{1}{D-3}}\,
\left(\ft{Q_1}{\sqrt{N_1}}\right)^{\fft12 N_1}
\left(\ft{Q_2}{\sqrt{N_2}}\right)^{\fft12 N_2}\,.
\ee
%%%%%
Thus the entropy product is independent of the mass.

    There exists an extremal limit in which we send
$\mu\rightarrow 0$ while keeping the charges $Q_i$ non-vanishing.  In
this limit, the inner and outer horizons coalesce and the near-horizon
geometry becomes AdS$_{D-2}\times S^2$.  The mass now
depend only on the charges, and is given by
%%%%%
\begin{equation}
M_{\rm ext}=\sqrt{N_1}\, Q_1 + \sqrt{N_2}\,  Q_2\,.\label{massentropyextr}
\end{equation}
%%%%%
It is useful to define
%%%%%
\be
\widetilde M=\fft{16\pi}{(D-2)\omega_{D-2}}M\,,\qquad \widetilde Q_i = \fft{8\pi}{(D-3)\omega_{D-2}\sqrt{N_i}} Q_i\,,\qquad
\widetilde S=\fft{1}{\omega_{D-2}} S\,,
\ee
%%%%%
and then we have
%%%%%
\be
s_i^2=\fft{\sqrt{\widetilde Q_i^2 +16\mu^2}}{2\mu} - \fft12\,.
\ee
%%%%%

   Some specific examples are as follows:
\medskip

\noindent{\bf Case 1: $D=4$, $N_1=N_2=2$}:

%%%%%
\be
\widetilde M^2-\frac{4\left(\widetilde Q_1^2+\widetilde S\right)
\left(Q_2^2+\widetilde S\right)}{ \widetilde S}=0\,.
\ee
%%%%%
We can define
%%%%%
\be
\hat S=\fft{\widetilde S}{\widetilde Q_1 \widetilde Q_2}\,,
\ee
%%%%
and then
%%%%%
\be
\widetilde M^2 - 4 (\widetilde Q_1^2 + \widetilde Q_2^2) - 4
\widetilde Q_1 \widetilde Q_2 \Big( \hat S + \fft{1}{\hat S}\Big)=0\,.
\ee
%%%%%

\bigskip

\noindent{\bf Case 2: $D=4$, $N_1=1, N_2=3$:}

%%%%%
\bea
\widetilde M^6&+&\frac{\widetilde M^4 \left(\widetilde S^4-3
\widetilde S^2 \widetilde Q_1^4-15 \widetilde S^2 \widetilde Q_1^2
\widetilde Q_2^2+\widetilde Q_1^2
\widetilde Q_2^6\right)}{\widetilde S^2 \widetilde Q_1^2}
\nn\\
&-&\frac{\left(4 \widetilde S^4+\widetilde S^2 \widetilde Q_1^4-
6 \widetilde S^2 \widetilde Q_1^2 \widetilde Q_2^2-
3 \widetilde S^2 \widetilde Q_2^4+
4\widetilde Q_1^2 \widetilde Q_2^6\right)^2}{\widetilde S^4 \widetilde Q_1^2}
\nn\\
&-&\frac{\widetilde M^2}{\widetilde S^2 \widetilde Q_1^2}
\Big(20 \widetilde S^4 \widetilde Q_1^2+12 \widetilde S^4
\widetilde Q_2^2-3 \widetilde S^2 \widetilde Q_1^6-3 \widetilde S^2
\widetilde Q_1^4 \widetilde Q_2^2
  -57 \widetilde S^2 \widetilde Q_1^2 \widetilde Q_2^4-
\widetilde S^2 \widetilde Q_2^6   \nn\\
&& \qquad\qquad  +20 \widetilde Q_1^4 \widetilde Q_2^6+
12 \widetilde Q_1^2 \widetilde Q_2^8\Big)=0
\,.
\eea
%%%%%

\bigskip

\noindent{\bf Case 3: $D=5$, $N_1=1, N_2=2$:}

%%%%%
\bea
0&=&\widetilde M^4+\frac{\widetilde M^3
\left(4 \widetilde S^4+
\widetilde Q_1^2 \widetilde Q_2^4\right)}{3 \widetilde S^2 \widetilde Q_1^2}-
\frac{4 \widetilde M^2 \left(8 \widetilde Q_1^4+
20 \widetilde Q_1^2 \widetilde Q_2^2-
           \widetilde Q_2^4\right)}{9 \widetilde Q_1^2}\nn\\
&&-\frac{8 \widetilde M \left(2 \widetilde Q_1^2+\widetilde Q_2^2\right)
\left(4 \widetilde S^4+\widetilde Q_1^2
\widetilde Q_2^4\right)}{3 \widetilde S^2 \widetilde Q_1^2}\\
&&
-\frac{4 \left(432 \widetilde S^8-64 \widetilde S^4 {\widetilde Q_1}^6+192
\widetilde S^4 \widetilde Q_1^4 {\widetilde Q_2}^2+24 \widetilde S^4
\widetilde Q_1^2 \widetilde Q_2^4+64 \widetilde S^4 \widetilde Q_2^6+
27 \widetilde Q_1^4 \widetilde Q_2^8\right)}{81 \widetilde S^4
\widetilde Q_1^2}\,.\nn
\eea
%%%%%

\bigskip

\noindent{\bf Case 4: General $D$, but with $N_1=N_2=(D-2)/(D-3)$}

  These cases lie, in general, outside the realm of supergravity theories.
We have
%%%%%
\be
\widetilde M^2 -4(\widetilde Q_1^2 + \widetilde Q_2^2) -
\Big(16^{\frac{1}{D-2}} \widetilde Q_1^2 \, \widetilde Q_2^2
\widetilde S^{\frac{2}{D-2}-2}+16^{\frac{D-3}{D-2}}
\widetilde S^{2-\frac{2}{D-2}}
\Big)=0\,.
\ee
%%%%%

   Entropy super-additivity is difficult to prove in general, but we can
at least look at the case of extremal black holes, for which
%%%%%
\be
S_{\rm ext} \sim \sqrt{Q_1^{N_1} Q_2^{N_2}}\,.
\ee
%%%%%
It seems that super-additivity will be satisfied if $N_1 + N_2 \ge 2$, and
in fact, from (\ref{Niident}), we have $N_1+N_2>$ in all dimensions.

\section{Entropy Product and Inversion Laws}

  It is well known from many examples that if a black hole has two horizons
then the product of the areas, or equivalently entropies, of these
horizons is equal to an expression written purely in terms of the
conserved charges and angular momenta
\cite{Cvetic:2010mn,Cvetic:2013eda}. Thus we may write
%%%%%
\be
S_+\, S_- = K({\bf Q},{\bf J})\,,\label{entropyprod}
\ee
%%%%%
where ${\bf Q}$ represents the complete set of charges carried by the
black hole, and ${\bf J}$ represents the set of angular momenta.
(Generalisations arise
also if there are more than two horizons or ``pseudo-horizons'' (see,
for example, \cite{Cvetic:2010mn}.))  We also saw various examples in the
previous section where there is a Christodoulou-Ruffini formula relating
the entropy to the mass, charges and angular momenta, of the form
%%%%%
\be
W(S,M,{\bf Q},{\bf J})=0\,,\label{CRgen}
\ee
%%%%%
for which there was a symmetry under a certain inversion of the entropy,
$S\rightarrow S'\sim 1/S$.

   Here, we make some observations about the relation between these
properties of the black hole entropy.  First, we note that when one
derives a Christodoulou-Ruffini formula of the form (\ref{CRgen}), one
uses properties of the metric functions that determine the horizon radius
in terms of the metric parameters, and hence implicitly they determine
the horizon radius in terms of $M$, ${\bf Q}$ and ${\bf J}$.  This means
that when one arrives at the Christodoulou-Ruffini relation (\ref{CRgen}),
the expression will necessarily be valid not only when $S=S_+$, but also
when instead $S=S_-$.  Since $S_+$ and $S_-$ are related by the
product formula (\ref{entropyprod}), this means that if $S$, the entropy
of the outer horizon, obeys (\ref{CRgen}) then we will also have
%%%%%
\be
W\Big(\fft{K({\bf Q},{\bf J})}{S}, {\bf Q},{\bf J}\Big) =0\,.
\ee
%%%%%
In other words, the Christodoulou-Ruffini formula will be
invariant\footnote{Or {\it conformally} invariant, depending on how one
chooses the overall multiplicative factor when defining
$W(S,M,{\bf Q},{\bf J})$.} under the inversion symmetry
%%%%%
\be
S\rightarrow \fft{K({\bf Q},{\bf J})}{S}\,,\label{inversiongen}
\ee
%%%%%
where $K({\bf Q},{\bf J})$ is the right-hand side of the entropy-product
formula (\ref{entropyprod}).

   In some cases, for example in the case of STU black holes where
$J=0$ and insufficiently many charges are turned on, there is only one
horizon and so there is no entropy-product formula.  In such cases the
argument above demonstrating the existence of an inversion symmetry of
the Christodoulou-Ruffini relation breaks down.  Indeed, in section
\ref{EMDsec} we saw examples where, for this reason,
the Christodoulou-Ruffini relation had no inversion symmetry.

   One important consequence of the inversion symmetry of the
Christodoulou-Ruffini relation $M=M(S,{\bf Q},{\bf J})$ is that the
relation $S_+\, T_+ + S_-\, T_-=0$, seen, for example, for the
STU black holes in  (\ref{STsum}), 
is true quite generally.  Since the
temperature is given by $\del M/\del S$ at fixed ${\bf Q}$ and ${\bf J}$ we
have
%%%%%
\bea
T &=& \fft{\del M(S,{\bf Q},{\bf J})}{\del S}=\fft{\del}{\del S}\,
  M\Big(\fft{K}{S},{\bf Q},{\bf J}\Big)\nn\\
&=& -\fft{K}{S^2} \,\fft{\del M(S',{\bf Q},{\bf J})}{\del S'}\Big|_{S'=K/S}
\,,\label{TSderiv}
\eea
%%%%%
where $K= K({\bf Q},{\bf J})$ is the numerator in the inversion formula
(\ref{inversiongen}).  Taking $S=S_+$ we therefore have $S'=S_-$, and so
we find from (\ref{TSderiv}) that
%%%%%
\be
  T_+\, S_+ + T_-\, S_-=0\label{TSrelgen}
\ee
%%%%%
whenever there is an entropy-product rule of the form (\ref{entropyprod})
and the related inversion symmetry under (\ref{inversiongen}).

\section{Asymptotically AdS and dS Black Holes}

   In this section we shall extend the previous discussion to the
case of a non-vanishing cosmological constant.  If the cosmological
constant is negative, the situation is similar to the case when it
vanishes.  However, if the cosmological constant is positive a new
feature arises, namely the occurrence of an additional
``cosmological'' horizon outside the black hole event horizon.  Typically
the surface gravity at the cosmological horizon is negative.

\subsection{Kottler}

Either we regard $\Lambda$ as a fixed constant
or  as an intensive variable which may be
varied, in which case we obtain an analogy with a gas with
positive  pressure
\ben
P=-\frac{\Lambda}{8 \pi} \,.
\een
In the first case we should think of
the Abbott-Deser mass $M$ as the total energy.
In the second  case, we should instead think of it as the total enthalpy
\cite{Kastor:2009wy,Cvetic:2010jb}.
In both cases we have
%%%%%
\ben
2M=  \bigl (
\frac{S}{\pi}
  \bigr ) ^\half
 -\frac{\Lambda}{3}
 \bigl (\frac{S}{\pi} \bigr
 ) ^{ \frac{3}{2} }\,,
\een
%%%%%
and in both cases
%%%%%
\ben
T= \frac{\p M}{\p S }\Big | _\Lambda  = \frac{1}{4 \pi} \Bigl[
  \sqrt{\frac{\pi}{S}} - \Lambda \sqrt{\frac{S}{\pi} } \Bigr ]
\een
%%%%%
and the heat capacity at constant pressure is given by
 \ben
 C_\Lambda = T\, \Big(\frac{\p T}{\p S} \Big | _\Lambda\Big)^{-1}
= \fft{2S(\Lambda S -\pi)}{\Lambda S +\pi}
  \,.
 \een
%%%%%
We now consider the two cases where $\Lambda<0$ and $\Lambda>0$.

 \subsubsection{$\Lambda <0$}

The temperature $T$ is a positive, monotonic-increasing
function of entropy $S$ at fixed pressure $P$.
The isobaric curve in the $S-M$ plane   has a point of inflection
at which the heat capacity changes sign when
\ben
\frac{S}{\pi}= -\frac{1}{\Lambda}\,, \qquad M= \frac{2}{3 \sqrt{-\Lambda}}\,,
\een
where  the slope,  and hence the temperature, has
a minimum value;
\ben
 T=T_{\rm min}= \frac{1}{2 \pi} \sqrt{-\Lambda}   \,.
\een
It follows that for  fixed negative $\Lambda$
there are no black holes with temperatures less than  $T_{\rm min} $.
For temperatures above $T_{\min}$ there are two black holes, one with
a mass smaller  than $ \frac{2}{3 \sqrt{-\Lambda}}$ and the other with a
mass greater than $\frac{2}{3 \sqrt{-\Lambda}}$.

The radius $r_H$ of the critical black hole, where the two branches
coalesce, is given by
%%%%%
\ben
r_H= \frac{3}{2}M\,.
\een
%%%%%
This is the location where the  heat capacity diverges. It
is connected with the
Hawking-Page phase transition \cite{Hawking:1982dh,Witten:1998qj}.
There is actually a region of masses $M_{HP}>M>M_{cr}$  where the $AdS_4$
space is entropically favoured; however the black hole still has a
positive heat capacity. As with the Reissner-Nordstr\"om black hole,
it has been shown that  the sign of the lowest
eigenvalue of the Lichnerowicz  operator changes sign
as the heat capacity changes sign \cite{Prestidge:1999uq}.

\subsubsection{$\Lambda>0$}
We have a negative pressure, $P<0$.
If $M$ is assumed  positive
we have two horizons, a black hole horizon with
%%%%%
\ben
0< S \le \fft{\pi}{\Lambda} \,,
\een
%%%%%
and positive temperature $T=\del M/\del S$,
and a cosmological horizon with
\ben
\frac{\pi}{\Lambda} \le S \le  \frac{3\pi}{\Lambda} \,,
\een
for which $T=\del M/\del S<0$, and hence the temperature is negative.
The heat capacity is therefore always negative.
The temperature vanishes when the two horizons coincide,  that is if
%%%%%
\ben
\frac{S}{\pi} = \Lambda\,,
\een
%%%%%
at which the mass has a maximum of
%%%%%
\ben
M= \frac{1}{3 \sqrt{\Lambda}}\,.
\een
%%%%%

In summary, we
have two horizons; a black hole horizon and a cosmological horizon.
The entropy of the former is smaller then or equal to the entropy of
the latter.
It seems most appropriate to regard $M$ as the enthalpy.
In this case the black hole horizon  has positive temperature and the
cosmological  horizon has negative temperature.
This differs from the usual interpretation in which both
temperatures
are taken to be positive. In effect one takes
$T_C = \frac{|\kappa_C|}{2 \pi}$  where $\kappa_C$ , where $\kappa_C$ is
the surface gravity of the event horizon
\cite{Gibbons:1977mu,Frolov:2002va,Dolan:2013ft, Gregory:2017sor}.
However, even if one follows the conventional  interpretation
it should be borne in mind that it is not an equilibrium system and there
is no period in imaginary  time which would produce an everywhere non-singular
gravitational instanton, except when the black hole is absent
as in \cite{Gibbons:1977mu,Gibbons:1976ue}.

\subsection{Reissner-Nordstr\"om-de Sitter}

\subsubsection{$\Lambda <0$}

If $r = \sqrt{\frac{S}{\pi} }$ is the radius in the area
coordinate, we have
\ben
2M = r + \frac{Q^2}{r}  + g^2  r^3 \,.
\een
where $\frac{\Lambda}{3}=-g^2 $.
using the fact that
%%%%%
\ben
\frac{\p }{\p S}= \frac{1}{2 \pi r} \frac{\p }{\p r}
\een
%%%%%
one finds that
%%%%%
\ben
T= \frac{\p M}{\p S}  = \frac{1}{4 \pi r}
\bigl ( 1- \frac{Q^2}{r^2} + 3 g^2 r^2 \bigr)
\een
%%%%%
and thus $T$ vanishes at $r=r_{\rm extreme} $ where
\ben
r^2_{\rm extreme}= \frac{1}{6g^2} \bigl( \sqrt{1+12 Q^2 g^2 } -1 \bigr ) \,.
\een
One has
%%%%%
\ben
\frac{\p ^2 M}{\p S ^2} = \frac{1}{4 \pi^2 }
\bigl(- \frac{1}{r^3} + \frac{Q^2}{r^5} + \frac{3 g^2}{r}   \bigr )
\een
%%%%%
If $6|gQ| <1$ there are two inflection points at
which the heat capacity changes sign at $r= r_{\rm inflection}$ where
%%%%%
\ben
r^2_{\rm inflection} = \frac{1}{6 g^2} \bigl( 1 +\pm \sqrt {1- 36 Q^2 g^2   } \bigr )  \een
%%%%%
If we take the limit that $Q^2 \rightarrow 0$ we obtain
the spinodal curve of the
the Hawking-Page  phase transition \cite{Hawking:1982dh}
and if we take the limit
$g^2 \rightarrow 0$ we obtain the spinodal curve of the
Davies phase transition  \cite{Davies:1978mf}.
The two curves meet at the critical point $6 |gQ|=1$.

\subsubsection{$\Lambda>0$}
This case admits new qualitatively different phenomena
since both a black hole and a cosmological
horizon are present. This was extensively investigated
in  1989 \cite{Mellor:1989ac,Mellor:1989wc,Mellor:1989gi,Moss:1989js, Davies:1989ew, Davies:1989ey}. In all these references the absolute value
of the surface gravity was taken and the  and so the temperature
of both horizons was take to be positive.
For the choice $M=|Q|$
the temperatures of the black hole and cosmological horizon
were observed to be equal. This allowed the construction of a
gravitational instanton. To ensure that the electromagnetic
field is real on the Euclidean section it is most convenient to assume
that the electro-magnetic field is purely magnetic which can be arranged
by a duality rotation. In order to  avoid  confusion with pressure
in what follows we replace $Q$ by $Z$ and take $Z$ to be real and positive.
We have
\ben
-r^2 g_{tt}= (r- M )^2 + Z^2-M^2 - \frac{r^4}{l^2} \,,
\een
and
\ben
2 M= r + \frac{Z^2}{r}   - \frac{r^3}{l^2}  \,,
\een
with $l^2 = \frac{3}{\Lambda}$.

If $M^2=Z^2$  there are three  positive values of $r$
for which $g_{tt}=0$:
\bea
r_1  &=& \frac{l}{2} \bigl ( 1 + \sqrt{1- 4\frac{M}{l}} \bigr) \,, \\
r_2   &=& \frac{l}{2} \bigl ( 1 - \sqrt{1- 4\frac{M}{l}}\bigr ) \,, \\
r_3  &=& \frac{l}{2} \bigl ( \sqrt{1+4\frac{M}{l}}  -1  \bigr ) \,.
\eea
which correspond to the cosmological event horizon, the black hole horizon
and its inner horizon respectively.
From the Gibbsian point of view one has
$T=\frac{\kappa}{2 \pi }$  and therefore
\bea
T_1 &=&  -  \frac{1}{2 \pi l} \sqrt{ 1- 4 Ml }   \,,\\
T_2 & =& \quad  \frac{1}{2 \pi l}\sqrt{ 1- 4 Ml}      \,,\\
T_3 &= & - \frac{1}{2\pi l} \sqrt{ 1+ 4 Ml  }    \,.
\eea
Because $|T_1|$=$T_2$ we obtain a gravitational instanton by setting $t=i\tau$
and identifying $\tau$ modulo $ \beta = \frac{1}{T_2}$
\cite{Mellor:1989wc} . The sign used for the period appears
to have no geometrical significance  and proceeding
in the standard way one may argue that the two horizons are in equilibrium
with respect to the exchange of thermal Hawking quanta.

It was also argued that  if $|\kappa_3|\ge |\kappa_1$, then the Cauchy horizon should be stable.

\subsection{Kerr-Newman-de Sitter black holes}

From \cite{Caldarelli:1999xj} we take the formula
\ben
M= \half \sqrt{{\tilde S}} \, \sqrt{\bigl(1- \frac{\Lambda {\tilde S}}{3} +
  \frac{ Q^2}{{\tilde S}}\Bigr)^2 + \frac{4  J^2}{{\tilde S}^2}
  \bigl(1-\frac{\Lambda {\tilde S}}{3} \bigr )} \, \label{CCK}
\een
where ${\tilde S}=\frac{S}{\pi}$. Writing
 $\Lambda=-3g^2$,  the formula takes the form

\ben
M^2=\frac{\pi}{4S}\Bigl\{\Bigl[\frac{S}{\pi}\Bigl(1+g^2 \frac{S}{\pi}\Bigr) +
  { Q^2}\Bigr]^2 + {4  J^2}
 \Bigl(1+g^2\frac{S}{\pi}\Bigr) \Bigr\}\, .
\een

For $\Lambda=0$ the result reduces to that of the Kerr-Newman black hole.

\subsection{Pairwise-equal charge anti-de Sitter black hole}

These solutions were obtained in \cite{Chong}, and they are special cases of
solutions in the gauged STU supergravity model. (Those are also solutions of maximally supersymmetric four-dimensional theory, which is a consistent truncation of a Kaluza-Klein compactified eleven-dimensional supergravity on $S^7$.) The theory is specified by mass $M$, angular momentum $J$, two charges, i.e., equating $Q_1=Q_3$ and $Q_2=Q_4$, and cosmological constant $\Lambda=-3g^2$. In \cite{Chong} the solution was parameterised by  the non-extremality parameter $m$,  rotational parameter $a$, two boost parameters $\delta_{1,2}$ and $g^2$. The thermodynamic quantities  are of the following form:
\bea
M&=&\frac{m(1+s_1^2 +s_2^2)}{\Xi^2}\,  , \\
J&=&\frac{a\,m(1+s_1^2 +s_2^2)}{\Xi^2}\, , \\
Q_i&=&\frac{m s_ic_i}{2\Xi}\, , \quad i=1,2\, ,
\eea
where $s_i=\sinh\delta_i$, $c_i=\cosh\delta_i$ ($i=1,2$). and $\Xi=1-g^2a^2$.
The entropy is of the form:
\be
S=\frac{\pi}{\Xi} \, (r_1r_2+a^2)\, ,
\ee
where $r_i=r_++m s_i^2$ ($i=1,2$) and $r_+$ is a location of a horizon, which is a solution of the equation:
\be
r^2-2mr+a^2+g^2r_1r_2(r_1r_2+a^2)=0\, .
\ee

Manipulation of the horizon equation, along with the expressions for the $M$, $J$, $Q_i$ and $S$, allows one to derive the following explicit Christodoulou-Ruffini mass:
\be
M^2=\frac{\pi}{4S}\Bigl\{\Bigl[ \frac{S}{\pi}\Bigl(1+g^2\frac{S}{\pi}\Bigr)+
 16 Q_1^2\Bigr]\Bigl[\frac{S}{\pi}\Bigl(1+g^2\frac{S}{\pi}\Bigr)+16 Q_2^2
\Bigr] +4J^2\Bigl(1+g^2\frac{S}{\pi}\Bigr)\Bigr\}\,.
\ee

\subsection{Wu black hole}

The Wu black hole \cite{WuD5}  is 5D,
three charge rotating solution with negative cosmological
constant ($\propto g^2$).
Employing expressions from \cite{Birkandan:2014vga} for
a product of the entropy and temperature of this black hole, associated with all three horizons we obtain the following interesting expression:
\ben
n_1+n_2+n_3+\frac{1}{2}\left(\frac{n_1n_2}{n_3}+\frac{n_1n_3}{n_2}+
\frac{n_2n_3}{n_1}\right)=0\,,
\een
where
\ben
n_1=\frac{4\xi_a\xi_b}{g^2\pi}\, T_1S_1=(u_1-u_2)(u_2-u_3) \quad
\& \hbox{cyclic permutations}\,.
\een
%%%%%
Here $\xi_{a}=1-g^2a^2$, $\xi_b=1-g^2b^2$ and $u_i$ is the root of the
horizon equation $X=g^2 (u-u_1)(u-u_2)(u-u_3)$.
Note that as $g^2\to 0$, $u_3\to -1/g^2\to -\infty$, and in this case the above equation reduces to the standard equation $T_1S_1+T_2S_2=0$.

\section{Entropy and Super-Additivity}

The thermodynamics of equilibrium systems with a  substantial
contribution to the total energy from  their
gravitational  self energy differs  significantly from that
of ordinary substances encountered in the laboratory.
This is because of the long range nature
of the Newtonian gravitational force, which cannot be screened.
As a consequence the total entropy $S$ of a gravitating
system need not be proportional to the total energy $M$.
A consequence of this is that {\sl negative}
heat capacities are possible, and indeed these
have long been encountered in the theory of stellar structure
\cite{LyndenBell:1998fr}.

In the case of black holes, the long range nature of gravitational
interaction expresses itself in the fact that while the individual
extensive variables may be added,  they do not not necessarily scale.
Even if they do, they do not scale
with the same power  as the total energy $M$.
In the case of ungauged
supergravity black holes, the scaling behaviour
is guaranteed, but the fact that the scaling behaviour is
not {\sl homogeneous}, that is,
not the same for all extensive variables, leads to a modification of the
standard form of the Gibbs-Duhem relation for ordinary homogeneous  substances
%%%%%
\be
G=M-TS-PV =0\,,
\ee
%%%%%
where $G$ is the Gibbs free energy, $V$ the volume and
$P$ the pressure.  By contrast, for black holes the Smarr relation
(\ref{smarrrel}) gives rise to the Gibbs function (\ref{gibbsbh}).

    The requirement  of homogeneous scaling
 plays such an important  role
in the thermodynamics of ordinary substances that it has been
been suggested that it be called the {\sl Fourth Law of Thermodynamics}
\cite{Landsberg0,Landsberg4}.
It certainly fails for systems with significant self-gravitation
and,  {\it a fortiori}, for black holes.  In fact if the matter sector is
sufficiently non-linear
such as in Einstein's theory  coupled to non-linear electrodynamics,
even the property of weighted homogeneity ceases to hold.\footnote{A function
$f(x_1,x_2,\,\dots,x_n)$ of $n$ variables
is said to be weighted homogeneous of weights $w_1$,$w_2,\,\dots,w_n $ if
$f(\lambda ^{w_1}  x_1 , ,\lambda ^{w_2}  x_2,\,\dots,\lambda ^{w_n}  x_n)=
\lambda f(x_1,x_2,\,\dots,x_n)$. If $ w_i = 1$ for all $i$,
the function is said to be homogeneous of weight one. The Fourth Law
is the statement that all extensive variables have weight one and
thus all intensive variables have weight zero.}
As a consequence, while
the first law of black hole thermodynamics holds there is no analogue of
a Smarr formula \cite{Rasheed:1997ns}.

 In the thermodynamics of ordinary substances it is usually assumed that
the total energy $M$ is a convex function\footnote{A function $f(\bx)$
is said to be {\sl convex}
 if
 $f( \lambda \bx_1  +  (1-\lambda) \bx_2) \le \lambda  f(\bx  _1) +
 (1-\lambda) f(\bx_2)\,\, \forall \,\, 1\le \lambda \le 1 $ and
{\sl concave }if $\le$ is changed to $\ge$. Subject to suitable
differentiability this is equivalent to negative (positive)
definiteness of the
Hessian $\frac{\partial ^2 f}{\partial  x^i \partial x^j}$.
In other words, if $M$ is the total energy then
the graph of the Gibbs surface along a straight line
joining two equilibrium  states $\bx_1$ and $\bx_2$  never lies above the
straight line joining  these points on the Gibbs surface.} of the
extensive variables or that the $S$ is a concave function of
the other extensive variables. This guarantees that the
heat capacity and other susceptibilities are positive,
and that the Hessians have the correct signs to render
the Weinhold and Ruppeiner metrics positive definite.

Now if the extensive quantities scale in a uniform fashion,
the property of concavity  is equivalent to that of
super-additivity,\footnote{A function $f(\bx)$ is super-additive
if $ f(\bx_1 + \bx_2) \ge   f(\bx_1) + f(\bx _2)$
and sub-additive if we replace $\ge$ by $\le $.} but not necessarily if
uniform scaling ceases to hold \cite{Landsberg1,Landsberg2,Tranah,Landsberg3}.
Remarkably, it was shown long ago
in a little noticed paper by Tranah and Landsberg
\cite{Tranah}\footnote{Apparently not accessible on-line. The only paper
we know of that has followed up on this is \cite{CurirC}.} that
while concavity fails for the entropy
of Kerr-Newman black holes, super-additivity remains true.
In other words
\be
S(M_1+M_2,J_1+J_2,Q_1+Q_2) \ge   S(M_1,J_1,Q_1)  + S(M_2,J_2,Q_3) \,.
\label{super}
\ee
%%%%%

   The super-additivity inequality (\ref{super}) is related to Hawking's
area theorem \cite{Hawking:1971tu,Hawking:1971vc}.  If two black holes of areas $A_1$ and $A_2$ can merge to 
form a single black hole
of area $A_3$, then, subject to the assumption of cosmic censorship,
%%%%%
\be
A_3 \ge A_1 + A_2 \,. \label{hawking}
\ee
%%%%
If the angular momentum and charge of the final black hole are equal to the
sums of the angular momenta and charges of the initial black holes, one
has in addition
%%%%%
\be
S(M_3,J_1+J_2,Q_1+Q_2) \ge   S(M_1,J_1,Q_1)  + S(M_2,J_2,Q_3) \,,
\label{hawking2}
\ee
%%%%%
where $M_3$, the mass of the black hole final state after the merger, obeys
%%%%%
\be
M_3 <M_1 + M_2\,,
\ee
%%%%
since energy will be lost by gravitational radiation.  It follows from the
first law that at fixed charge and angular momentum, $dM=T dS$ and
so provided that the temperature is positive,
%%%%%
\be
S(M_1+M_2,J_1+J_2,Q_1+Q_2) > S(M_3,J_1+J_2,Q_1+Q_2)\,.
\ee
%%%%%
The assumption that $Q_3=Q_1+Q_2$ is reasonable for theories like 
Einstein-Maxwell or ungauged supergravity, where there are no particles
that carry charge.  The assumption that $J_3=J_1+J_2$, however, is less
reasonable, because both electromagnetic and gravitational waves can
carry angular momentum.

In the following subsections we shall obtain generalisations of the
Kerr-Newman super-additivity result of Tranah and Landsberg for
various more complicated black hole solutions. We also obtain a
counter-example in the case of dyonic Kaluza-Klein black holes.

\subsection{STU black holes with pairwise-equal charges}

  From the formula expressing $M$ in terms of $S$, $Q_1$, $Q_2$ and $J$
for pairwise-equal charged STU black holes, we have
%%%%%
\be
\fft1{\pi}\, S(M,Q_1,Q_2,J) =  Y+\sqrt{X}\,,\quad
  Y = 2M^2 -\ft12(Q_1^2+Q_2^2)\,,\quad
X=Y^2 -Q_1^2\, Q_2^2 - 4 J^2
\,.\label{Sform}
\ee
%%%%%
For regular black holes we must have $X\ge0$ and hence
$Y\ge\sqrt{4 Q_1^2\, Q_2^2 + 16 J^2}$, thus implying
%%%%%
\be
4 M^2 \ge Q_1^2+Q_2^2 + \sqrt{4 Q_1^2\, Q_2^2 + 16 J^2}\,.\label{bound2}
\ee
%%%%%
Without loss of generality, we shall assume $Q_1$, $Q_2$ and $J$ are
all non-negative.  Note that we also have the weaker inequality
%%%%%
\be
M\ge \ft12(Q_1+Q_2)\,,\label{bpsineq}
\ee
%%%%%
which we shall use frequently in the following.

 We wish to check whether the entropy of these pairwise-equal charged
black holes obey the super-additivity inequality
%%%%%
\be
S_{\rm tot}\ge S +S'\,,\label{superadd}
\ee
%%%%%
where
%%%%%
\bea
S_{\rm tot} &\equiv& S(M+M',Q_1+Q_1', Q_2 + Q_2', J+J')\,,\cr
S &\equiv& S(M,Q_1,Q_2,J)\,,\qquad S'\equiv S(M',Q_1',Q_2',J')\,.
\eea
%%%%%
With analogous definitions for the quantities $X$ and $Y$, proving
super-additivity requires proving that
%%%%%
\be
Y_{\rm tot} -Y -Y' + \sqrt{X_{\rm tot}} - \sqrt{X} - \sqrt{X'}\ge 0\,.
\label{Sineq2}
\ee
%%%%

   We first note that the $Y$ functions are non-negative, and that
they obey
%%%%%
\bea
Y_{\rm tot} - Y -Y' &=& 4M\, M' - Q_1\, Q_1' -Q_2\, Q_2'\nn\\
&\ge& (Q_1+Q_2)(Q_1'+Q_2')-Q_1\, Q_1' -Q_2\, Q_2'\nn\\
&=& Q_1\, Q_2' + Q_2\, Q_1'\nn\\
  &\ge& 0\,.\label{Yineq}
\eea
%%%%%
Thus, if we can show that
%%%%%
\be
\sqrt{X_{\rm tot}} -\sqrt{X}-\sqrt{X'}\ge0\label{Xineq}
\ee
%%%%%
then the super-additivity inequality (\ref{superadd}) will be established.
To prove this, we first note that is can be re-expressed as
%%%%%
\be
X_{\rm tot} -(\sqrt{X}+ \sqrt{X'})^2\ge0\,.\label{rtpX}
\ee
%%%%%
We now observe that the following identity holds:
%%%%%
\bea
P &\coloneqq& \Big(c\, \sqrt{X} -\fft1{c}\, \sqrt{X'}\Big)^2 +
  4 \Big(c\, J - \fft1{c}\, J'\Big)^2\nn\\
&=& -2\sqrt{X}\, \sqrt{X'} -
8J\, J' + 8M^2\, {M'}^2 - 2M^2\, ({Q_1'}^2 +{Q_2'}^2) -
  2 {M'}^2\, (Q_1^2 + Q_2^2) \nn\\
&& \ \  -2Q_1\, Q_2\, Q_1'\, Q_2' +
\ft12 (Q_1^2+Q_2^2)({Q_1'}^2 + {Q_2'}^2)\,,\label{id0}
\eea
%%%%%
where we have defined
%%%%%
\be
c^2 = \fft{4{M'}^2  - (Q_1'-Q_2')^2}{4M^2 -(Q_1-Q_2)^2}\,.
\ee
%%%%%
We can use (\ref{id0}) to substitute for $\sqrt{X}\, \sqrt{X'}$ in
(\ref{rtpX}), thus yielding
%%%%%
\bea
&&X_{\rm tot} -X-X' - 2\sqrt{X}\, \sqrt{X'} = \nn\\
&&\qquad P +
 8(M\, M' - Q_-\, Q_-') (M^2+{M'}^2 - Q_+^2 - {Q_+'}^2) \nn\\
&&\qquad  + 8[(M+M')^2 -(Q_- + Q_-')^2](M\, M' - Q_+\, Q_+')\,,\label{Xrel}
\eea
%%%%%
where we have defined
%%%%%
\be
Q_\pm= \ft12(Q_1 \pm Q_2)\,,\qquad Q_\pm' = \ft12(Q_1' \pm Q_2')\,.
\ee
%%%%%%

   The inequality (\ref{bpsineq}) implies $M\ge Q_+$ and $M'\ge Q_+'$,
and {\it a fortiori} $M\ge |Q_-|$ and $M'\ge |Q_-'|$ (recall that we
are taking all charges to be non-negative).  Since $P$, defined
in (\ref{id0}), is manifestly non-negative it follows from
(\ref{Xrel}) that the left-hand side must be non-negative, and
hence the required inequality (\ref{Xineq}) is satisfied.  Thus we have
proven that the super-additivity property (\ref{superadd}) is indeed
obeyed by the entropy of the pairwise-equal charged black holes of
STU supergravity.

\subsection{STU black holes with three equal non-zero charges}

One can also show analytically  that the super-additivity
property of the entropy is true for the case of  STU black holes with
three equal non-zero charges, say,  $Q_1=Q_2=Q_3=q$, with $Q_4=0$.
In this case $S=\pi (Y+\sqrt{X})$  with:
%%%%%
\ben
Y^2=\frac{1}{64}(3z-2M)(z+2M)^3\,,
\ee
%%%%%
where
%%%%%
\be
z=\sqrt{4M^2-2q^2}\,,
\ee
%%%%%
and
%%%%%
\ben
X=Y^2-J^2\,.
\een
%%%%%

It is straightforward to show that
%%%%%
\ben
z_{tot}^2-(z+z')^2=8M\, M'\left(1-ww'-\sqrt{1-w^2}\sqrt{1-w'^2}\right)\ge 0
\,,\label{zineq}
\ee
%%%%%
where $w=\frac{q}{\sqrt{2}M}$ and   $w'=\frac{q'}{\sqrt{2}M'}$.
The second inequality in (\ref{zineq}) is true for any value of $\{w,w'\}\le1$.
This result  implies
%%%%%
\be
Y_{tot}-Y-Y'\ge 0\,.
\ee
%%%%%
It is now straightforward to show that
%%%%%
\be
\sqrt{X_{tot}} -\sqrt{X}-\sqrt{X'}\ge 0\,,
\ee
%%%%%
thus proving the super-additivity of the entropy in this case as well.

An analytic proof of the  super-additivity of the entropy for the case of
one non-zero charge follows analogous steps.

While a numerical analysis indicates that the super-additivity is true for the
STU black holes with four arbitrary electric charges, it would be interesting
to prove this result analytically.

\subsection{Dyonic Reissner-Nordstr\"om}

   In the explicit examples we have studied so far, the black hole is
supported by one or more field strengths that each carry a single complexion of
field (pure electric charge, or instead and equivalently, one could consider
pure magnetic charge).  The details of the entropy super-additivity
inequality are different if we consider a case where one or more
field strengths carries both electric and magnetic charge. In this subsection,
we shall study the dyonic Reissner-Nordstr\"om black hole, and show that in
this case too the super-additivity property is satisfied.  This case,
where the Lagrangian is just that of the pure Einstein-Maxwell system,
can be view as STU black holes where all four field strengths are equal. By
contrast, in the next subsection we shall see that in the case of STU
black holes where only a single field strength is non-zero, the dyonic
black holes have an entropy that violates the super-additivity property.

  The Einstein-Maxwell Lagrangian ${\cal L}=\sqrt{-g}(R-F^2)$ admits
static dyonic black hole solutions given by
%%%%%
\bea
ds^2&=& -h dt^2 + \fft{dr^2}{h} + r^2\, (d\theta^2 + \sin^2\theta\, d\varphi^2)
\,,\nn\\
A&=& \fft{Q}{r}\, dt + P \sin\theta\, d\varphi\,,\qquad
  h= 1 - \fft{2M}{r} + \fft{Q^2+P^2}{r^2}\,,
\eea
%%%%%
with mass $M$, electric charge $Q$ and magnetic charge $P$.  To have
a black hole, these quantities must obey the inequality
%%%%%
\be
 M\ge \sqrt{Q^2+P^2}\,,\label{MQPineq}
\ee
%%%%%
with extremality being attained when the inequality is saturated.
The entropy is
given by
%%%%%
\be
  S(M,Q,P)= \pi\, \Big[ 2M^2 -Q^2 -P^2 + 2M\, \sqrt{M^2-Q^2-P^2}\Big]\,.
\label{dyonicS}
\ee
%%%%%

    For super-additivity, one must have
%%%%%
\be
S(M+M',Q+Q',P+P')-S(M,Q,P) - S(M',Q',P')\ge0\,,\label{superadddyon}
\ee
%%%%%
where, as usual, we assume, without loss of generality, that the charges
are all non-negative.  Substituting (\ref{dyonicS}) into this, we see
that super-additivity is satisfied if
%%%%%
\bea
&&4M M' - 2Q Q' -2P P' +(M+M')\sqrt{(M+M')^2 - (Q+Q')^2 -(P+P')^2} \nn\\
&&-
    M\sqrt{M^2-Q^2-P^2} - M'\sqrt{{M'}^2 -{Q'}^2 -{P'}^2}\ge0\,.
\label{ineqdyon}
\eea
%%%%%
First, we note that the argument of the first square root is non-negative,
since, after using (\ref{MQPineq}) for the unprimed and primed
quantities we have
%%%%%
\be
(M+M')^2 - (Q+Q')^2 -(P+P')^2\ge 2(M M' -Q Q' -P P')\,,
\ee
%%%%%
and since
%%%%%
\be
(M M')^2 -(Q Q' + P P')^2 \ge (Q^2+P^2)({Q'}^2+{P'}^2) -
  (Q Q' + P P')^2  = (Q P'- P Q')^2\ge 0\,,
\ee
%%%%%
the non-negativity is proven.

   Returning to the inequality (\ref{ineqdyon}) that we wish to establish,
we see that the terms $4M M' - 2Q Q' -2P P'$ are themselves certainly
non-negative,
since $2M M' -2Q Q' -2 P P'\ge0$ as
we just demonstrated.  The inequality is therefore established if we can show
that
%%%%%
\be
 M(\sqrt{(M+M')^2 - (Q+Q')^2 -(P+P')^2} -\sqrt{M^2-Q^2-P^2} )\ge0\,,
\ee
%%%%%
together with the analogous expression with the primes and unprimed variables
exchanged.  The expression in parentheses is non-negative if
%%%%%
\be
(M+M')^2 - (Q+Q')^2 -(P+P')^2 -(M^2-Q^2-P^2)\label{dyon3}
\ee
%%%%%
is non-negative.  After using (\ref{MQPineq}) again we see that (\ref{dyon3})
is greater than or equal to $2(M M'-Q Q' -P P')$, and we have already
shown that this is non-negative.  Thus the super-additivity property
(\ref{superadddyon}) is established for the dyonic Reissner-Nordstr\"om
black holes.

\subsection{A counterexample: The dyonic Kaluza-Klein black hole}

  Here, we demonstrate that dyonic Kaluza-Klein black holes that we
discussed in section \ref{kkdyonsec} provide counterexamples where
entropy super-additivity breaks down.
    The phase space for checking entropy super-additivity for these
dyonic black holes is rather large, so we shall just focus on a restricted
subspace within which we are able to exhibit violations.  Specifically,
we shall consider two black holes with the following $(M,Q,P)$ values:
%%%%%
\be
(P,0,P)\qquad \hbox{and}\qquad (M',Q',0)\,,
\ee
%%%%%
so the unprimed case is an extremal black hole with purely magnetic
charge,\footnote{Strictly speaking, the extremal configuration
$(P,0,P)$ is not a black hole, but rather a naked singularity.  However,
one can make an infinitesimal deformation away from extremality,
to a configuration with parameters $(P+\delta,0,P)$, and this will describe
a genuine black hole. The results that we shall derive here, including
the bound (\ref{PQ'bound}) on $P$ versus $Q'$ for obtaining violations
of entropy super-additivity, are thus valid.}
and the primed case is a (sub-extremal) black hole with purely electric
charge.  The masses and charges will be chosen so that the black hole
with the summed mass and charges will be an extremal dyonic black hole,
for which $M_{\rm tot}=(Q_{\rm tot}^{2/3} + P_{\rm tot}^{2/3})^{3/2}$.  Thus
%%%%%
\be
M_{\rm tot} = M + M' = P+ M'\,,\qquad
  Q_{\rm tot}= Q'\,,\qquad P_{\rm tot} = P\,,
\ee
%%%%%
with
%%%%%
\be
 P+ M'= \Big({Q'}^{\ft23} + P^{\ft23}\Big)^{\ft32} \,.\label{dyonext}
\ee
%%%%%
We shall characterise the ratio $P/Q'$ by means of a constant $x$, such that
%%%%%
\be
P= x^{\ft32}\, Q'\,.\label{xdef}
\ee
%%%%%
We therefore have
%%%%%
\be
S=0\,,\qquad S'= \fft{\pi\, {m}^2}{\sqrt{1-2\beta_1}}\,,\qquad
  S_{\rm tot}= 8\pi\, x^{\ft32}\, {Q'}^2\,,
\ee
%%%%%
where the primed black hole defined above has metric parameters $m$
and $\beta_1$, with $\beta_2=0$.  This means that
%%%%%
\be
M'=\fft{(1-\beta_1)\, m}{2(1-2\beta_1)}\,,\qquad
    Q' = \fft{\sqrt{2\beta_1}\, m}{4(1-2\beta_1)}\,,
\ee
%%%%%
the entropy is given by
%%%%%
\be
S' = 8\pi\, \fft{(1-2\beta_1)^{\ft32}}{\beta_1}\, {Q'}^2\,,
\ee
%%%%%
and from (\ref{dyonext}) $\beta_1$ is given in terms of $x$ by
%%%%%
\be
\fft{2(1-\beta_1)}{\sqrt{2\beta_1}} = (1+x)^{\ft32} -x^{\ft32}\,.
\label{betax}
\ee
%%%%%

   Let us first consider the case where $x$ is very small,
$x=\epsilon^{\ft23}$.  From (\ref{betax}) we find at leading order
$\beta_1=\ft12(1-\epsilon^{\ft23})$,
and so $S'= 16\pi \epsilon\, {Q'}^2$.  Thus we have
%%%%%
\be
S_{\rm tot} -S-S' = 8\pi \epsilon \, {Q'}^2 - 0 - 16\pi \epsilon\, {Q'}^2
 = -8\pi \epsilon \, {Q'}^2\,,
\ee
%%%%%
and so super-additivity does not hold in this region of the parameter space.

   When $x$ becomes larger, we find from numerical analysis that the
ratio $S_{\rm tot}/(S+S')$, which equals 2 in the limit as $x$ goes to zero,
falls monotonically.  The ratio reaches unity when $S'=S_{\rm tot}$, which
implies
%%%%%
\be
   x = (1-2\beta_1)\, \beta_1^{-\ft23}\,.
\ee
%%%%%
Substituting into (\ref{betax}), we find that this occurs when $\beta_1=y^3$
and $y$ is the single real root of the 9th-order polynomial
%%%%%
\be
17 y^9 -12 y^8 + 42 y^7 - 80 y^6 + 39 y^5 -48 y^4 + 54 y^3 -12 y^2 + 9y -8=0\,.
\ee
%%%%%
This root is given approximately by $y=0.698234$, implying $\beta_1
=0.340411$, and hence $x=0.654681$.  Thus the parameter range where we
find a violation of entropy super-additivity is when
%%%%%
\be
0< P < 0.529718 \, Q'\,.\label{PQ'bound}
\ee
%%%%%
In other words, we have found super-additivity violation when we add
an extremal purely magnetic black hole and a non-extremal purely electric
black hole, with parameters arranged such that the ``total'' dyonic black
hole is extremal,
provided that the magnetic charge of the original extremal black hole is
sufficiently small in comparison to the electric charge of the
original non-extremal
black hole.

  We can give a more complete treatment by choosing two black holes
with parameters $(M,Q,P)$ of the form $(M,0,P)$ and $(M',Q',0)$, subject
to the assumption that the total black hole $(M_{\rm tot},Q_{\rm tot},
P_{\rm tot})$ is again extremal, obeying
%%%%%
\be
M_{\rm tot}= [Q_{\rm tot}^{2/3} +
 P_{\rm tot}^{2/3}]^{3/2}\,.\label{kkdyonextr}
\ee
%%%%%%
Thus
%%%%%
\be
M_{\rm tot}= M+M'\,,\qquad Q_{\rm tot}= Q'\,,\qquad P_{\rm tot} =P\,.
\label{sums}
\ee
%%%%%
It is straightforward to show from the formulae in section 3.4.5 that
for the individual black holes that carry purely electric or
purely magnetic charge, one has
%%%%%
\bea
S &=& \sqrt8\,\pi\sqrt{M^4 -20 M^2 P^2 - 8P^4 + M(M^2+8 P^2)^{3/2}}\,,\nn\\
S' &=& \sqrt8\,\pi\sqrt{{M'}^4 -20 {M'}^2 {Q'}^2 - 8{Q'}^4
   + M'({M'}^2+8 {Q'}^2)^{3/2}}\,.
\eea
%%%%%
One can then use (\ref{kkdyonextr}), together with (\ref{sums}), to
solve for $M'$, and hence one can express $Y\equiv S_{\rm tot} - S-S'$, where
$S_{\rm tot} = 8 \pi P_{\rm tot}\, Q_{\rm tot}$, as a function of
$M$, $P$ and $Q'$.  One can then explore the regions in the space of these
parameters for which $Y$ is negative, signifying a violation of entropy
super-additivity.

   Of course, by continuity we expect that super-additivity violations will
occur at least in some neighbourhood of the region found above when all the
masses and charges are allowed to be adjusted.  In other words, there
will also be super-additivity violations if we consider
cases where all three black holes are non-extremal, for appropriate
ranges of the various masses and charges.

   In our earlier remarks relating super-additivity to the Hawking area
theorem, we assumed not only cosmic censorship but also that the coalescence
of the two black holes was allowed physically.  In the case of dyons, it should
be recalled that they carry angular momentum, and moreover it is not localised
within the event horizon.  This, as suggested in \cite{larsen}, may lead to
restrictions on what coalescences are allowed, and thus the non-super
additivity of the entropy in this counter-example need not
imply any conflict with Hawking's area theorem.  This is
an interesting problem worthy of further study.

\section{Conclusions and Future Prospects}

   We shall turn in this section to a consideration of the significance of
negative surface gravities, and negative Gibbsian temperatures.   We shall
begin by recalling the most physically convincing argument that
Schwarzschild black holes have a temperature, and hence entropy.
This was given by Hawking \cite{Hawking:1974rv,Hawking:1974sw},
who coupled a collapsing
black-hole metric in an asymptotically-flat spacetime to a quantum field,
and showed that if the quantum field was initially in its vacuum state, then
at late times it would emit particles with a thermal spectrum and
temperature given by (\ref{Tfromkappa}).  The term ``vacuum state'' implied
that it contained no particles having positive frequency with respect to
the standard retarded time coordinate on past null infinity.  This required
his considering the behaviour of the quantum field as it passed through the
time-dependent spacetime generated by the collapse.  However, one may
dispense with that region, and work with the exact vacuum Schwarzschild
solution, obtaining the same result, by choosing an appropriate boundary
condition for the quantum field on the past horizon.  The appropriate
boundary condition, which reproduces Hawking's result, in the
exterior region of the Schwarzschild solution, corresponds to requiring that
the state contains no particles defined as having positive frequency with
respect to a Kruskal null coordinate on the past horizon.  This state is
now referred to as the Unruh vacuum state.  This is obviously not a
state in thermal equilibrium.  A different state, introduced by Hartle and
Hawking, is defined on the past horizon in the same way, but at past null
infinity the definition of positive frequency is such that it describes an
ingoing flux of particles at the Hawking temperature.  Thus the Hartle-Hawking
state should be regarded as a state in thermal equilibrium.

   The situation with two event horizons is more complicated. In order
to discuss quantum fields between the horizons, one needs to specify a notion
of positive frequency on each past horizon.  If the region is static, and
one interprets positive frequency as being with respect to a local
Kruskal coordinate on the horizons, the resulting quantum state will
describe thermal radiation entering the static region at temperatures given
by $\ft1{2\pi}\, |\kappa_\pm|$.  This is not a state in thermal
equilibrium.  If the region between the two horizons is not static, as
for example in the Reissner-Nordstr\"om solution, one may define a
similar state which would also not be in thermal equilibrium.  If, on the
other hand, one considers the static region behind the inner horizon in
the Reissner-Nordstr\"om, one needs to specify boundary consitions on the
singularity at $r=0$.  If one chose the notation of positive frequency on
the past inner horizon, then whatever boundary conditions were chosen on the
singularity, the quantum state would contain radiation coming from the
inner horizon with a temperature $\ft1{2\pi}\, |\kappa_\pm|$.  Thus
if we adopt this procedure, we see in all cases that the temperature
we associate with particles coming from the horizons is given by
the absolute value of the surface gravity, divided by $2\pi$.

   An alternative way of establishing the temperature and entropy of an
asymptotically-flat black hole is to follow the procedure of
\cite{hartlehawking,Gibbons:1976ue}, in which one analytically continues the
metric to imaginary time, and discovers that the metric is periodic in
imaginary time with a period given by $2\pi/|\kappa|$, which
is what one expects for a state in thermal equilibrium at temperature
$\ft1{2\pi}\, |\kappa_\pm|$.  Of course, the period itself can have either
sign, but the quantum state would not necessarily exist if one chose a
negative sign for the temperature.  This procedure will work when one
has a single horizon, including an asymptotically anti-de Sitter spacetime
\cite{Hawking:1982dh,Witten:1998qj}.  However, this procedure will not
work for a
spacetime with two horizons having differing values of $|\kappa|$.  The
conclusion seems to be that classically, the sign of the temperature can only
be determined by appealing to the first law, and this provides us with a
Gibbsian temperature.  Quantum mechanically, which seems to be the only
physically reliable argument provided one is prepared to contemplate
non-equilibrium situations, the temperature should be taken to be positive.
In other words, the temperature is not unquely defined by the metric,
a conclusion also reached in \cite{mipark1}.

   The original suggestion that inner horizons should be assigned a
negative temperature \cite{Curir} was based not quantum field theoretic
considerations, but rather on a consideration of quantum mechanical
systems, such as spin systems, exhibiting population inversion \cite{Ramsey}.
Thus one might regard the total energy of a black hole as receiving
contributions both from the outer and inner horizons.  The inner system
would then be thought of as the analogue of a spin system.  This
viewpoint was supported by the existence for the Kerr-Newman
black hole of the modified Smarr formula
(\ref{modSmarrKerr}), and its variation, which may be written as
%%%%%
\be
dM= \ft12( T_+\, dS_+ + \Omega_+\, dJ + \Phi_+\, dQ) +
   \ft12( T_-\, dS_- + \Omega_+\, dJ + \Phi_+\, dQ)\,.\label{hh}
\ee
%%%%%
As we saw, these formulae generalise to the case of STU black holes with
four electric charges.  The addition of electric charges, which were not
included in the discussion in \cite{Curir}, suggest that the posited
spin system inverted population should be supplemented by the inclusion
of charged states.

   In the case of four-dimensional STU black holes, the generalisation of 
equation (\ref{hh}) may be rewitten in terms of the left-moving and
right-moving sectors (see (\ref{LRfirstlaws})) as
%%%%%
\be
dM= (T_L\, dS_L + \Omega_L\, dJ + \Phi^i_L\, dQ_i +
   \Psi_{L, i}\, dP^i) +
  (T_R\, dS_R + \Omega_R\, dJ + \Phi^i_R\, dQ_i +
   \Psi_{R, i}\, dP^i) \,,
\ee
%%%%%
with each sector contributing equally to $dM$.  In contrast to the
proposal in \cite{Curir}, which attempted to give a microscopic interpretation
to the negative temperature on the inner horizon, here the left-moving
and right-moving sectors both have positive temperatures, consistent with
the proposed microscopic interpretation in 
terms of D-brane states \cite{CYI,horlowmal}.  An analogous interpretation
for five-dimensional STU black holes has also been given \cite{CLIII}.

    This paper has been concerned exclusively with time-independent solutions;
we have not discussed what happens to inner horizons when perturbations are 
considered.  There is a widespread belief that in classical general
relativity, generic perturbations will 
render Cauchy horizons, of the sort one finds inside black holes, singular.
This is referred to as the Cosmic Censorship Hypothesis.  There are various 
forms of this hypothesis, and the literature is at present rather inconclusive.
A recent discussion can be found in \cite{dafluk}.  Our 
motivation is largely quantum mechanical, and the relevance of these 
classical results to a full quantum gravitational treatment is unclear.

\section*{Acknowledgements}

G.W.G.~thanks Kei-Ichi Madeda for helpful  discussions about the Hawking-Page
phase transition. M.C.~is supported in part by DOE Grant Award de-sc0013528,
the Fay R. and Eugene L. Langberg Endowed Chair and the Slovenian Research Agency (ARRS) (M.C.).
H.L.~is supported in part by NSFC grants No.~11475024 and No.~11235003.
C.N.P.~is supported in part by DOE grant DE-FG02-13ER42020.

\appendix

\section{Carter-Penrose Diagram for Two Horizons}

  In this appendix, we summarise some facts about the Carter-Penrose diagram of
asymptotically-flat spherically symmetric spacetimes with an inner and outer
horizon. Consider a suitable metric of the form
%%%%%
\be
ds^2 = -A(r) dt^2 + \fft{dr^2}{f^2(r)\, A(r)} + R(r)^2 d\Omega^2\,.
\ee
%%%%%
Introducing an advanced time coordinate $v$ by defining
%%%%%
\be
dv= dt + \fft{dr}{f\, A}\,,
\ee
%%%%%
the metric takes the Eddington-Finkelstein form
%%%%%
\be
ds^2 = -A dv^2 + 2 f^{-1}\, dr dv + R^2 d\Omega^2\,.
\ee
%%%%%
The metric will be regular as long as $A$, $f$ and $R^2$ are 
real, bounded, and twice differentiable, and in addition $f$ and
$R$ are non-zero.  We may take $f$, without loss of generality, to be 
positive.  In particular, the metric is well-behaved regardless of 
whether $A$ is positive, zero or negative.  
Asymptotic flatness requires
that $A$ and $f$ tend to 1 as $R^2$ tends to infinity.  In the cases we shall 
consider, $R$ tends to $r$ at infinity.  We shall assume that $A$ is positive
in the interval $r_+ <r \le \infty$, and negative in the the
interval $r_-< r < r_+$, and that it vanishes on the outer horizon $r=r_+$
and the inner horizon $r=r_-$.  We shall also assume that $A$ has a smooth
positive extension for values of $r <r_-$. 
The Killing vector $K=\del/\del v$ is thus timelike for $r_+<r <\infty$, 
lightlike
at $r=r_+$, spacelike for $r_-< r< r_+$, lightlike at $r=r_-$ and timelike
for $r<r_-$.  It becomes lightlike as $v$ tends to $\pm\infty$, and 
also as $r$ tends to infinity.

  If $r_+ < r <\infty$, then as $v$ tends to $+\infty$ we obtain future
null infinity, $\scri^+$.  For $v$ instead tending to $-\infty$, we
obtain past null infinity $\scri^-$.  As $v$ tends to $-\infty$ and $r$
tends to $r_+$ we obtain the past null horizon.  The Killing vector
$K$ is future-directed inside and on the boundary of this region.  The
inner region is bounded by a past Cauchy horizon at $v=-\infty$ and $r=r_+$,
and a future Cauchy horizon at $v=+\infty$ and $r=r_-$.  It has a further
boundary on the inner horizon at $r=r_-$, with $-\infty < v < +\infty$.  
Thus the Killing vector $K$ is future directed both on this inner horizon
and on the outer horizon.

  If one looks at radial geodesics in this spacetime, there are two
conserved quantities $p_v$ and $k$, where
%%%%%
\be
p_v = A  \dot v- f^{-1}\, \dot r\,,\qquad 
-A \dot v^2 + 2 f^{-1}\, \dot r \,\dot v =-k\,,
\ee
%%%%%
and a dot denotes a derivative with respect to an affine parameter 
$\lambda$.  Thus radially-infalling geodesics obey
%%%%%
\be
\dot r = -f\, \sqrt{p_v^2 - k A}\,,
\ee
%%%%%
with $k>0$ and $p_v^2>k$ for timlike geodesics that originate at
large $r$.  The constant $p_v$ is positive.  The infalling particle
passes through the outer and the inner horizons before reaching a 
turning point at a radius $\bar r <r_-$ at which $p_v^2 =k A(\bar r)$.  

  Solving for $\dot v$ one finds
%%%%%
\be
\dot v =  \fft{p_v - \sqrt{p_v^2 - k A}}{A}\,,
\ee
%%%%%
and so
%%%%%
\be
\fft{dv}{dr} = \fft1{f\, A}\, \Big[ 1 - \fft{p_v}{\sqrt{p_v^2 - k A}}\Big]\,.
\ee
%%%%%
Thus one finds that $\dot v$, $dv/dr$ and $v$ all remain finite as the 
particle falls in from infinity to $\bar r$.  Note that $\dot v$ is
always positive.

   In conclusion, we note that the Killing vector $K=\del/\del v$ is future
directed and lightlike on both the future event horizon of the exterior
region, $r=r_+$ with
$-\infty < v < +\infty$, and on the inner horizon, $r=r_-$ 
with $-\infty < v < +\infty$.  

   For the four-charge STU black holes considered 
in this paper, the situation when they are non-rotating 
is qualitatively similar to that for the Reissner-Nordstr\"om solution.  
The metric takes the form
%%%%%
\be
ds^2= -(H_1 H_2 H_3 H_4)^{-1/2}\, W\, dt^2 + (H_1 H_2 H_3 H_4)^{1/2}\, 
  (W^{-1}\, dr^2 + r^2\, d\Omega^2)\,,
\ee
%%%%%
where 
%%%%%
\be
  H_i = 1 + \fft{\mu\, \sinh^2\delta_i}{r}\,,\qquad W= 1 - \fft{\mu}{r}\,.
\ee
%%%%%
The outer horizon is located at $r_+=\mu$, and the inner horizon at
$r_-=0$.  There are curvature singularities at the four locations
$r=- \mu\, \sinh^2 \delta_i$, and the Carter-Penrose diagram will
be similar to that for Reissner-Nordstr\"om, with the curvature singularity
in the diagram occurring at the least negative of the four locations.

\section{STU Supergravity}\label{STUlagsec}

  The Lagrangian of the bosonic sector of four-dimensional
ungauged STU supergravity can be written in the relatively simple
form
%%%%%
\bea
{\cal L}_4 &=& R\, {*\oneone} - \ft12 {*d\varphi_i}\wedge d\varphi_i 
   - \ft12 e^{2\varphi_i}\, {*d\chi_i}\wedge d\chi_i - \ft12 e^{-\varphi_1}\,
\Big( e^{\varphi_2-\varphi_3}\, {*F_{\2 1}}\wedge F_{\2 1}\nn\\
&& + e^{\varphi_2+\varphi_3}\, {*  F_{\2 2}}\wedge F_{\2 2}
   + e^{-\varphi_2 + \varphi_3}\, {*\cF_\2^1 }\wedge\cF_\2^1 + 
     e^{-\varphi_2 -\varphi_3}\, {*\cF_\2^2}\wedge \cF_\2^2\Big)\nn\\
&& - \chi_1\, ( F_{\2 1}\wedge \cF_\2^1 + 
                  F_{\2 2}\wedge \cF_\2^2)\,,
\label{d4lag}
\eea
%%%%%
where the index $i$ labelling the dilatons $\varphi_i$ and axions $\chi_i$
ranges over $1\le i \le 3$.  The four field strengths can be written in 
terms of potentials as
%%%%%
\bea
F_{\2 1} &=& d A_{\1 1} - \chi_2\, d\cA_\1^2\,,\nn\\
F_{\2 2} &=& d A_{\1 2} + \chi_2\, d \cA_\1^1 - 
    \chi_3\, d A_{\1 1} +
      \chi_2\, \chi_3\, d\cA_\1^2\,,\nn\\
\cF_\2^1 &=& d\cA_\1^1 + \chi_3\, d\cA_\1^2\,,\nn\\
\cF_\2^2 &=& d\cA_\1^2\,.\label{4Fs}
\eea
%%%%%
The field strengths here are not in the same duality frame as the one we
have assumed in our discussions in this paper however.  To convert from
(\ref{d4lag}) and (\ref{4Fs}) to the frame we are using, one would need
to dualise the field strengths $\cF_\2^1$ and $\cF_\2^2$, and if then 
written explicitly, the resulting Lagrangian would be rather cumbersome.
Alternatively, one could simply exchange the roles of the electric and
magnetic charges for the field strengths $\cF_\2^1$ and $\cF_\2^2$,
and work with (\ref{d4lag}) without performing any dualisations.  For
example, the 4-charge black hole solutions that we refer to in this paper
as having four electric charges would, as solutions in terms of the
fields in (\ref{d4lag}), instead comprise two electric and two magnetic
charges.  (As for example, in the presentation of these solution in
\cite{Chong}.)


\begin{thebibliography}{99}

%%CITATION = doi:10.1007/BF02345020, 10.1007/BF01608497;%%

\bibitem{Curir} A. Curir, {\it Spin entropy of a rotating black hole,}
 {\it  Il Nuovo Cimento B } {\bf 52} (1979) 262-266.

\bibitem{CFrancaviglia1} A. Curir and M. Francaviglia, {\it Spin thermodynamics
  of a Kerr black hole, } {\it  Il Nuovo Cimento B }  {\bf 52} (1979) 165-176.

\bibitem{CFrancavilia2} A. Curir and M. Francaviglia,
{\it   On certain transformations for black-holes energetics,}
{\it Atti Acad Naz. Lincei }Ser VIII {\bf 61} (1976) 448.


\bibitem{CalvaniFrancaviglia}M. Calvani and M. Francaviglia,
 {\it Irreducible mass, unincreasable angular momentum and isoareal
 transformations for black hole physics,}
  {\it Acta Phys Polonica B}{\bf 9} (1978) 11-14.


\bibitem{CFrancaviglia3} A. Curir and M. Francaviglia,
{\it  Isoareal transformations of the Kerr-Newman black holes,}
 {\it Acta Phys Polonica B} {\bf 9} (1978) 3-10.


\bibitem{Curir:1985wn}
  A.~Curir,
{\it On the energy emission by a Kerr black hole in the superradiant range,}
  Phys.\ Lett.\  {\bf 161B} (1985) 310.
  doi:10.1016/0370-2693(85)90768-3

\bibitem{Curir:1986irp}
  A.~Curir,
 {\it On the generalized second law for rotating black holes,}
  Phys.\ Lett.\ B {\bf 176} (1986) 26.
  doi:10.1016/0370-2693(86)90918-4

\bibitem{CurirC}   A.~Curir,
{\it Convexity of thermodynamic functions and thermodynamics of rotating black holes,}
Europhys.\ Lett.\ {\bf 9} (1989) 609-612.


\bibitem{Curir:2011zza}
  A.~Curir,
 {\it Entropic forces in a Kerr geometry: a link with rotational properties,}
  Commun.\ Theor.\ Phys.\  {\bf 55} (2011) 594.
  doi:10.1088/0253-6102/55/4/12

 \bibitem{OK}I.~Okamoto and O.~Kaburaki, {\it The €˜inner-horizon
  thermodynamics€?of Kerr black holes,}
Mon.\ Not\. R.\ Ast.\ Soc.
{\bf 255} (1992), 539 (1992)

\bibitem{CYI} M.~Cveti\v c and D.~Youm,
{\it Entropy of non-extreme charged rotating black holes in string theory},
Phys.\ Rev.\ D {\bf 54}, 2612 (1996)
  doi:10.1103/PhysRevD.54.2612
  [hep-th/9603147].

\bibitem{CYII}
 M.~Cveti\v c and D.~Youm,
{\it General rotating five dimensional black holes of toroidally compactified
heterotic string},
Nucl. Phys. {\bf B476}, 118 (1996)
doi:10.1016/0550-3213(96)00355-0
  [hep-th/9603100].

\bibitem{CveticTseytlin}
M.~Cveti\v c and A.A.~Tseytlin,
 {\it Solitonic strings and BPS saturated dyonic black holes,}
  Phys.\ Rev.\ D {\bf 53}, 5619 (1996)
  Erratum: [Phys.\ Rev.\ D {\bf 55}, 3907 (1997)]
  doi:10.1103/PhysRevD.53.5619, 10.1103/PhysRevD.55.3907
  [hep-th/9512031].

\bibitem{Larsen}
  F.~Larsen,
{\it A string model of black hole microstates},
Phys. Rev. {\bf D56} (1997) 1005
doi:10.1103/PhysRevD.56.1005
  [hep-th/9702153].


\bibitem{CLII} M.~Cveti\v c and F.~Larsen,
  {\it General rotating black holes in string theory: Grey body factors and event horizons,}
  Phys.\ Rev.\ D {\bf 56}, 4994 (1997)
  doi:10.1103/PhysRevD.56.4994
  [hep-th/9705192].

\bibitem{CLIII}
M.~Cveti\v c and F.~Larsen,
{\it Grey body factors for rotating black holes in four-dimensions,}
  Nucl.\ Phys.\ B {\bf 506}, 107 (1997)
  doi:10.1016/S0550-3213(97)00541-5
  [hep-th/9706071].

\bibitem{Castro:2010fd}
  A.~Castro, A.~Maloney and A.~Strominger,
 Hidden conformal symmetry of the Kerr black hole,
  Phys.\ Rev.\ D {\bf 82} (2010) 024008
  doi:10.1103/PhysRevD.82.024008
  [arXiv:1004.0996 [hep-th]].

\bibitem{Cvetic:2010mn}
  M.~Cveti\v c, G.W.~Gibbons and C.N.~Pope,
 {\it  Universal area product formulae for
  rotating and charged black holes in four and higher dimensions,}
  Phys.\ Rev.\ Lett.\  {\bf 106} (2011) 121301
  doi:10.1103/PhysRevLett.106.121301
  [arXiv:1011.0008 [hep-th]].

\bibitem{Visser:2012wu}
  M.~Visser,
 {\it Area products for stationary black hole horizons,}
  Phys.\ Rev.\ D {\bf 88} (2013) no.4,  044014
  doi:10.1103/PhysRevD.88.044014
  [arXiv:1205.6814 [hep-th]].

\bibitem{Cvetic:2013eda}
  M.~Cveti\v c, H.~L{\"u} and C.N.~Pope,
 {\it Entropy-product rules for charged rotating black holes,}
  Phys.\ Rev.\ D {\bf 88} (2013) 044046
  doi:10.1103/PhysRevD.88.044046
  [arXiv:1306.4522 [hep-th]].

\bibitem{Page:2015gia}
  D.N.~Page and A.A.~Shoom,
 {\it Universal area product for black holes: A heuristic argument,}
  Phys.\ Rev.\ D {\bf 92} (2015) no.4,  044039
  doi:10.1103/PhysRevD.92.044039
  [arXiv:1504.05581 [hep-th]].



\bibitem{Castro:2012av}
  A.~Castro and M.J.~Rodriguez,
{\it  Universal properties and the first law of black hole inner mechanics,}
  Phys.\ Rev.\ D {\bf 86} (2012) 024008
  doi:10.1103/PhysRevD.86.024008
  [arXiv:1204.1284 [hep-th]].

\bibitem{sqwuneg} S.Q.~Wu,
{\it New formulations of first law of black hole thermodynamics: 
A ``Stringy'' analogy},''
  Phys.\ Lett.\ B {\bf 608}, 251 (2005)
  doi:10.1016/j.physletb.2005.01.018
  [gr-qc/0405029].
  %%CITATION = doi:10.1016/j.physletb.2005.01.018;%%

\bibitem{huan} Y.H.~Wei,
{\it Effective first law of thermodynamics of black holes with two horizons},
  Chin.\ Phys.\ B {\bf 18}, 821 (2009).
  doi:10.1088/1674-1056/18/2/068

\bibitem{mipark1} M.I.~Park,
{\it Can Hawking temperatures be negative?},
  Phys.\ Lett.\ B {\bf 663}, 259 (2008)
  doi:10.1016/j.physletb.2008.04.009
  [hep-th/0610140].
  %%CITATION = doi:10.1016/j.physletb.2008.04.009;%%

\bibitem{mipark2} M.I.~Park,
{\it Thermodynamics of exotic black holes, negative temperature, and 
Bekenstein-Hawking entropy},
  Phys.\ Lett.\ B {\bf 647}, 472 (2007)
  doi:10.1016/j.physletb.2007.02.036
  [hep-th/0602114].

\bibitem{mipark3} M.I.~Park,
{\it Note on the area theorem},
  Int.\ J.\ Mod.\ Phys.\ A {\bf 24S1}, 3111 (2009).
  doi:10.1142/S0217751X09044267

\bibitem{Gibbons:1977mu}
  G.W.~Gibbons and S.W.~Hawking,
{\it Cosmological wvent horizons, thermodynamics, and particle creation,}
  Phys.\ Rev.\ D {\bf 15} (1977) 2738.
  doi:10.1103/PhysRevD.15.2738


\bibitem{Frolov:2002va}
  A.V.~Frolov and L.~Kofman,
{\it Inflation and de Sitter thermodynamics,}
  JCAP {\bf 0305} (2003) 009
  doi:10.1088/1475-7516/2003/05/009
  [hep-th/0212327].

\bibitem{Dolan:2013ft}
  B.P.~Dolan, D.~Kastor, D.~Kubiznak, R.B.~Mann and J.~Traschen,
{\it Thermodynamic volumes and isoperimetric inequalities for de Sitter black holes,}
  Phys.\ Rev.\ D {\bf 87} (2013) no.10,  104017
  doi:10.1103/PhysRevD.87.104017

\bibitem{Gregory:2017sor}
  R.~Gregory, D.~Kastor and J.~Traschen,
{\it black hole thermodynamics with dynamical lambda,}
  arXiv:1707.06586 [hep-th].

\bibitem{Bousso:1996au}
  R.~Bousso and S.W.~Hawking,
{\it Pair creation of black holes during inflation},
  Phys.\ Rev.\ D {\bf 54}, 6312 (1996),
  doi:10.1103/PhysRevD.54.6312,
gr-qc/9606052.

\bibitem{Gibbs}J.W.~Gibbs, 
{\it On the equilibrium of heterogeneous substances,}
  Transactions of the Connecticut Academy of Arts and Sciences, {\bf 3}
  (1874-1878), pages 104-248 and 343-524.

\bibitem{Christodoulou}
  D.~Christodoulou, {\it Reversible and irreversible
  transformations in black hole physics,}
  Phys.\ Rev.\ Lett.\  {\bf 25}  (1970) 1596-1597,
 http://dx.doi.org/10.1103/PhysRevLett.25.1596

\bibitem{Hawking:1971tu}
  S.W.~Hawking,
{\it Gravitational radiation from colliding black holes},
  Phys.\ Rev.\ Lett.\  {\bf 26} (1971) 1344.
  doi:10.1103/PhysRevLett.26.1344

\bibitem{Hawking:1971vc}
  S.W.~Hawking,
{\it Black holes in general relativity},
  Commun.\ Math.\ Phys.\  {\bf 25} (1972) 152.
  doi:10.1007/BF01877517

\bibitem{Christodoulou:1972kt}
  D.~Christodoulou and R.~Ruffini,
 {\it Reversible transformations of a charged black hole,}
  Phys.\ Rev.\ D {\bf 4} (1971) 3552.
  doi:10.1103/PhysRevD.4.3552

\bibitem{Smarr:1972kt}
  L.~Smarr,
{\it  Mass formula for Kerr black holes,}
  Phys.\ Rev.\ Lett.\  {\bf 30} (1973) 71
   Erratum: [Phys.\ Rev.\ Lett.\  {\bf 30} (1973) 521].
  doi:10.1103/PhysRevLett.30.71

\bibitem{Bardeen:1973gs}
  J.M.~Bardeen, B.~Carter and S.W.~Hawking,
{\it The four laws of black hole mechanics},
  Commun.\ Math.\ Phys.\  {\bf 31}, 161 (1973).
doi:10.1007/BF01645742
%%CITATION = doi:10.1007/BF01645742;%%


\bibitem{hawkingvar} S.W.~Hawking,
{\it A variational principle for black holes},
  Commun.\ Math.\ Phys.\  {\bf 33}, 323 (1973).
doi:10.1007/BF01646744
%%CITATION = doi:10.1007/BF01646744;%%

\bibitem{dain} S.~Dain,
{\it A variational principle for stationary, axisymmetric solutions of
Einstein's equations},
  Class.\ Quant.\ Grav.\  {\bf 23}, 6857 (2006),
  doi:10.1088/0264-9381/23/23/016,
gr-qc/0508061.
  %%CITATION = doi:10.1088/0264-9381/23/23/016;%%

\bibitem{ancvpa} O.S.~An, M.~Cveti\v c and I.~Papadimitriou,
{\it Black hole thermodynamics from a variational principle: Asymptotically
conical backgrounds},
  JHEP {\bf 1603}, 086 (2016),
  doi:10.1007/JHEP03(2016)086,
arXiv:1602.01508 [hep-th].
%%CITATION = doi:10.1007/JHEP03(2016)086;%%

\bibitem{Bardeen:1970vja}
  J.M.~Bardeen,
{\it A variational principle for rotating stars in general relativity},
  Astrophys.\ J.\  {\bf 162}, 71 (1970).
doi:10.1086/150635
%%CITATION = doi:10.1086/150635;%%

\bibitem{Geroch} Colloquium at Princeton University (Dec. 1971).

\bibitem{Bekenstein:1972tm}
J.D.~Bekenstein,
{\it Black holes and the second law},
  Lett.\ Nuovo Cim.\  {\bf 4}, 737 (1972).
  doi:10.1007/BF02757029
%%CITATION = doi:10.1007/BF02757029;%%


\bibitem{Hawking:1974rv}
S.W.~Hawking,
{\it Black hole explosions},
  Nature {\bf 248}, 30 (1974).
  doi:10.1038/248030a0
%%CITATION = doi:10.1038/248030a0;%%

\bibitem{Hawking:1974sw}
S.W.~Hawking,
{\it Particle creation by black holes},
  Commun.\ Math.\ Phys.\  {\bf 43}, 199 (1975),
  Erratum: [Commun.\ Math.\ Phys.\  {\bf 46}, 206 (1976)].
doi:10.1007/BF02345020, 10.1007/BF01608497


\bibitem{Gibbons:1996af}
  G.W.~Gibbons, R.~Kallosh and B.~Kol,
{\it Moduli, scalar charges, and the first law of black hole thermodynamics,}
  Phys.\ Rev.\ Lett.\  {\bf 77} (1996) 4992
  doi:10.1103/PhysRevLett. 77.4992
  [hep-th/9607108].

\bibitem{Weinhold} F.~Weinhold, 
{\it Metric geometry of equilibrium thermodynamics},
J.\ Chem. \ Phys. \ {\bf 63} (1975) 2479,
doi:10.1063/1.431689.


\bibitem{Pagepd} D.N.~Page, {\it Thermodynamic paradoxes},
Physics Today {\bf 30}, 1, 11 (1977), doi:10.1063/ 1.3037360.

\bibitem{hongthermo}
 H.~Liu, H.~L\"u, M.~Luo and K.N.~Shao,
{\it Thermodynamical metrics and black hole phase transitions},
  JHEP {\bf 1012}, 054 (2010),
  doi:10.1007/JHEP12(2010)054
  [arXiv:1008.4482 [hep-th]].
  %%CITATION = doi:10.1007/JHEP12(2010)054;%%
  %33 citations counted in INSPIRE as of 24 Jan 2018

\bibitem{Ferrara:1997tw}
  S.~Ferrara, G.W.~Gibbons and R.~Kallosh,
{\it  Black holes and critical points in moduli space,}
 Nucl.\ Phys.\ B {\bf 500} (1997) 75
 doi:10.1016/S0550-3213(97)00324-6
 [hep-th/9702103].

\bibitem{Aman:2015wsa}
  J.~Aman, I.~Bengtsson and N.~Pidokrajt,
{\it  Thermodynamic metrics and black hole physics,}
  Entropy {\bf 17} (2015) 6503
  doi:10.3390/e17096503
  [arXiv:1507.06097 [gr-qc]].

\bibitem{Davies:1978mf}
  P.C.W.~Davies,
{\it  Thermodynamics of black holes,}
  Proc.\ Roy.\ Soc.\ Lond.\ A {\bf 353} (1977) 499.
  doi:10.1098/rspa.1977.0047

\bibitem{Monteiro:2008wr}
  R.~Monteiro and J.E.~Santos,
{\it  Negative modes and the thermodynamics of Reissner-Nordstrom black holes,}
  Phys.\ Rev.\ D {\bf 79} (2009) 064006
  doi:10.1103/PhysRevD.79. 064006
  [arXiv:0812.1767 [gr-qc]].

\bibitem{1411.2582}
S.A.H.~Mansoori, B.~Mirza and M.~Fazel,
{\it Hessian matrix, specific heats, Nambu brackets, and thermodynamic
geometry},
  JHEP {\bf 1504}, 115 (2015)
  doi:10.1007/JHEP04(2015) 115
  [arXiv:1411.2582 [gr-qc]].
  %%CITATION = doi:10.1007/JHEP04(2015)115;%%

%\cite{Huang:2016fks}
\bibitem{Huang:2016fks}
  H.~Huang, X.H.~Feng and H.~L\"u,
``Holographic complexity and two identities of action growth,''
  Phys.\ Lett.\ B {\bf 769}, 357 (2017)
  doi:10.1016/j.physletb.2017.04.011
  [arXiv:1611.02321 [hep-th]].
  %%CITATION = doi:10.1016/j.physletb.2017.04.011;%%
  %17 citations counted in INSPIRE as of 15 Jun 2018

\bibitem{Ramsey} N.F.~Ramsey, {\it Thermodynamics \& statistical mechanics
at negative absolute temperatures,} Phys. \ Rev.\ {\bf 103} (1956) 20-28.

\bibitem{Abram} M.A.~Abramowicz, private communication.

\bibitem{Gibbons:1982fy}
  G.W.~Gibbons and C.M.~Hull,
{\it A Bogomolnyi bound for general relativity and solitons in $N=2$ supergravity,}
  Phys.\ Lett.\  {\bf 109B} (1982) 190.
  doi:10.1016/0370-2693(82)90751-1

\bibitem{Chow:2013tia}
  D.D.K.~Chow and G.~Comp{\` e}re,
 {\it  Seed for general rotating non-extremal black holes of
  $\mathcal {N}= 8$ supergravity,}
  Class.\ Quant.\ Grav.\  {\bf 31} (2014) 022001
  doi:10.1088/0264-9381/31/2/022001
  [arXiv:1310.1925 [hep-th]].

\bibitem{horlowmal} G.T.~Horowitz, D.A.~Lowe and J.M.~Maldacena,
{\it Statistical entropy of nonextremal four-dimensional black holes and 
U duality},
  Phys.\ Rev.\ Lett.\  {\bf 77}, 430 (1996)
  doi:10.1103/PhysRevLett.77.430
  [hep-th/9603195].

\bibitem{Sarosi:2015nja}
  G.~S\'arosi,
{\it Entropy of nonextremal STU black holes: The F-invariant unveiled,}
  Phys.\ Rev.\ D {\bf 93} (2016) no.2,  024036
  doi:10.1103/PhysRevD.93.024036
  [arXiv:1508.06667 [hep-th]].

\bibitem{Chong}
  Z.-W.~Chong, M.~Cveti\v c, H.~L\"u and C.N.~Pope,
  {\it Charged rotating black holes in four-dimensional gauged and ungauged supergravities,}
  Nucl.\ Phys.\ B {\bf 717}, 246 (2005)
  doi:10.1016/j.nuclphysb.2005.03.034
  [hep-th/0411045].


%\cite{Lu:2013ura}
\bibitem{Lu:2013ura}
  H.~L\"u, Y.~Pang and C.N.~Pope,
  {\it AdS dyonic black hole and its thermodynamics,}
  JHEP {\bf 1311}, 033 (2013)
  doi:10.1007/JHEP11(2013)033
  [arXiv:1307.6243 [hep-th]].
  %%CITATION = doi:10.1007/JHEP11(2013)033;%%
  %67 citations counted in INSPIRE as of 15 Jun 2018

\bibitem{gibbemd} G.~W.~Gibbons,
{\it Antigravitating black hole solitons with scalar hair in $N=4$ 
supergravity}, 
  Nucl.\ Phys.\ B {\bf 207}, 337 (1982).
  doi:10.1016/0550-3213(82)90170-5
  %%CITATION = doi:10.1016/0550-3213(82)90170-5;%%

\bibitem{Gibbons:1987ps}
  G.W.~Gibbons and K.i.~Maeda,
  {\it Black holes and membranes in higher dimensional
  theories with dilaton fields,}
  Nucl.\ Phys.\ B {\bf 298} (1988) 741.
  doi:10.1016/0550-3213(88)90006-5


\bibitem{gahost} D.~Garfinkle, G.~T.~Horowitz and A.~Strominger,
{\it Charged black holes in string theory},
  Phys.\ Rev.\ D {\bf 43}, 3140 (1991),
  Erratum: [Phys.\ Rev.\ D {\bf 45}, 3888 (1992)].
  doi:10.1103/PhysRevD.43.3140, 10.1103/PhysRevD.45.3888
  %%CITATION = doi:10.1103/PhysRevD.43.3140, 10.1103/PhysRevD.45.3888;%%

\bibitem{gikalototr} G.~W.~Gibbons, D.~Kastor, L.~A.~J.~London, 
P.~K.~Townsend and J.~H.~Traschen,
{\it Supersymmetric selfgravitating solitons},
  Nucl.\ Phys.\ B {\bf 416}, 850 (1994),
  doi:10.1016/0550-3213(94)90558-4,
  [hep-th/9310118].
  %%CITATION = doi:10.1016/0550-3213(94)90558-4;%%

\bibitem{ped} J.~E.~Aman, N.~Pidokrajt and J.~Ward,
{\it On geometro-thermodynamics of dilaton black holes}, 
  EAS Publ.\ Ser.\  {\bf 30}, 279 (2008),
  doi:10.1051/eas:0830044,
  [arXiv:0711.2201 [hep-th]].
  %%CITATION = doi:10.1051/eas:0830044;%%

%\cite{Lu:2013eoa}
\bibitem{Lu:2013eoa}
  H.~L\"u,
{\it Charged dilatonic ads black holes and magnetic AdS$_{D-2} \times R^{2}$ vacua,}
  JHEP {\bf 1309}, 112 (2013)
  doi:10.1007/JHEP09(2013)112
  [arXiv:1306.2386 [hep-th]].
  %%CITATION = doi:10.1007/JHEP09(2013)112;%%
  %25 citations counted in INSPIRE as of 14 Jun 2018


\bibitem{Kastor:2009wy}
  D.~Kastor, S.~Ray and J.~Traschen,
 {\it Enthalpy and the mechanics of AdS black holes,}
  Class.\ Quant.\ Grav.\  {\bf 26} (2009) 195011
  doi:10.1088/0264-9381/26/19/195011
  [arXiv:0904.2765 [hep-th]].

\bibitem{Cvetic:2010jb}
  M.~Cveti{\v c}, G.W.~Gibbons, D.~Kubiznak and C.N.~Pope,
{\it Black Hole Enthalpy and an Entropy Inequality for the Thermodynamic Volume,}
  Phys.\ Rev.\ D {\bf 84} (2011) 024037
  doi:10.1103/PhysRevD.84.024037
  [arXiv:1012.2888 [hep-th]].

\bibitem{Hawking:1982dh}
  S.W.~Hawking and D.N.~Page,
{\it  Thermodynamics of black holes in anti-de sitter space,}
  Commun.\ Math.\ Phys.\  {\bf 87} (1983) 577.
  doi:10.1007/BF01208266

\bibitem{Witten:1998qj}
  E.~Witten,
{\it  Anti-de Sitter space and holography},
  Adv.\ Theor.\ Math.\ Phys.\  {\bf 2} (1998) 253
  [hep-th/9802150].

\bibitem{Prestidge:1999uq}
  T.~Prestidge,
{\it  Dynamic and thermodynamic stability and negative modes in Schwarz- schild-anti-de Sitter,}
  Phys.\ Rev.\ D {\bf 61} (2000) 084002
  doi:10.1103/PhysRevD. 61.084002
  [hep-th/9907163].

\bibitem{Gibbons:1976ue}
  G.W.~Gibbons and S.W.~Hawking,
{\it Action integrals and partition functions in quantum gravity,}
  Phys.\ Rev.\ D {\bf 15} (1977) 2752.
  doi:10.1103/PhysRevD.15.2752

\bibitem{Mellor:1989ac}
  F.~Mellor and I.~Moss,
{\it Stability of black holes in de Sitter space,}
  Phys.\ Rev.\ D {\bf 41} (1990) 403.
  doi:10.1103/PhysRevD.41.403

 \bibitem{Mellor:1989wc}
  F.~Mellor and I.~Moss,
{\it Black holes and gravitational instantons,}
  Class.\ Quant.\ Grav.\  {\bf 6} (1989) 1379.
  doi:10.1088/0264-9381/6/10/008

\bibitem{Mellor:1989gi}
  F.~Mellor and I.~Moss,
{\it  Black holes and quantum wormholes}
  Phys.\ Lett.\ B {\bf 222} (1989) 361.
  doi:10.1016/0370-2693(89)90324-9

\bibitem{Moss:1989js}
 I.~Moss,
{\it Journeys beyond the Cauchy horizon,}
 NCL-89-TP6.

\bibitem{Davies:1989ew}
  P.C.W.~Davies and I.G.~Moss,
{\it  Journey through a black hole,}
  Class.\ Quant.\ Grav.\  {\bf 6} (1989) L173.
  doi:10.1088/0264-9381/6/9/004

\bibitem{Davies:1989ey}
  P.C.W.~Davies,
{\it Thermodynamic phase transitions of {Kerr-Newman} black holes in
  de Sitter space,}
  Class.\ Quant.\ Grav.\  {\bf 6} (1989) 1909.
  doi:10.1088/0264-9381/6/12/018

\bibitem{Caldarelli:1999xj}
  M.M.~Caldarelli, G.~Cognola and D.~Klemm,
 {\it Thermodynamics of Kerr-Newman-AdS black holes and conformal field theories,}
  Class.\ Quant.\ Grav.\  {\bf 17} (2000) 399
  doi:10.1088/0264-9381/17/2/310


\bibitem{WuD5}
S.Q.~Wu,
{\it General nonextremal rotating charged AdS black holes in 
five-dimensional $U(1)^3$ gauged supergravity: A simple construction method},
  Phys.\ Lett.\ B {\bf 707}, 286 (2012)
  doi:10.1016/j.physletb.2011.12.031
  [arXiv:1108.4159 [hep-th]].
  %%CITATION = doi:10.1016/j.physletb.2011.12.031;%%


\bibitem{Birkandan:2014vga}
T.~Birkandan and M.~Cveti\v c,
{\it Wave equation for the Wu black hole,}
JHEP {\bf 1409}, 121 (2014)
doi:10.1007/JHEP09(2014)121

\bibitem{LyndenBell:1998fr}
  D.~Lynden-Bell,
{\it Negative specific heat in astronomy, physics and chemistry,}
  Physica A {\bf 263} (1999) 293
  doi:10.1016/S0378-4371(98)00518-4
  [cond-mat/9812172].

\bibitem{Landsberg0} P.T. Landsberg, {\it The fourth law of thermodynamics},
 Nature {\bf 238} (1972) 229.

\bibitem{Landsberg4}  P.T.~Landsberg, {\it Thermodynamics and black holes}, in
Black Hole Physics,
  edited by Venzo de Sabbata and Zhenjiu Zhang. Dordrecht,
  Netherlands, Kluwer Academic, (1992)  99-146
  (NATO Advanced Study Institute (ASI),
  Series C: Mathematics and Physical Sciences, v. 364)

\bibitem{Rasheed:1997ns}
  D.A.~Rasheed,
{\it Nonlinear electrodynamics: Zeroth and first laws of black hole mechanics},
  hep-th/9702087.

\bibitem{Landsberg1} P.T.~Landsberg and D.~Tranah,
{\it Entropies need not be convex},
Phys. Lett. A {\bf 78}(1980) 219-220

\bibitem{Landsberg2}  P.T.~Landsberg and D. Tranah,
{\it Thermodynamics of non-extensive Entropies I},
Collective Phenomena {\bf 3} (1980) 73-80.

\bibitem{Tranah} D.`Tranah and P.T.~Landsberg,  {\it Thermodynamics of
  non-extensive entropies II},
 Collective Phenomena {\bf 3} (1980) 81-88.

\bibitem{Landsberg3}   P.T.~Landsberg,
{\it Is Equilibrium Always an Entropy Maximum}, J. Statistical Physics
{\bf 35} (1984) 159-169.

\bibitem{larsen} F.~Larsen,
{\it Rotating Kaluza-Klein black holes},
  Nucl.\ Phys.\ B {\bf 575}, 211 (2000)
  doi:10.1016/S0550-3213(00)00064-X
  [hep-th/9909102].
  %%CITATION = doi:10.1016/S0550-3213(00)00064-X;%%

\bibitem{hartlehawking} J.B.~Hartle and S.W.~Hawking,
{\it Path integral derivation of black hole radiance},
  Phys.\ Rev.\ D {\bf 13}, 2188 (1976).
  doi:10.1103/PhysRevD.13.2188.
  %%CITATION = doi:10.1103/PhysRevD.13.2188;%%
  %829 citations counted in INSPIRE as of 01 Jun 2018

\bibitem{dafluk} M.~Dafermos and J.~Luk,
{\it The interior of dynamical vacuum black holes I: The $C^0$-stability 
of the Kerr Cauchy horizon},''
  arXiv:1710.01722 [gr-qc].
  %%CITATION = ARXIV:1710.01722;%%

\end{thebibliography}
\end{document}